\documentclass[
        parskip = half, 
        twoside,
        footinclude = false, 
        ]{scrartcl}

\usepackage{ifthen} 

\newboolean{boldeqnitalic}
\setboolean{boldeqnitalic}{true}
\usepackage{stylefiles/defdb1} 

\usepackage{stylefiles/rublkm_paper} 

\usepackage{stylefiles/defet1} 
\usepackage{stylefiles/nmstyle} 
\usepackage{relsize}

\makeatletter
\newcommand*{\authormark}{}
\newcommand*{\markauthor}[1]{%
  \renewcommand{\authormark}{#1}%
  \ignorespaces
}
\newcommand*{\titlemark}{}
\newcommand*{\marktitle}[1]{%
  \renewcommand{\titlemark}{#1}%
  \ignorespaces
}
\makeatother

\markauthor{Klemens Uhlmann, Daniel Balzani}
\marktitle{Homeostatic Kinematic Growth Model for Arteries - Residual Stresses and Active Response}

\clearpairofpagestyles
\rohead{\pagemark}
\rehead{\pagemark}

\lohead{\titlemark}
\lehead{\authormark}

\KOMAoptions{
   headsepline = true
}
\setkomafont{pagehead}{\normalfont\normalcolor\small}

\newcommand{\figpath}{figures}       

\definecolor{colorbalzani}{rgb}{0.0,0.0,1.0}
\definecolor{coloruhlmann}{rgb}{0.0,0.4,0.0}
\definecolor{colorcyron}{rgb}{0.0,0.8,1.0}
\definecolor{colorcomment}{rgb}{1.0,0.0,0.0}

\AfterEndEnvironment{figure}{\noindent\ignorespaces}

\definecolor{dia6}{RGB}{136,46,114}
\definecolor{dia5}{rgb}{0.0, 0.75, 1.0} 
\definecolor{dia4}{cmyk}{0.75,0.0,1.0,0}
\definecolor{dia3}{RGB}{241,147,45}
\definecolor{dia2}{RGB}{220,5,12}
\definecolor{dia1}{rgb}{0.0,0.2070,0.3750}

\usepackage{pgfplots}
    \pgfplotsset{
        layers/my layer set/.define layer set={
            background,
            main,
            foreground
        }{
        },
        set layers=my layer set,
    } 
\definecolor{capri}{rgb}{0.0, 0.75, 1.0}             

\begin{document}


\clearpage
\setcounter{page}{1}
\begin{center}

{\LARGE Homeostatic Kinematic Growth Model for Arteries - Simulation of Residual Stresses and Active Response}

\vspace{5mm}

Klemens Uhlmann$^{1}$, Daniel Balzani$^{1\star}$

\vspace{3mm}

{\small $^1 \,$Chair of Continuum Mechanics, Ruhr University Bochum,\\ 
Universit\"atsstra{\ss}e~150, 44801~Bochum, Germany}\\[3mm]

\vspace{3mm}

{\small ${}^{\star}$E-mail address of corresponding author: daniel.balzani@rub.de}

\vspace{10mm}

\begin{minipage}{15.0cm}
\textbf{Abstract}\hspace{3mm}
A simple kinematic growth model for muscular arteries is presented which allows the
incorporation of residual stresses such that a homeostatic in-vivo stress state under physiological loading is obtained.
To this end, new evolution equations for growth are proposed, which avoid issues with instability of final growth states known from other kinematric growth models.
These evolution equations are connected to a new set of growth driving forces.
By introducing a formulation
using the principle Cauchy stresses, reasonable in vivo stress distributions can be obtained while ensuring realistic geometries of arteries after growth.
To incorporate realistic Cauchy stresses for muscular arteries under varying loading conditions, which appear in vivo, e.g., due to
physical activity, the growth model is combined with a newly proposed stretch-dependent model for smooth muscle activation.
To account for the changes of geometry and mechanical behavior during the growth process, an optimization procedure is described which leads to a more accurate representation
of an arterial ring and its mechanical behavior after the growth-related homogenization of the stresses is reached.
Based thereon, parameters are adjusted to experimental data of a healthy middle cerebral artery of a rat showing that the proposed model accurately describes real material behavior.
The successful combination of growth and active response indicates that the new growth model can 
be used without problems for modeling active tissues under various conditions.
\end{minipage}
\end{center}

\medskip{}

\section{Introduction}
\label{sec:intro}

Over the last few decades, computational techniques and models to simulate the mechanical behavior of arterial walls have been improved.
However, further developments are needed to put accurate mechanical simulations of diseased arteries on the clinical map.
Since mechanical stresses are believed to significantly contribute to the nucleation and progression of cardiovascular diseases, their accurate description throughout the arterial wall is essential.
This does not only require a reliable model for the passive and active material response, but also for the fibrous microstructure and growth-related processes dominating the residual stress state.
Growth processes are important for the preservation of a healthy mechanical state of the tissue which is understood as mechanical homeostasis~\cite{EicHaePauAydHumCyr:2021:mhi}.
As a consequence, the tissue of the arterial wall maintains a stress distribution which is 
to a large extent homogeneous and
can only be realized by residual stress states, i.e. stresses which are still present in the tissue even
after unloading.
Without such residual stresses, the circumferential stresses, especially close to the lumen, would reach values which are by far not bearable by the cells in the tissue.
There, however, the stress state is particularly important as it strongly influences the progression of atherosclerosis.
Especially in case of degenerative alterations of the endothelium, the stresses in the tissue may influence relevant time-averaged wall shear stresses (TAWSS) \cite{HeKu:1996:pfi,WanUhlVedBalVar:2022:fsi}, which are assumed to severly influence atherosclerosis.
Increased values of TAWSS resulting from damage-induced stress-softening \cite{AntDeeBalMacKem:2019:cod,SchBal:2016:riv} were also found to potentially contribute to a self-amplification of aneurysms \cite{WanBalVedUhlVar:2021:otp}.
Therefore, not including realistic residual stresses in numerical simulations implies completely unphysical results, which emphasizes the great importance of residual stresses.
Also changes of the microstructure in terms of directions of collagen and muscle fibers are considered as remodeling processes, furthermore allowing for almost homogeneous stress states in healthy arteries under varying loading conditions.
For the simulations in this publication, focus is set on the simultaneous application of the growth process and the active contraction of smooth muscle cells (SMCs) in muscular arteries to describe a reliable mechanical state of the arterial wall.
The contraction mechanism of SMCs was described as stretch-dependent in a previous publication~\cite{UhlBal:2023:cmm} which is based on the activation of membrane receptors during mechanical loading of the cells.
Although remodeling processes of the microstructure in terms of fiber reorientation will not be taken into account, realistic fiber directions of collagen and SMCs will be considered based on experimental measurements from the literature.
The influence of variations
of the microstructure on the growth process will be
discussed in Section~\ref{sec:num}.

Growth and remodeling processes of SMCs and the extracellular matrix (ECM), including elastin and collagen, are mainly governed by long-term changes in the mechanical loading as e.g. happening in humans when growing up.
But they can also be triggered by various circumstances such as damage or inflammation of the tissue caused by infectious pathogens, hypertension or intake of antihypertensive medication.
Depending on the cause of the growth and remodeling process, the proliferation of the tissue cells and protein synthesis are activated differently.
In a healthy state of the artery, vascular SMCs actively synthesize, secrete, modulate and maintain the ECM to provide elasticity to the blood vessel.
The maintenance of the tissue is supported by fibroblasts which mainly exist in the adventitia.
When damage occurs to the tissue, various cell types can transition into myofibroblasts which generate collagen-rich scar tissue to repair the wound~\cite{BagMohPun:2017:mmo}.
Under the influence of hypertension, contractile SMCs can change their phenotype into fibroblast-like and mesenchymal-like SMCs in which these cells lose their ability to actively contract~\cite{YapMieVriMicWaa:2021:sso}.
Instead, fibroblast-like SMCs synthesize more proteins for the ECM, while mesenchymal-like SMCs are characterized by the ability to proliferate and self-renew which increases the amount of SMCs in the tissue.
In a state of inflammation of the tissue, contractile SMCs can change their phenotype to macrophage-like SMCs which play a central role in healing and recruiting other immune cells to initiate an appropriate immune response.
Especially in muscular arteries where the concentration of nitric oxide is lower than in large, elastic arteries, macrophage-like SMCs can lead to the formation of foam cells in atherosclerotic plaques~\cite{BonCorLerPomGau:2021:pmo}.
The process of growth and remodeling of the tissue of the arterial wall is complex and also strongly dependent on the size and type of artery.
Large arteries, such as the aorta, contain more elastin and stiffen with age due to the production of collagen~\cite{KelHayAvoOro:1989:ndo}.
In contrast, the ratio of elastin to collagen might possibly even increase in aging muscular arteries which could be connected to the higher density of SMCs and their ability to synthesize elastin~\cite{HaySteMeiGreRac:2001:tit}.
The complexity of these processes is obviously high and should be premeditated when making simplifying assumptions to obtain a reasonable tradeoff between model accuracy, simulation efficiency, and model complexity with regard to the availability of data for the parameters.
Especially if the main interest is 
only in obtaining a reasonably grown state of the artery as it is important for incorporating a reasonable residual stress state, rather than an accurate description of the time-evolution of biological processes, complicated processes as described above may not be included in the model and focus should be put on mechanical and/or chemo-mechanical, phenomenological descriptions.\\
A general overview of growth and remodeling models was given in several papers over the last decade~\cite{AmbAmaCyrDesGorHumKuh:2019:gar,CyrHum:2017:gar,CyrWilHum:2017:cff,Sae:2016:ott,Kuh:2014:gma,MenKuh:2012:fig}.
In principle, two different growth theories have been established as widely accepted: the constrained mixture model and the kinematic growth theory.
The constrained mixture model was proposed by Humphrey and Rayagopal in 2002~\cite{HumRaj:2002:acm} which assumes that each volume element contains a mixture of several structurally significant constituents where each constituent was deposited within the body at a different time and possesses an individual stress-free configuration.
Many authors focused on this growth theory recently, since this approach is 
suitable to capture the details of the biological growth processes~\cite{LatHum:2018:ame,FamVasFehMaeVanRegMouAvr:2018:nso,MouAvr:2017:pss,ValHum:2009:eof}.
However, complex models in general require comprehensive and complex data for a meaningful calibration.
In addition to that, tracking the configurations of all constituents increases the computational cost and the effort considerably even if massive parallel solution strategies as e.g. proposed in~\cite{BalBraKlaRheSch:2010:otm} are considered, which may limit practical applications.
%

To reduce computational cost, different concepts have been proposed~\cite{FamVasFehMaeVanRegMouAvr:2018:nso,CyrAydHum:2016:ahc,BraSeiAydCyr:2017:hcm}.
Specifically, homogenized constrained mixtured models \cite{BraSeiAydCyr:2017:hcm,CyrAydHum:2016:ahc} form a bridge to the second major branch of growth and remodeling models, the so-called kinematic growth models suggested by Rodriguez et al.~\cite{RodHogMcc:1994:sdf} in 1994.
In the kinematic growth approach, the multiplicative split of the deformation gradient into an elastic and a growth and remodeling part is not applied to different constituents but to the entire tissue at a material point at once.
This theory has been applied and extended by many authors, see e.g.~\cite{ImaMau:2002:acm,LubHog:2002:otm,HimKuhMenSte:2005:cmo,KuhMaaHimMen:2007:cmo,GokAbiKuh:2010:aga,GokAbiKuh:2010:aga,SaePenMarKuh:2014:cmo,LeeGenAceOrdGucKuhl:2015:acm,ZahBal:2018:acg,LamHolBreJocRee:2022:ama}.
The research on kinematic growth has especially focused on the correct definition of the driving forces and the growth directions.
It is still not guaranteed whether stretches or stresses of the tissue have to be considered as the mechanical quantity which triggers tissue growth.
While stretch 
has been observed to play an important role for the activation of growth factors such as TGF-$\beta$~\cite{Hin:2015:tem},
using stress-related quantities as driving forces~\cite{KuhHol:2007:acm,TabHum:2001:smg,ComCarGioPerShe:2018:mti} is motivated by the fact that it remains unclear how and to which extent cells can sense stretch on the microscale.
In concern of the growth direction, an anisotropic growth process can be assumed where the intensity depends on the type of artery and the corresponding homeostatic stress state.
In this context, based on the general formulation given in \cite{ZahBal:2018:acg}, Anna Zahn focused her dissertation~\cite{Zah:2020:mog} on the identification of the mathematically optimal kinematic growth mechanism by investigating various combinations of growth directions which depended on the eigenvectors of the principal stresses of the elastic part of the Mandel stress.
The objective function in this optimization process was defined to ensure minimal volumetric change of the tissue and a homogeneous stress distribution.
As a result, growth into the direction of the second and third principal stress was identified as advantageous, however, growth only in the direction of the third principle eigenstress was found insufficient.
When considering a hollow cylinder, which is often considered as basic geometry of a healthy artery in numerical simulations, the directions of the first, second, and third principle stress coincide with the circumferential, axial, and radial direction, respectively.
%
%
In contrast to \cite{ZahBal:2018:acg}, the growth model proposed here relies on the Cauchy stress as foundation for the driving forces to more appropriately allow for a model in line with the homeostasis hypothesis.
Initially, we adopt the results presented in~\cite{Zah:2020:mog} to apply growth in the directions of the second and third principal stress of a material point which is explained in section~\ref{sec:model}.
By applying a novel optimization procedure, we are able to automatically address three major aspects of the mechanical simulation to describe the structural problem of an artery: (i)~an accurate identification of material parameters to fit experimental data for a middle cerebral artery of a rat~\cite{JohElyTakWalWalCol:2009:csv}, (ii)~an accurate identification
of growth parameters to achieve homeostatic stress distributions, and (iii)~a resulting geometry of the artery after growth which matches experimental measurements~\cite{GanVanJerGriHamDru:2008:ipi,BelKunMon:2013:baf}.
The objective function is designed to find the minimal volumetric change by the growth process while a homogeneous stress distribution is applied during the loading of an artery at a static pressure level.
In addition, the reference geometry of the arterial ring is not pre-defined but part of the optimization.
After finishing the growth process, the loaded geometry of the arterial ring is evaluated by the objective function to match the inner and outer radii measured by~\cite{GanVanJerGriHamDru:2008:ipi}.
The optimization procedure is explained in detail in section~\ref{sec:opt} where results for a first example, which includes only the passive material response, are discussed.
In Section~\ref{sec:num}, we show that the combination of the proposed growth model with the optimization procedure is applicable to varying directions of collagen fibers and loading scenarios in axial direction.
To limit the computational time of the optimization processes, the model for SMCs is not included in the analysis of the growth model in Section~\ref{sec:num} yet.
Based on these simulation results, small changes to the growth model are discussed and applied in Section~\ref{sec:act} to enable more appropriate simulations where growth and active SMC response are activated simultaneously.
The optimization procedure for this simulation protocol includes a fitting of the axial prestretch of the arterial wall.
In consequence, the final simulation results show a reasonable mechanical description of a healthy middle cerebral of a rat.

\section{Mechanical Model of the Arterial Wall}
\label{sec:model}

In this Section, the new kinematic growth model is introduced which bases on the simple idea that there exists a specific value of homeostatic stress level in axial and circumferential direction.
Accordingly, these values constitute the limit scenarios in the driving forces of the growth evolution.
Based on the concept of kinematic growth theory, the deformation gradient~$\bF = \bF_{\text{e}} \bF_{\text{g}}$ is multiplicatively split into an elastic part $\bF_{\text{e}}$ and a growth part~$\bF_{\text{g}}$.
Hence, growth described by~$\bF_{\text{g}}$ increases the reference volume at the material point by a factor of~$J_{\text{g}} = \text{det} \left( \bF_{\text{g}} \right)$.
By mapping vector elements from the reference configuration using~$\bF_{\text{g}}$, an intermediate configuration is obtained, which represents the fictitious state that would be obtained if growth could evolve without any resistance.
Due to mechanical loading and kinematic constraints resulting from e.g. Neumann and Dirichlet boundary conditions in a boundary value problem, growth is restricted and elastic strains occur inducing mechanical stresses.
Therefore, the strain-energy density function~$\Psi (\bC_{\text{e}}) = \Psi (\bC, \bF_{\text{g}})$ is defined to depend on the elastic part of the right Cauchy-Green tensor~$\bC_{\text{e}} = \bF_{\text{e}}^{\text{T}} \bF_{\text{e}}$.
Note that in this section we focus on describing the details of the model for growth and the passive response, although the combination with the smooth muscle model from~\cite{UhlBal:2023:cmm} is analyzed in Section~\ref{sec:act}.
Recapitulating the smooth muscle model here would make the paper unnecessarily long and thus, for details we refer the reader to the original paper.

\subsection{Anisotropic, Kinematic Growth Model based on Cauchy Stresses}
\label{subsec:grow}

Similar to~\cite{ZahBal:2018:acg}, we consider growth to happen mainly in directions defined by the eigenvectors of the stress tensor.
However, here we focus on the eigenvectors of the Cauchy stress tensor~$\Bsigma$ instead of the Mandel stresses as they are given in the real physical loading state and may thus more appropriately define suitable comparative measures for homeostatic stresses.
In principle, following \cite{ZahBal:2018:acg}, we could thus formulate an anisotropic growth model which considers a different growth intensity for different directions and/or planes described by the three stress eigenvectors.
Therefore, the growth tensor itself is already defined as multiplicative split into three parts
\eb
\bF_{\text{g}} = \bF_{\text{g}}^{(3)} \bF_{\text{g}}^{(2)} \bF_{\text{g}}^{(1)} \, ,
\label{eq:defoGrad_growth}
\ee
where~$\bF_{\text{g}}^{(1)}$, $\bF_{\text{g}}^{(2)}$ and~$\bF_{\text{g}}^{(3)}$ define growth in the direction of the first, second, and third principal direction of the Cauchy stress tensor, which are as usual ordered by their values from the maximum to the minimum.
Note that, based on the formulation of growth in principal directions of the Cauchy stress, all parts~$\bF_{\text{g}}^{(a)}$ with~$a = 1,2,3$ are symmetric and equally diagonalizable, and hence, the tensor multiplication in Eq.~\eqref{eq:defoGrad_growth} is commutative.
In the dissertation of Zahn~\cite{Zah:2020:mog}, various combinations of growth directions and driving forces were investigated including isotropic growth and growth in the directions of and perpendicular to the eigenvectors.
It was found that growth in the directions of the second and third principal stress is a numerically and biologically advantageous combination as stresses could be homogenized best with a minimum effort of growth and minimum sensitivity to axial loading.
The advantages of this growth mechanism can conceptually be explained by considering the dominant load scenario of an arterial wall.
In an idealized arterial geometry as a hollow cylinder, which is also used in the simulations shown later, the directions of the second and third principal stresses are equal to the axial and radial directions, respectively.
The arterial wall is loaded with an intravascular pressure mainly leading to the highest principle stress to appear in circumferential direction.
In addition, it is known that arteries can shorten substantially in axial direction after being cut out of the body, cf.~\cite{HorAdaGulZitVesChlKon:2011:cba}.
This implies that arteries stand under a prestretch in axial direction which influences the axial stress value.
In addition to that and depending on the kind of artery, arteries can be stretched in axial direction resulting from kinematic constraints of the surrounding, e.g. carotid arteries when moving the head sideways.
This mainly corresponds to a displacement-driven process leading to stresses in axial direction whereas the intravascular pressure represents a force-driven process resulting in circumferential stresses.
From a mechanical point of view, a displacement-controlled stress value can be reduced by growth into the stretched direction which can be identified as the axial direction.
In contrast, a force-controlled stress value in circumferential direction can be reduced by growth in perpendicular direction, i.e. in axial and radial direction.
This also explains that growth purely in radial direction or isotropic growth, as often considered in the literature, are not sufficient to suitably reduce stress peaks and homogenize the stress distribution.
Therefore, here we focus on growth in the directions of principle stresses described by
\eb
\bF_{\text{g}}^{(a)} = \bI + \left( \vartheta^{(a)} - 1 \right) \bn^{(a)} \otimes \bn^{(a)} \, ,
\ee
where~$\vartheta^{(a)}$ is the growth factor and~$\bn^{(a)}$ is the eigenvector of the Cauchy stress tensor~$\Bsigma$ with~$a = 1,2,3$.
By definition, $\bn^{(a)}$ are perpendicular to each other which ensures the commutativity of the parts of the growth tensor and its symmetry.
The growth factors~$\vartheta^{(a)}$ are defined by a set of evolution equations which depend on the Cauchy stress due to the driving force~$\phi^{(a)}(\Bsigma)$.
Based on the results from~\cite{Zah:2020:mog} and the reasoning described above, growth in the direction of the first principal stress is not considered for the investigation of the growth model and hence, the evolution equation for~$\vartheta^{(1)}$ is assumed to be~$\dot{\vartheta}^{(1)} = 0$.
The evolution equations for the second and third principal direction are defined as
\eb
\label{eq:groFac}
\dot{\vartheta}^{(a)} = \kappa_\vartheta^{(a)} \left(\phi^{(a)} - \phi^{(a)}_{\text{con}} \right) \quad \text{with} \quad a = 2,3 \, .
\ee
Here, $\phi^{(a)}_{\text{con}}$ defines the specific value which the driving force~$\phi^{(a)}$ converges to during the growth process, and~$\kappa_\vartheta^{(a)}$ defines a growth velocity factor.
Since the driving force will be formulated in terms of the Cauchy stresses, these evolution equations will already try to accomodate a certain homeostatic stress state given by~$\phi{(2)}_{\text{con}}$ and~$\phi{(3)}_{\text{con}}$.
However, the first principal stress is not yet directly controlled and thus, a suitable definition of the driving forces is key.
As explained above, the growth induced by~$\bF_{\text{g}}^{(2)}$ aims to adjust the displacement-influenced part of the stresses in axial direction.
Therefore, we define the axial Cauchy stresses as the driving force for the growth process in this direction.
Since growth in circumferential direction is not considered for the reasons mentioned above, only the growth process described by~$\bF_{\text{g}}^{(3)}$ is left to reduce the first principal stress value (i.e. circumferential stresses in the idealized geometry).
However, to allow an adaption of the circumferential stress state, the circumferential stresses should be included in the driving force.
Therefore, the driving force for the growth tensor~$\bF_{\text{g}}^{(3)}$ is defined as the isotropic value of the Cauchy stress as suggested in~\cite{Zah:2020:mog}.
Consequently, this leads to defining the driving forces~$\phi^{(2)}(\Bsigma)$ and~$\phi^{(3)}(\Bsigma)$ as
\eb
\label{eq:drivingForce}
\phi^{(2)}(\Bsigma) = \Bsigma : \left( \bn^{(2)} \otimes \bn^{(2)} \right) \qquad \text{and} \qquad
\phi^{(3)}(\Bsigma) = \text{tr}(\Bsigma) \, ,
\ee
where~$\bn^{(2)}$ is the second eigenvector and~$\text{tr}(\BSigma)$ the trace of the Cauchy stress tensor.
The driving force for growth in the direction of the third principal stress is chosen this way in this section because it is rather established in the literature to consider growth in radial direction depending on the first invariant of the stress tensor.

{\itshape Remark:} Note that although the reasoning behind this set of evolution equations and driving forces appears well suited for a homeostatic kinematic growth model, it may be questioned from a biological perspective.
Already the media of an artery is made of many concentric layers with helically arranged fibers crosswise changing their fiber angle from layer to layer.
Therefore, the microstructural composition in radial direction differs from the circumferential and axial direction.
Furthermore, in contrast to the circumferential-axial plane, the principal stress value in radial direction will always be negative for equlibrium reasons in arteries under intravascular pressure.
It is thus only natural to assume a specific growth mechanism in this radial direction.
For the axial and circumferential directions, however, the microscopic composition is actually quite similar in these two directions and so is the character of the principal stress values.
Hence, it is not clear why the growth mechanism in each of these two directions should be different.
In addition to that, growth in radial direction should probably not depend on the stress in radial direction as it has a negative value and is dominated by equilibrium, not by the potentially artery-specific composition and properties.
As the particular growth model introduced here did also not perform optimally when combining growth with the active response from SMCs, an alternative formulation will be introduced and analyzed in Section~\ref{sec:act}.

\subsection{Passive Material Response}
\label{subsec:pass}

For the description of the material behavior we consider the model proposed in \cite{UhlBal:2023:cmm}.
Therein, an additive decomposition of the strain energy density function~$\Psi$ into an isotropic part~$\Psi_{\text{isot}}$ for the elastin-rich matrix, two anisotropic parts~$\Psi_{\text{ti}}^{(1)}$ and~$\Psi_{\text{ti}}^{(2)}$ for the embedded collagen fibers and two additional anisotropic parts for the active SMC response.
For now, we exclude the active material response to reduce the computing time when first analyzing the growth model itself in Sections~\ref{sec:opt} and~\ref{sec:num}.
The inclusion of the active model and its implications on the growth model are described and investigated in Section~\ref{sec:act}.
The two fiber directions are described with the concept of structural tensors, cf. e.g.~\cite{Boe:1987:itt}, where the inner structure of the fiber-reinforced material is described by so-called structural tensors~$\bM^{(f)}=\ba^{(f)} \otimes \ba^{(f)}$ with the fiber direction~$\ba^{(f)}$ arranged helically around the vessel wall.
In addition, a gradient of the fiber angle along the radial direction will be considered in the structural problems to obtain a qualitatively more realistic distribution of fiber orientation.
%
The additive decomposition of~$\Psi$ is written as
\eb
\label{eq:stain-energy}
\Psi = \Psi_{\text{p,isot}} + \sum_{f=1}^2 \Psi_{\text{p}, \, \text{ti}}^{(f)}  \, .
\ee
For a more specific definition of the strain-energy density, a coordinate-invariant representation in terms of the principal and mixed invariants
\eb
I_1 = \text{tr} (\bC_{\text{e}}) \, , \quad I_3 = \text{det} (\bC_{\text{e}}) \, , \quad I_{4}^{(f)} = \bC_{\text{e}} \bM^{(f)} \, , \quad I_{5}^{(f)} = \bC_{\text{e}}^2\colon \bM^{(f)} \,
\label{eq:invar}
\ee
is considered.
To a priori ensure material stability in the sense that only real wave speeds occur in the material and no artificial, uncontrolled singular surface nucleate, we consider a polyconvex strain energy density.
Thereby, we also guarantee weak sequential lower semincontinuity, a mathematical property essentially needed for appropriate solutions of discretized forms of the equilibrium equations.
For the definition of polyconvexity and its implications see \cite{Bal:1977:cca}.
Since the fivth invariant is not polyconvex, the alternative invariant for the transversely isotropic part introduced in~\cite{SchNef:2003:ifo} given as
\eb
K_{3}^{(f)} = I_1 I_{4}^{(f)} - I_{5}^{(f)} \,
\label{eq:K3}
\ee
is considered.
For the isotropic part of the energy, we use the isochoric Neo-Hooke model where an additional volumetric penalty function is applied to ensure a nearly incompressible behavior of the arterial tissue.
Deviations from the incompressible state are punished by the term~$\alpha_2 ( I_3^{\alpha_3} + I_3^{-\alpha_3} - 2)$.
The isotropic and the transversely isotropic parts of the passive material model are then given by
\eb
\Psi_{\text{p}, \, \text{isot}} = \alpha_1 \left( I_1 I_3^{-1/3} - 3 \right) + \alpha_2 \left( I_3^{\alpha_3} + I_3^{-\alpha_3} - 2 \right)  \quad \text{and} \quad \Psi_{\text{p}, \, \text{ti}}^{(f)} = \alpha_4 \left< K_3^{(f)} - 2 \right>^{\alpha_5} \, ,
\label{eq:psi}
\ee
where~$\langle\bullet\rangle=\half\left(\bullet + |\bullet|\right)$ denotes the Macaulay brackets.
The corresponding material parameters are restricted to~$\alpha_1 > 0$, $\alpha_2 > 0$, $\alpha_3 > 1$, $\alpha_4 > 0$ and $\alpha_5 > 2$. \\
Based on the strain-energy density function, we can derive the elastic part of the second Piola-Kirchhoff stress~$\bS_{\text{e}}$ from the dissipation inequality which arises from the elastic part of the deformation gradient.
In this context, the second Piola-Kirchhoff stress can be computed from the pull back operation by
\eb
\bS = \bF_{\text{g}}^{-1} \bS_{\text{e}} \bF_{\text{g}}^{- \text{T}} \qquad \text{with} \qquad \bS_{\text{e}} = 2 \frac{\partial \Psi}{\partial \bC_{\text{e}}}   \, .
\ee

\subsection{Implementation}
For the analysis of structural boundary value problems, we consider the Finite-Element Method, the Newton-Raphson scheme together with discrete time stepping to solve the nonlinear equations of the balance of linear momentum.
For the linearized system of equations to be solved in each time step, the total derivative of~$\bS$ with respect to~$\bC$ is required for the determination of the tangent modulus~$\IC = 2\partial_C\bS$ which we obtain from
\eb
\IC = \underbrace{2 \frac{\partial \bS}{\partial \bC}}_{\IC^{\text{e}}} + \underbrace{2 \sum_{a=2}^3 \frac{\partial \bS}{\partial \vartheta^{(a)}} \otimes \frac{\partial \vartheta^{(a)}}{\partial \bC}}_{\IC^{\text{g}}} \, ,
\ee
where~$\IC^{\text{e}}$ constitutes the elastic part and~$\IC^{\text{g}}$ is the growth part.
The elastic part of the tangent modulus can be build in the intermediate configuration as~$\IC^{\text{e}}_{\text{i}}$ and pulled back into the reference configuration by
\eb
\IC^{\text{e}} = \left( \bF_{\text{g}}^{-1} \boxtimes \bF_{\text{g}}^{-1} \right) : \IC^{\text{e}}_{\text{i}} : \left( \bF_{\text{g}}^{- \text{T}} \boxtimes \bF_{\text{g}}^{- \text{T}} \right) \qquad \text{with} \qquad
\IC^{\text{e}}_{\text{i}} := 2 \frac{\partial \bS_{\text{e}}}{\bC_{\text{e}}}
 \, .
\ee
The growth part of the tangent modulus $\IC^{\text{g}}$ has to be considered for $\IC$ only if the solution scheme for the evolution equations of the growth factors in Eq.~(\ref{eq:groFac}) depends on the current values of $\dot{\vartheta}^{(a)}$ as it is the case when using implicit integration schemes.
However, in this paper we are not interested in the realistic description of the time-process of growth, but rather on the resulting, converged grown state to obtain a suitable automatized procedure to include residual stresses which allow for a realistic in vivo stress distribution in line with the homeostasis hypothesis.
Therefore, a high accuracy in integrating the evolution equations is not required and thus, an explicit forward Euler is used here.
Furthermore, preliminary studies have shown that the time step size required for a converging Newton-Raphson scheme in structural finite element problems is anyhow smaller than the time step size needed for an accurate time integration using the forward Euler.
Since the forward Euler requires significantly less computational effort compared to implicit schemes, potential benefits of the latter vanish.
Therefore, we have the derivative~$\partial \vartheta^{(a)} / \partial \bC = \bzero$ and thus, the tangent moduli reduce to~$\IC = \IC^{\text{e}}$.
Therefore, the analytic representation of the elastic moduli has been implemented in our code.
However, connecting the growth model with the active SMC model requires a more sophisticated treatment as the muscle activation happens at higher rates and requires more accurate time integration of the associated evolution equations.
Therefore, a backward Euler is used therein to not be forced to unnecessarily reduce the time step size, requiring the full tangent moduli from the muscle model.
They have been implemented making use of complex-step derivative approximations~\cite{BalGanTanSch:2015:nco}, however, further reductions in implementation effort could be realized by taking into account second-order derivatives using hyper-dual numbers~\cite{TanBalSch:2016:ioi}.

\section{Optimization of Parameters}
\label{sec:opt}

To analyze the performance of the proposed model in characteristic simulations of idealized arteries, suitable parameter values have to be identified.
To this end, the parameters can be adjusted to experiments, e.g. performed on whole artery segments.
Then, finite element simulations of the arteries with known undeformed (reference) geometries can be performed repeatedly while updating the parameter values by an automated optimization procedure until the difference of computed quantities of interest and the experimental counterparts become as small as possible.
However, in case of growth, the reference geometry may change significantly during the growth process, which is in turn needed to include realistic (homeostatic) stress distributions.
Therefore, in this section we propose an adopted optimization procedure to also account for minimizing the mismatch between given reference geometry and the one resulting from growth.
%
In Subsection~\ref{subsec:opt_pro}, the optimization procedure is described in detail.
Afterwards, results for a first application of the optimization are shown and discussed in Subsection~\ref{subsec:sim}.
Due to the optimized identification of parameters related to geometry and material models, only two additional assumptions are required to describe the structural problem adequately because data on these aspects is not available in the experiments: firstly, the directions of the main fiber families, and secondly, the loading in axial direction when fixating the artery in the experimental device.
In the simulations of this Section and the following Section, the converging value~$\phi^{(2)}_{\text{con}}$ in axial direction is predefined which results in different corresponding values for the axial prestretch.
Based on a variation of the fiber directions as well as the convergence value~$\phi^{(2)}_{\text{con}}$ in different scenarios, the efficiency of the optimization can be investigated~(see Section~\ref{sec:num}).
For a precise estimation of the axial prestretch, an additional part of the optimization procedure is defined in the application shown in Section~\ref{sec:act} where also the active material response is considered.
In Subsection~\ref{subsec:mandel}, it is shown that the new kinematic growth model does not depend on a specific run-time or values for the growth velocity factors to produce an almost homogeneous stress distribution.
A comparison to a simulation using a more classical version of the growth model (cf. \cite{ZahBal:2018:acg}), which depends on the elastic part of the Mandel stress~$\BSigma_{\text{e}}$ as driving force for the growth process, shows the advantages of the proposed approach.

\subsection{Description of the Optimization Procedure}
\label{subsec:opt_pro}
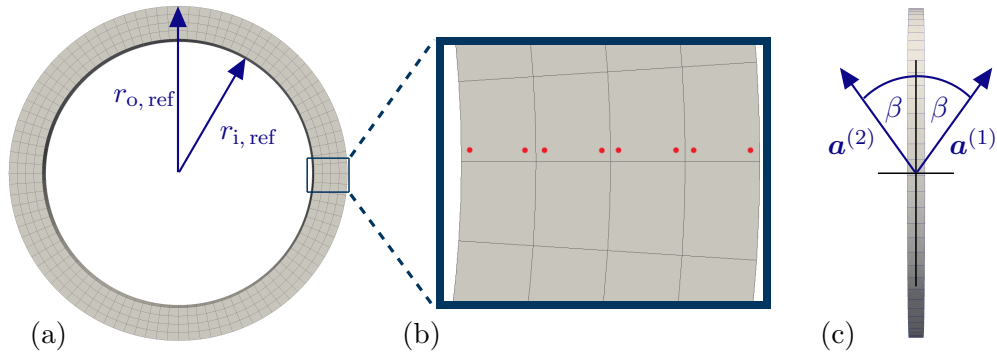
\begin{figure}[!t]
\unitlength1cm
\begin{picture}(15.5,4.5)
\put(7.0,0.0){
\linethickness{0.2mm}
\put(6.2,0.0){\includegraphics[height=4.5cm]{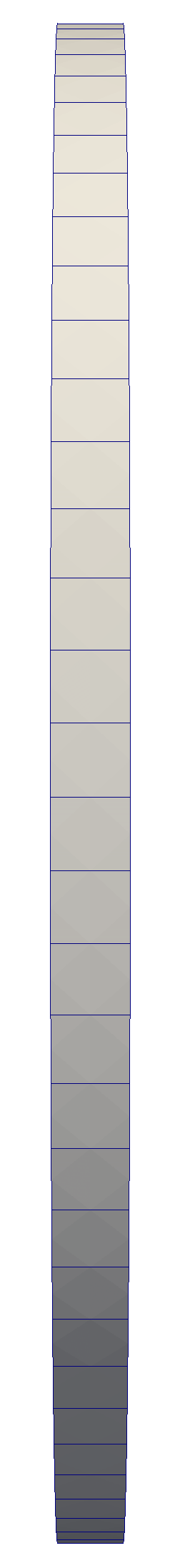}}
\put(6.46,0.75){\color{black}\line(0,1){3}}
\put(5.96,2.25){\color{black}\line(1,0){1}}
\put(6.46,2.25){\begin{tikzpicture}[thick, scale=0.6, transform shape, color=\tcolor]
\draw[>=triangle 45, ->] (0,0) -- (1.65,2.31);
\end{tikzpicture}
}
\put(5.445,2.25){\begin{tikzpicture}[thick, scale=0.6, transform shape, color=\tcolor]
\draw[>=triangle 45, ->] (0,0) -- (-1.65,2.31);
\end{tikzpicture}
}
\put(5.76,3.23){\begin{tikzpicture}
\draw [\tcolor,thick,domain=-45:45] plot ({sin(\x)}, {cos(\x)});
\end{tikzpicture}
}
\put(5.3,2.5){$\color{\tcolor} \ba^{(2)}$}
\put(6.9,2.5){$\color{\tcolor} \ba^{(1)}$}

\put(6.05,3.0){$\color{\tcolor} \beta$}
\put(6.65,3.0){$\color{\tcolor} \beta$}

}

\put(1.0, 0.0){
\put( 0.5, 0.0){\includegraphics[height=4.5cm]{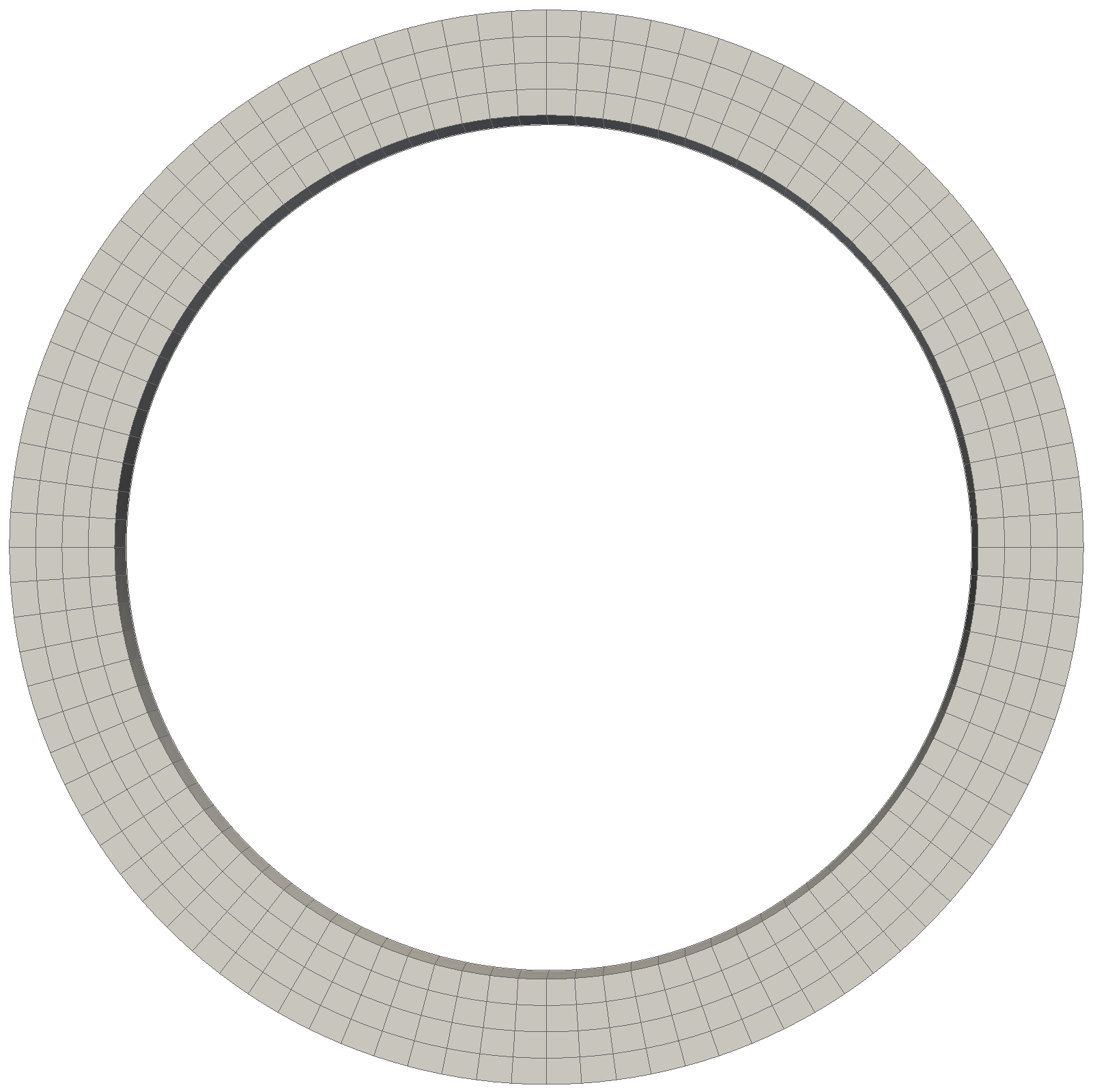}}

\put(2.45,2.25){\begin{tikzpicture}[thick, scale=1.0, transform shape, color=\tcolor]
\draw[>=triangle 45, ->] (0,0) -- (0.0,2.2);
\useasboundingbox (-0.3,0) rectangle (0.3,2.2);
\end{tikzpicture}
}
\put(2.75,2.25){\begin{tikzpicture}[thick, scale=1.0, transform shape, color=\tcolor]
\draw[>=triangle 45, ->] (0,0) -- (0.875,1.515544);
\end{tikzpicture}
}
\put(1.9,3.2){$\color{\tcolor} r_{\text{o,} \, \text{ref}}$}
\put(3.3,2.7){$\color{\tcolor} r_{\text{i,} \, \text{ref}}$}


\linethickness{0.2mm}
\put(4.45,2.45){\color{rub_blue}\line(1,0){0.551}}
\put(5.0,2.45){\color{rub_blue}\line(0,-1){0.451}}
\put(5.0,2.0){\color{rub_blue}\line(-1,0){0.551}}
\put(4.45,2.0){\color{rub_blue}\line(0,1){0.451}}

}

\put(-3.0,0.0){
\put(9.0,2.45){\color{rub_blue}\tikz\draw [very thick,dashed] (0,0) -- (1.10,1.55);}
\put(9.0,0.45){\color{rub_blue}\tikz\draw [very thick,dashed] (0,0.0) -- (1.10,-1.50);}

\put( 10.2, 0.5){\includegraphics[height=3.5cm]{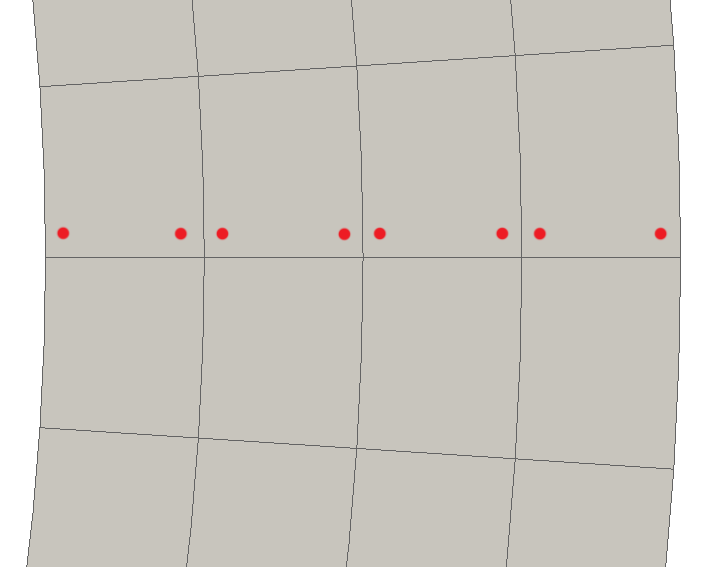}}
\linethickness{1mm}
\put(10.2,4.0){\color{rub_blue}\line(1,0){4.35}}
\put(14.5,4.0){\color{rub_blue}\line(0,-1){3.55}}
\put(14.5,0.5){\color{rub_blue}\line(-1,0){4.35}}
\put(10.2,0.5){\color{rub_blue}\line(0,1){3.55}}
}

\put( 1.8, 0.0){(a)}
\put( 6.7, 0.0){(b)}
\put( 12.2, 0.0){(c)}

\end{picture}
\caption{Visualization of the mesh of the arterial ring:
(a)~the mesh in the reference configuration with the inner and outer radii $r_{\text{i,} \, \text{ref}}$ and $r_{\text{o,} \, \text{ref}}$.
(b)~A close-up of the wall with marks at the positions of the Gauss points which are used to illustrate results in Fig.~\ref{fig:opt1}, \ref{fig:progress}, \ref{fig:variation_beta}, \ref{fig:opt_act}, \ref{fig:opt_mandel},  \ref{fig:variation_lambda} and \ref{fig:variation_phi}.
(c)~The fiber vectors $\ba^{(1)}$ and $\ba^{(2)}$ and the corresponding fiber angle $\beta$ illustrated in the plane of longitudinal and circumferential direction.  \label{fig:mesh}}
\end{figure}

%
For the simulations, which are used to test different data sets during the optimization procedure, the idealized geometry of a hollow cylinder is considered which is illustrated in Fig.~\ref{fig:mesh}a.
Note that only the media is taken into account for simulations in this paper since results of various previously published simulations have illustrated that the mechanical contribution of the media dominates the wall response.
To minimize the volumetric change of the artery during growth, the convergence value~$\phi^{(3)}_{\text{con}}$ of the growth method has to be selected appropriately.
First test simulations showed already promising results when the convergence value~$\phi^{(3)}_{\text{con}}$ was set to the mean value of the driving force~$\phi^{(3)}$ at the loaded state of the artery before the growth process is started.
As described in Eq.~\eqref{eq:drivingForce}, the driving force~$\phi^{(3)}$ is the trace of the Cauchy stress tensor~$\sigma_{\text{tr}} = \Bsigma : \bI$.
Based thereon, we define the convergence value~$\phi^{(3)}_{\text{con}}$ as a multiple of the mean value~$\bar{\sigma}_{\text{tr}}$ modified by a factor~$w_{\text{min}}$.
This factor~$w_{\text{min}}$ is the first variable of the optimization process which is mainly responsible for the minimization of the volumetric change based on the growth dependent on~$\vartheta^{(3)}$ and, hence, is called the volumetric minimizer.
Accordingly, $\phi^{(3)}_{\text{con}}$ can be written as
\eb
\label{eq:conv3}
\phi^{(3)}_{\text{con}} = w_{\text{min}} \bar{\sigma}_{\text{tr}} \qquad \text{with} \qquad \bar{\sigma}_{\text{tr}} = \frac{1}{n_{\text{gp}}}\sum_{g=1}^{n_{\text{gp}}} \text{tr}(\Bsigma)_g \, ,
\ee
where $\text{tr}(\Bsigma)_g$ is the trace of the Cauchy stresses at the Gauss point~$g$.
To obtain a reliable mean value~$\bar{\sigma}_{\text{tr}}$, eight Gauss points were selected as reference points.
The eight chosen Gauss points have the same position in axial direction and build an exact line from the inside to the outside of the arterial wall (see Fig.~\ref{fig:mesh}b).\\
As second objective of the optimization, the resulting geometry of the arterial ring after growth has to match realistic measurements.
Since material data from a middle cerebral artery of a rat are used in the fitting of the material parameters, measurements for a comparable artery have to be considered.
Accordingly, we adopt the data from~\cite{GanVanJerGriHamDru:2008:ipi} (see Table 1 in the original publication) who measured a ratio of 0.21 between wall thickness and outer radius for a middle cerebral artery of a mouse which is expected to be comparable to the ratio of the same artery type in a rat.
During this measurement, an intravascular pressure of~15$\,$mmHg was applied and the surrounding solution contained no calcium ions which led to a deactivation of the contraction mechanism of SMCs inside of the tissue.
Since the volumetric change of the wall thickness during the growth process is unknown, the inner and outer radius of the reference configuration of the arterial ring, $r_{\text{i,} \, \text{ref}}$ and~$r_{\text{o,} \, \text{ref}}$, are chosen as optimization parameters to accomplish this part of the procedure (see Fig.\ref{fig:mesh}a).
To replicate the experiment in~\cite{GanVanJerGriHamDru:2008:ipi}, the wall thickness of the resulting geometry is tested with a simulation in which the arterial wall is loaded with an intravascular pressure of~15$\,$mmHg.\\
As the third step of the optimization, the parameters of the material model are adjusted to the experimental data from~\cite{JohElyTakWalWalCol:2009:csv} (see Fig.~4a in the original publication) where the segment of a middle cerebral rat artery is set in Krebs solution and investigated by applying a sequence of intravascular pressure with increasing pressure values.
The temporal change of the outer diameter of the middle cerebral artery was measured during this process for the passive material response (no calcium ions in the surrounding Krebs solution), the fully active material response and a partially suppressed response of SMCs by including the ROCK inhibitor Y27632 into the solution.
To replicate the data of the experiments of the passive material with finite element simulations, we include the optimization of the parameters~$\alpha_1$, $\alpha_4$ and~$\alpha_5$ which describe the mechanical behavior of elastin and collagen fibers.
The parameters~$\alpha_2$ and~$\alpha_3$ are not considered as input parameters of the optimization, but are set prior the optimization to~100$\,$kPa and~2, respectively, to represent suitable penalty parameters for near incompressibility.
A more detailed description for the replication of the data for the active material response can be found in Section~\ref{sec:act}.\\
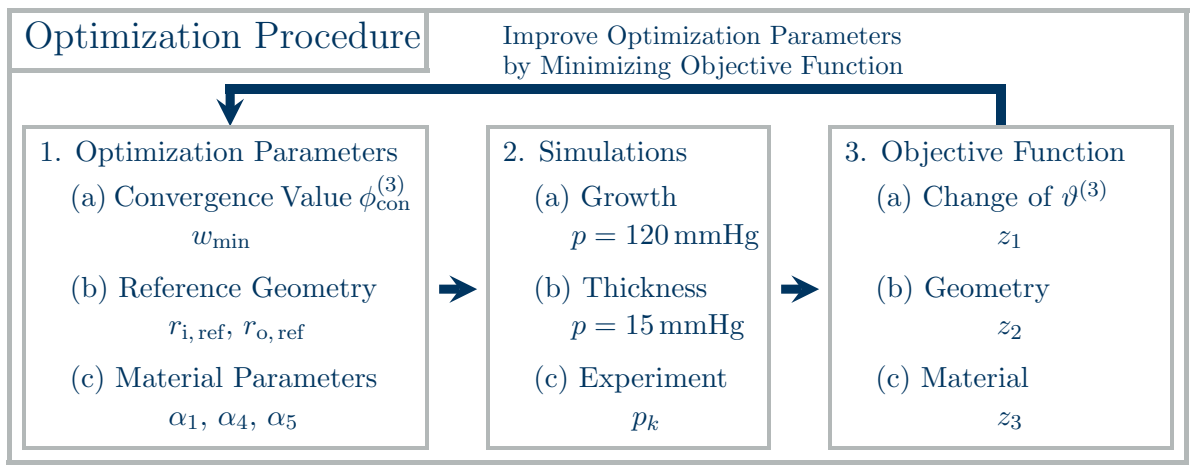
\begin{figure}[!b]
\unitlength1cm
\begin{picture}(15.5,6.0)

\put(0.2,5.5){\color{rub_blue} \Large Optimization Procedure}

\linethickness{0.6mm}
\put(0.0,5.2){\color{dia_grau_hell2}\line(1,0){5.53}}
\put(5.5,6.0){\color{dia_grau_hell2}\line(0,-1){0.83}}

\put(6.5,5.55){\color{rub_blue} \small Improve Optimization Parameters}
\put(6.5,5.15){\color{rub_blue} \small by Minimizing Objective Function}
\linethickness{1.2mm}
\put(13.05,4.45){\color{rub_blue}\line(0,1){0.56}}
\put(13.05,4.95){\color{rub_blue}\line(-1,0){10.2}}

\put(2.71,4.40){\begin{tikzpicture}[very thick, line width=1.2mm, scale=1.0, transform shape, color=rub_blue]
\draw [-stealth](0,0) -- (0,-0.5);
\useasboundingbox (-0.2,0) rectangle (0.2,0.5);
\end{tikzpicture}
}

\linethickness{0.6mm}
\put(0.0,6.0){\color{dia_grau_hell2}\line(1,0){15.53}}
\put(15.5,6.0){\color{dia_grau_hell2}\line(0,-1){6.03}}
\put(15.5,0.0){\color{dia_grau_hell2}\line(-1,0){15.53}}
\put(0.0,0.0){\color{dia_grau_hell2}\line(0,1){6.03}}

\put(0.4,0.0){
\put(0.0,4.0){\color{rub_blue} \parbox[t]{5cm}{1. Optimization Parameters}}
\put(0.4,3.4){\color{rub_blue} \parbox[t]{4.5cm}{(a) Convergence Value $\phi^{(3)}_{\text{con}}$}}
\put(2.0,2.9){\color{rub_blue} \parbox[t]{4cm}{$w_{\text{min}}$}}
\put(0.4,2.2){\color{rub_blue} \parbox[t]{4.5cm}{(b) Reference Geometry}}
\put(1.7,1.7){\color{rub_blue} \parbox[t]{4cm}{$r_{\text{i,} \, \text{ref}}$, $r_{\text{o,} \, \text{ref}}$}}
\put(0.4,1.0){\color{rub_blue} \parbox[t]{4.5cm}{(c) Material Parameters}}
\put(1.7,0.5){\color{rub_blue} \parbox[t]{4cm}{$\alpha_1$, $\alpha_4$, $\alpha_5$}}
}

\linethickness{0.4mm}
\put(0.2,4.45){\color{dia_grau_hell2}\line(1,0){5.32}}
\put(5.5,4.45){\color{dia_grau_hell2}\line(0,-1){4.27}}
\put(5.5,0.2){\color{dia_grau_hell2}\line(-1,0){5.32}}
\put(0.2,0.2){\color{dia_grau_hell2}\line(0,1){4.27}}

\put(5.6,2.1){\begin{tikzpicture}[very thick, line width=1.2mm, scale=1.0, transform shape, color=rub_blue]
\draw [-stealth](0,0) -- (0.5,0);
\useasboundingbox (0,-0.2) rectangle (0.5,0.2);
\end{tikzpicture}
}

\put(6.5,0.0){
\put(0.0,4.0){\color{rub_blue} \parbox[t]{5cm}{2. Simulations}}
\put(0.4,3.4){\color{rub_blue} \parbox[t]{4.5cm}{(a) Growth}}
\put(0.9,2.9){\color{rub_blue} \parbox[t]{4cm}{$p = 120\,$mmHg}}
\put(0.4,2.2){\color{rub_blue} \parbox[t]{4.5cm}{(b) Thickness}}
\put(0.9,1.7){\color{rub_blue} \parbox[t]{4cm}{$p = 15\,$mmHg}}
\put(0.4,1.0){\color{rub_blue} \parbox[t]{4.5cm}{(c) Experiment}}
\put(1.7,0.5){\color{rub_blue} \parbox[t]{4cm}{$p_k$}}
}

\linethickness{0.4mm}
\put(6.3,4.45){\color{dia_grau_hell2}\line(1,0){3.72}}
\put(10.0,4.45){\color{dia_grau_hell2}\line(0,-1){4.27}}
\put(10.0,0.2){\color{dia_grau_hell2}\line(-1,0){3.72}}
\put(6.3,0.2){\color{dia_grau_hell2}\line(0,1){4.27}}

\put(10.1,2.1){\begin{tikzpicture}[very thick, line width=1.2mm, scale=1.0, transform shape, color=rub_blue]
\draw [-stealth](0,0) -- (0.5,0);
\useasboundingbox (0,-0.2) rectangle (0.5,0.2);
\end{tikzpicture}
}

\put(11.0,0.0){
\put(0.0,4.0){\color{rub_blue} \parbox[t]{5cm}{3. Objective Function}}
\put(0.4,3.4){\color{rub_blue} \parbox[t]{4.5cm}{(a) Change of $\vartheta^{(3)}$}}
\put(2.0,2.9){\color{rub_blue} \parbox[t]{4cm}{$z_1$}}
\put(0.4,2.2){\color{rub_blue} \parbox[t]{4.5cm}{(b) Geometry}}
\put(2.0,1.7){\color{rub_blue} \parbox[t]{4cm}{$z_2$}}
\put(0.4,1.0){\color{rub_blue} \parbox[t]{4.5cm}{(c) Material}}
\put(2.0,0.5){\color{rub_blue} \parbox[t]{4cm}{$z_3$}}
}

\linethickness{0.4mm}
\put(10.8,4.45){\color{dia_grau_hell2}\line(1,0){4.52}}
\put(15.3,4.45){\color{dia_grau_hell2}\line(0,-1){4.27}}
\put(15.3,0.2){\color{dia_grau_hell2}\line(-1,0){4.52}}
\put(10.8,0.2){\color{dia_grau_hell2}\line(0,1){4.27}}

\end{picture}
\caption{Illustration of the optimization procedure based on a combined evolution-strategy and gradient method.
Three steps are executed to evaluate a new set of parameters: 
1.~Creation of new sample of parameters based on data from earlier generations; 
2.(a)~Execution of growth process; 
(b)~Checking the wall thickness of the grown geometry; 
(c)~Reproducing mechanical experiments with simulation;
3.~Evaluate the objective function for this sample.
\label{fig:optimization}}
\end{figure}
Consequently, while only the passive material response is considered, the optimization is enabled to adjust six different parameters, namely, the volumetric minimizer~$w_{\text{min}}$, the inner and outer radii of the reference configuration~$r_{\text{i,} \, \text{ref}}$ and~$r_{\text{o,} \, \text{ref}}$, and the material parameters~$\alpha_1$, $\alpha_4$ and $\alpha_5$ (see left side of Fig.~\ref{fig:optimization}).
The optimization was implemented in \textit{Python} by utilizing the library \textit{mystic}.
Mystic offers a mixture of evolution strategy and gradient methods.
Due to the parallelization features of mystic, we analyzed up to 40 children parameter sets at the same time.
For every set of input parameters generated by mystic, the finite element program with the model implementations in \textit{FEAP} is called by a \textit{Python} script to run three boundary value problems with the new parameter set.
These simulations are listed in the middle box of Fig.~\ref{fig:optimization} and contain: 2(a)~the growth process of the artery, 2(b)~the measurement of the inner and outer radius at an intravascular pressure of 15$\,$mmHg and 2(c)~the replication of the mechanical experiments from~\cite{JohElyTakWalWalCol:2009:csv}.
Since in simulation~2(b) and~2(c) the resulting geometry after growth has to be used, the growth factors~$\vartheta^{(a)}$ of every Gauss point are saved at the end of simulation~2(a).
Subsequently, these growth factors are applied as boundary condition during the first second of the simulations~2(b) and~2(c), in which the values of~$\vartheta^{(a)}$ are linearly increased over time and reach their final value at one second.
Further details considering the boundary value problems are explained in Subsection~\ref{subsec:sim}.
According to the objectives of the optimization, the objective function $z$ is split into three corresponding parts~$z = z_1 + z_2 + z_3$ with
\eb
\label{eq:objective}
\resizebox{\textwidth}{!}{$
z = \displaystyle \underbrace{\omega_{\vartheta} \displaystyle \sqrt{ \frac{1}{n_{\text{gp}}} \sum_{g=1}^{n_{\text{gp}}} \left( \vartheta^{(3)}_g-1 \right)^2 }}_{\mathlarger{z_1}} + \underbrace{\sqrt{\left( \frac{r_{\text{o}, \, 15}-r_{\text{i}, \, 15}}{r_{\text{o}, \, 15}} - 0.21 \right)^2 }}_{\mathlarger{z_2}} + \underbrace{\sqrt{ \frac{1}{n_{\text{data}}} \sum_{k=1}^{n_{\text{data}}} \left( \frac{d_{\text{exp}, \, k}-d_{\text{sim}, \, k}}{d_{\text{exp}, \, k}} \right)^2 }}_{\mathlarger{z_3}} \, .
$}
\ee
In the first part~$z_1$, the change of the growth factors~$\vartheta^{(3)}$ at the Gauss points~$g$ are evaluated which correspond to the main contributor to volumetric change due to growth.
In the second term~$z_2$, the inner and outer radii of the resulting geometry at an intravascular pressure of 15mmHg are described by~$r_{\text{o}, \, 15}$ and~$r_{\text{i}, \, 15}$ and compared to the desired ratio of the wall thickness of~0.21.
For the third term~$z_3$, the parameter~$d_{\text{exp}, \, k}$ is the measured outer diameter from the experiments at the intravascular pressure~$p_k$, and~$d_{\text{sim}, \, k}$ is the outer diameter of the resulting geometry from the simulation at the same load.
In our analysis, the weight for the objective function related to growth-induced changes in volume is set to $\omega_{\vartheta}=0.035$, which has been chosen from experience when running first series of optimizations.
The additive split of the objective function is advantageous because it enables the evaluation of the fitting of the mechanical experiments without significant influence on the other two parts.
This design is especially important when the optimization procedure is tested with varying conditions related to e.g. a change of the fiber directions or the convergence value~$\phi^{(2)}_{\text{con}}$.
These variations influence the magnitude of the growth factors notably which results in large differences between the final values of the objective function~$z_1$.
Due to the additive split, the quality of each optimization can still be estimated based on the accuracy of the fitted mechanical experiments via~$z_3$.

\subsection{Analysis of Growth Model using Optimization Procedure}
\label{subsec:sim}
In this section, a first analysis of the general behavior of the growth model and the optimization procedure is given.
To enable a good simulation of residual stresses, a realistic load scenario has to be chosen in the simulation while the growth model is active.
The artery in the simulation is supposed to describe the mechanical behavior of a middle cerebral artery of a rat.
The blood pressure of a healthy rat at rest is approximately 100/60$\,$mmHg, cf. e.g.~\cite{ParRav:2012:moi}.
Considering the activity of the rat over the day, an intravascular pressure of 120$\,$mmHg may be estimated as plausible mean value, which is why we apply the static value of 120$\,$mmHg as internal pressure in the simulations.
Note that this value will obviously not be precise and may differ from artery to artery, but it can be considered reasonable for the analysis of characteristic scenarios.
The axial prestretch is primarily influenced by the selected value for the convergence value~$\phi^{(2)}_{\text{con}}$ for growth in axial direction.
Investigations of large arteries such as the aorta, carotid artery, iliac artery or superficial femoral artery showed in vivo axial prestretches of~1.0 and lower for elderly donors, see e.g.~\cite{JadRazHabAntKam:2021:msa,HorAdaGulZitVesChlKon:2011:cba,SchMorHol:2003:pbm,SomRegKolHol:2010:bmp}.
Arteries in young donors illustrate values larger than~1.4, cf.~\cite{JadRazHabAntKam:2021:msa}.
A change of the axial prestretch from~1.4 to~1.0 requires considerable growth in axial direction for large arteries which possess a primarily passive material behavior.
To enable the possibility to obtain any type of artery by the application of the new kinematic growth model and the optimization procedure, the most expensive growth scenario should be investigated as the most challenging scenario.
Therefore, a convergence value~$\phi^{(2)}_{\text{con}} = 0 \,$kPa is chosen as standard value in the simulations considered in this Section and the following one.
Reducing the axial stresses to~0$\,$kPa in elastic arteries through growth will of course require large values for the growth factor~$\vartheta^{(2)}$.\\
\begin{figure}[!b]
\unitlength1cm
\begin{picture}(15.5,7.0)
\put( -0.1, 0.0){
\put( 0.0, 0.0){
\begin{tikzpicture}
\begin{axis}[
width=5.2cm,
height=7.5cm,
xlabel={{Radial position in $\mu$m}},
xmin=83, xmax=103,
xtick={83, 88, 93, 98, 103},
extra x ticks={0.0},
extra x tick style={%
    grid=major,
},
ylabel={{Stress in kPa}},
ymin=-1, ymax=225,
ytick={0, 25, 50, 75, 100, 125, 150, 175, 200, 225},
axis background/.style={fill=white!89.80392156862746!black},
axis line style={white!60.0!black},
axis x line*=bottom,
axis y line*=left,
tick align=outside,
tick pos=left,
scaled x ticks=false,
x grid style={white},
xmajorgrids,
x tick label style={
  font=\small,
  /pgf/number format/.cd,
  set decimal separator={.},
    fixed,
    fixed zerofill,
    precision=0,
  /tikz/.cd
},
xlabel style={at={(0.5,-0.1)},font=\small},
ylabel style={at={(-0.22,0.5)},font=\small},
y grid style={white},
ymajorgrids,
y tick label style={
  font=\small,
},
legend cell align={left},
legend style={at={(1.0,1.0)}, anchor=north east, draw=white!60.0!black, row sep=-0.08cm},
clip mode=individual,
]

\addplot [dashed, very thick,rub_green, mark options={solid, draw=\tcolorshade, line width=0.5pt}]
table {%
   83.5149455549278        202.252712552088
   87.3471346413345        161.326916056266     
   88.5650946596096        150.415469610495     
   92.3967801186199        121.814066034548     
   93.6148971035856        114.072470957680      
   97.4469020874785        93.6109110956246     
   98.6661032418199        88.0087833260855    
   102.499715567485        73.0097999469633              
};
\addlegendentry{{\scriptsize $\sigma^{\text{(cir)}}$}}

\addplot [dashed, very thick,red, mark options={solid, draw=\tcolorshade, line width=0.5pt}]
table {%
   83.5149455549278        103.033679235180 
   87.3471346413345        79.7129008119450     
   88.5650946596096        73.7212493545608     
   92.3967801186199        58.7472739415519     
   93.6148971035856        54.8245457404573     
   97.4469020874785        44.9568058597107     
   98.6661032418199        42.3348954826036     
   102.499715567485        35.6460745092288             
};
\addlegendentry{{\scriptsize $\sigma^{\text{(ax)}}$}}

\addplot [very thick,rub_green, mark options={solid, draw=\tcolorshade, line width=0.5pt}]
table {%
   83.5149455549278        154.152713432217 
   87.3471346413345        150.271450956373     
   88.5650946596096        149.159031137845       
   92.3967801186199        145.991115669419     
   93.6148971035856        145.087405829992   
   97.4469020874785        142.520200676633     
   98.6661032418199        141.834688864605      
   102.499715567485        139.909548187570          
};
\addlegendentry{{\scriptsize $\sigma^{\text{(cir)}}_{\text{gr}}$}}

\addplot [very thick,red, mark options={solid, draw=\tcolorshade, line width=0.5pt}]
table {%
   83.5149455549278       0.588781299974938  
   87.3471346413345       0.549498337489207     
   88.5650946596096       0.535973334461005   
   92.3967801186199       0.490604824346731     
   93.6148971035856       0.474667818547816      
   97.4469020874785       0.420914254268094     
   98.6661032418199       0.401220708317064      
   102.499715567485       0.336129628674836          
};
\addlegendentry{{\scriptsize $\sigma^{\text{(ax)}}_{\text{gr}}$}}

\end{axis}
\end{tikzpicture}}

\put( 0.6, 0.0){(a)}
\put( 5.15, 0.0){
\begin{tikzpicture}
\begin{axis}[
width=5.2cm,
height=7.5cm,
xlabel={{Radial position in $\mu$m}},
xmin=83, xmax=103,
xtick={83, 88, 93, 98, 103},
extra x ticks={0.0},
extra x tick style={%
    grid=major,
},
ylabel={{Driving Force in kPa}},
ymin=-1, ymax=300,
ytick={0, 50, 100, 150, 200, 250, 300},
axis background/.style={fill=white!89.80392156862746!black},
axis line style={white!60.0!black},
axis x line*=bottom,
axis y line*=left,
tick align=outside,
tick pos=left,
scaled x ticks=false,
x grid style={white},
xmajorgrids,
x tick label style={
  font=\small,
  /pgf/number format/.cd,
  set decimal separator={.},
    fixed,
    fixed zerofill,
    precision=0,
  /tikz/.cd
},
xlabel style={at={(0.5,-0.1)},font=\small},
ylabel style={at={(-0.20,0.5)},font=\small},
y grid style={white},
ymajorgrids,
y tick label style={
  font=\small,
},
legend cell align={left},
legend style={at={(1.0,1.0)}, anchor=north east, draw=white!60.0!black, row sep=-0.08cm},
clip mode=individual,
]

\addplot [dashed, very thick,rub_green, mark options={solid, draw=\tcolorshade, line width=0.5pt}]
table {%
   83.5149455549278        291.145270728808  
   87.3471346413345        231.184302892233     
   88.5650946596096        215.356963720047    
   92.3967801186199        174.989974693218     
   93.6148971035856        164.118503635117     
   97.4469020874785        136.186544932079     
   98.6661032418199        128.572240744374      
   102.499715567485        108.742572766723            
};
\addlegendentry{{\scriptsize $\phi^{(3)}$}}

\addplot [dashed, very thick,red, mark options={solid, draw=\tcolorshade, line width=0.5pt}]
table {%
   83.5149455549278        103.033679235180 
   87.3471346413345        79.7129008119450     
   88.5650946596096        73.7212493545608     
   92.3967801186199        58.7472739415519     
   93.6148971035856        54.8245457404573     
   97.4469020874785        44.9568058597107     
   98.6661032418199        42.3348954826036     
   102.499715567485        35.6460745092288         
};
\addlegendentry{{\scriptsize $\phi^{(2)}$}}

\addplot [very thick,rub_green, mark options={solid, draw=\tcolorshade, line width=0.5pt}]
table {%
   83.5149455549278        139.558567672673 
   87.3471346413345        139.604868092406     
   88.5650946596096        139.625469026024     
   92.3967801186199        139.709896901403     
   93.6148971035856        139.741750534692     
   97.4469020874785        139.849151562986     
   98.6661032418199        139.882844316777      
   102.499715567485        139.966977011818              
};
\addlegendentry{{\scriptsize $\phi^{(3)}_{\text{gr}}$}}

\addplot [very thick,red, mark options={solid, draw=\tcolorshade, line width=0.5pt}]
table {%
   83.5149455549278       0.588781299974938  
   87.3471346413345       0.549498337489207     
   88.5650946596096       0.535973334461005   
   92.3967801186199       0.490604824346731     
   93.6148971035856       0.474667818547816      
   97.4469020874785       0.420914254268094     
   98.6661032418199       0.401220708317064      
   102.499715567485       0.336129628674836     
};
\addlegendentry{{\scriptsize $\phi^{(2)}_{\text{gr}}$}}

\end{axis}
\end{tikzpicture}}

\put( 5.75, 0.0){(b)}
\put( 10.3, 0.0){
\begin{tikzpicture}
\begin{axis}[
width=5.2cm,
height=7.5cm,
xlabel={{Radial position in $\mu$m}},
xmin=83, xmax=103,
xtick={83, 88, 93, 98, 103},
extra x ticks={0.0},
extra x tick style={%
    grid=major,
},
ylabel={{Growth [-]}},
ymin=0.7, ymax=1.4,
ytick={0.7, 0.8, 0.9, 1.0, 1.1, 1.2, 1.3, 1.4},
axis background/.style={fill=white!89.80392156862746!black},
axis line style={white!60.0!black},
axis x line*=bottom,
axis y line*=left,
tick align=outside,
tick pos=left,
scaled x ticks=false,
x grid style={white},
xmajorgrids,
x tick label style={
  font=\small,
  /pgf/number format/.cd,
  set decimal separator={.},
    fixed,
    fixed zerofill,
    precision=0,
  /tikz/.cd
},
y tick label style={
  font=\small,
  /pgf/number format/.cd,
  set decimal separator={.},
    fixed,
    fixed zerofill,
    precision=1,
  /tikz/.cd
},
xlabel style={at={(0.5,-0.1)},font=\small},
ylabel style={at={(-0.18,0.5)},font=\small},
y grid style={white},
ymajorgrids,
y tick label style={
  font=\small,
},
legend cell align={left},
legend style={at={(0.0,0.0)}, anchor=south west, draw=white!60.0!black, row sep=-0.08cm},
clip mode=individual,
]

\addplot [very thick,rub_green, mark options={solid, draw=\tcolorshade, line width=0.5pt}]
table {%
   83.5149455549278        1.29190992782697       
   87.3471346413345        1.18852467248692     
   88.5650946596096        1.15393775386109     
   92.3967801186199        1.04026782013977     
   93.6148971035856        1.00287027887884      
   97.4469020874785       0.882400464982457     
   98.6661032418199       0.843280105967184     
   102.499715567485       0.720265924880499                   
};
\addlegendentry{{\footnotesize $\vartheta^{\text{(3)}}$}}

\addplot [very thick,red, mark options={solid, draw=\tcolorshade, line width=0.5pt}]
table {%
   83.5149455549278        1.34032526818106           
   87.3471346413345        1.33232719230759     
   88.5650946596096        1.32990158046196    
   92.3967801186199        1.32246899170047     
   93.6148971035856        1.32015409886314     
   97.4469020874785        1.31289977008467     
   98.6661032418199        1.31051127526333      
   102.499715567485        1.30285076462809            
};
\addlegendentry{{\footnotesize $\vartheta^{\text{(2)}}$}}

\end{axis}
\end{tikzpicture}}

\put( 10.9, 0.0){(c)}
}
\end{picture}
\caption{Distribution of 
(a)~Cauchy stresses~$\sigma^{\text{(ax)}}$ and~$\sigma^{\text{(cir)}}$, 
(b)~driving forces~$\phi^{(2)}$ and~$\phi^{(3)}$, and
(c)~growth factors~$\vartheta^{(2)}$ and~$\vartheta^{(3)}$
in circumferential (green) and axial (red) direction over the wall thickness. 
Artery is loaded with an intravascular pressure of 120mmHg. 
Dashed lines show results before growth, solid lines show results after growth.
The gradients of the stresses over the wall thickness are significantly reduced by growth. \label{fig:opt1}}
\end{figure}
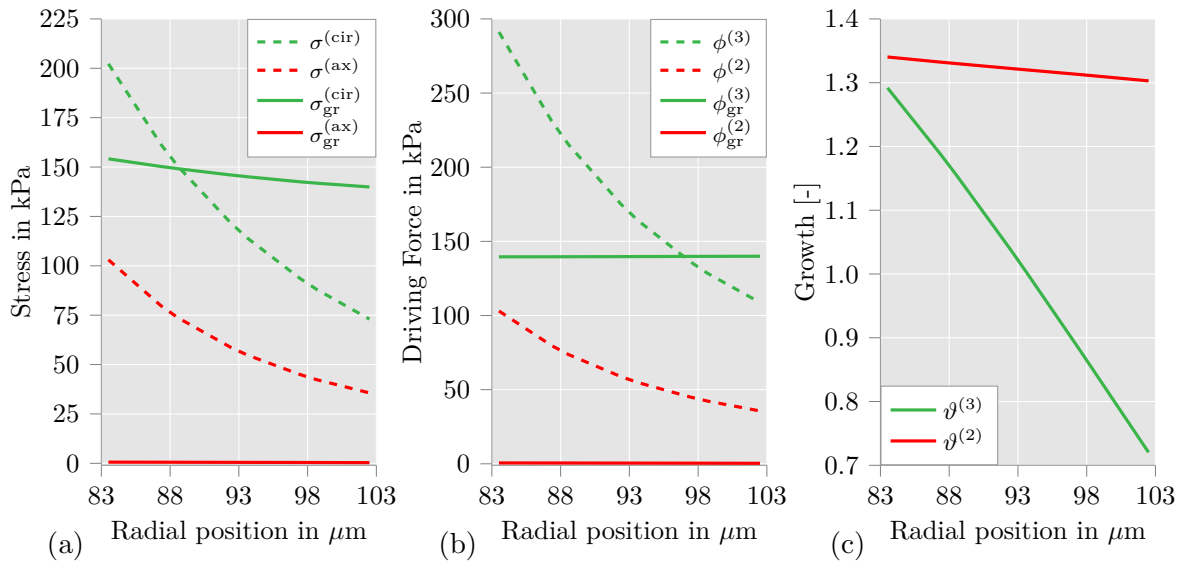
Here, the investigation will first only include the passive material response of the arterial wall.
Accordingly, only the experimental data for the passive material from~\cite{JohElyTakWalWalCol:2009:csv} will be fitted in the objective function~$z_3$.
An example of the reference geometry using tri-quadratic hexahedral elements is shown in Fig.~\ref{fig:mesh}a.
This mesh has been created with one element in longitudinal direction, four elements in radial direction and 96 elements in circumferential direction.
Accordingly, the mesh consists of 384 quadratic 20-node brick elements.
The axial displacements at the beginning and the end of the arterial ring is set to zero as boundary condition for all simulations here.
In all three simulations, the nodes of the arterial ring are able to freely move in radial direction.
Movement in circumferential direction is not expected due to the rotational symmetry of the problem.
Experimental investigations of the collagen fibers in arteries of the human brain in~\cite{FinMccCan:1995:tdc} show that the fiber direction of collagen is nearly circumferential in the media; the angle~$\beta$ between circumferential and fiber direction (cf. Fig.~\ref{fig:mesh}c for an illustration) may reach values up to~$13^{\circ}$.
In our simulations, $\beta$ is prescribed to be distributed from~10$^{\circ}$ at the inner side to~20$^{\circ}$ at the outer side of the arterial wall.
A higher fiber angle may be considered for arteries in a different position of the body, especially, when larger arteries such as the aorta are taken into account.
Due to the fact that different fiber angles may apparently occur, the impact of variations of~$\beta$ on the optimization procedure and the new kinematic growth model will be investigated in Section~\ref{sec:num}.
The growth velocity factors are chosen to be~$\kappa_\vartheta^{(2)} = 10^{-4} (\text{s} \cdot \text{kPa})^{-1}$ and~$\kappa_\vartheta^{(3)} = 10^{-4} (\text{s} \cdot \text{kPa})^{-1}$.
Note that the growth velocity parameters mainly modify the time process, but hardly the resulting grown state and thus, for the purposes of this paper they may be chosen rather arbitrarily.
However, for an improved of computational efficiency, they may be chosen as high as possible to have a reasonable tradeoff between large time steps and decent convergence of the Newton-Raphson scheme.\\
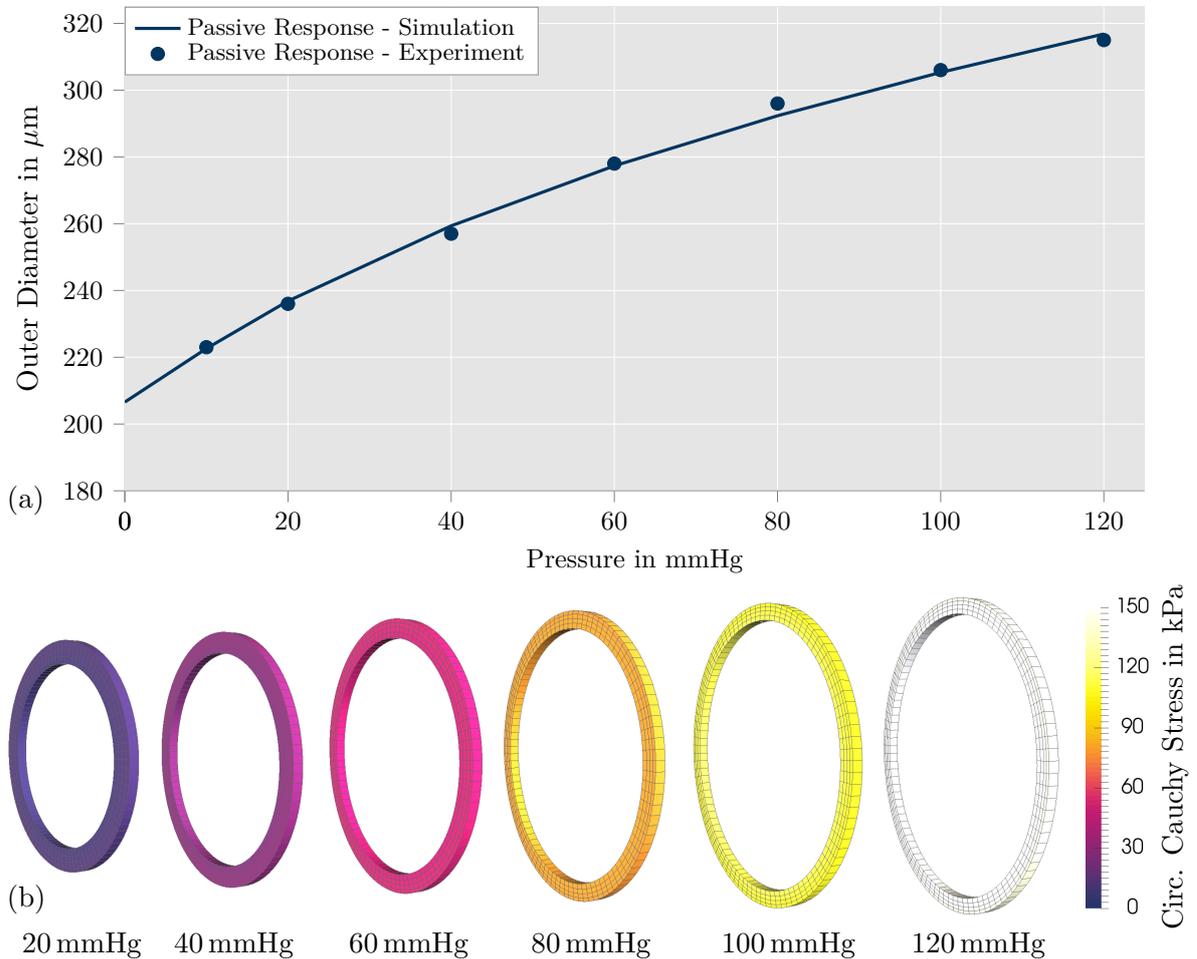
\begin{figure}[!b]
\unitlength1cm
\begin{picture}(15.5,12.5)
\put(0.0,4.9){
\begin{tikzpicture}
\begin{axis}[
width=15.0cm,
height=8.0cm,
xlabel={{Pressure in mmHg}},
xmin=0.0, xmax=125,
xtick={0, 20, 40, 60, 80, 100, 120},
extra x ticks={0.0},
extra x tick style={%
    grid=major,
},
ylabel={{Outer Diameter in $\mu$m}},
ymin=180, ymax=325,
ytick={180, 200, 220, 240, 260, 280, 300, 320},
axis background/.style={fill=white!89.80392156862746!black},
axis line style={white!60.0!black},
axis x line*=bottom,
axis y line*=left,
tick align=outside,
tick pos=left,
scaled x ticks=false,
x grid style={white},
xmajorgrids,
x tick label style={
  font=\small,
  /pgf/number format/.cd,
    fixed,
    fixed zerofill,
    precision=0,
  /tikz/.cd
},
xlabel style={at={(0.5,-0.10)},font=\small},
y grid style={white},
ymajorgrids,
y tick label style={
  font=\small,
},
ylabel near ticks,
legend cell align={left},
legend style={at={(0.0,1.0)}, font=\small, anchor=north west, draw=white!60.0!black, row sep=-0.08cm},
clip mode=individual,
]

\addplot [very thick, rub_blue, mark options={solid, draw=\tcolorshade, line width=0.5pt}]
table {%
0   206.566611519593
10  222.605987946333
20	236.8315794509
40	259.3547810777
60	277.3280635033
80 292.3328735742
100 305.3158832789
120 316.8252770306
};
\addlegendentry{{\footnotesize Passive Response - Simulation}}

\addplot [only marks, color=rub_blue, mark=otimes*, mark size=2.5,]
table {%
10	223
20	236
40	257
60	278
80	296
100	306
120	315
};
\addlegendentry{{\footnotesize Passive Response - Experiment}}

\end{axis}
\end{tikzpicture}}
\put(0.0,5.9){(a)}

\put( 0.0, 0.5){\includegraphics[height=4.2cm]{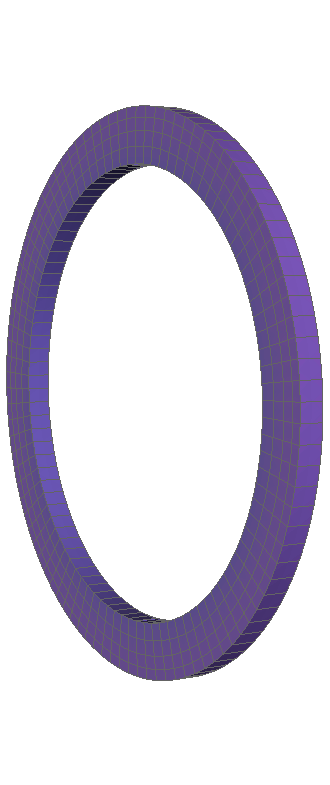}}
\put( 0.2, 0.0){20$\,$mmHg}
\put( 2.0, 0.5){\includegraphics[height=4.2cm]{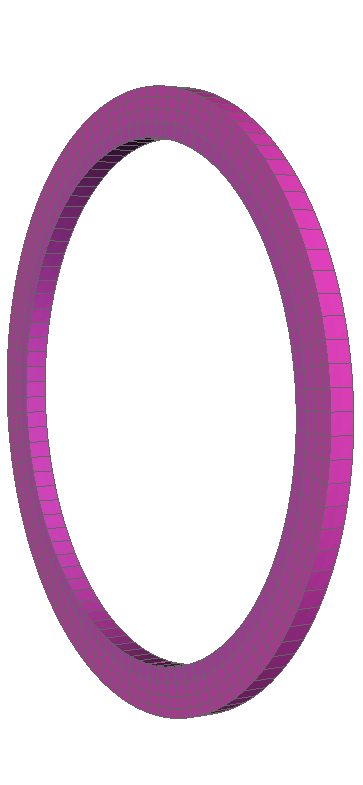}}
\put( 2.2, 0.0){40$\,$mmHg}
\put( 4.2, 0.5){\includegraphics[height=4.2cm]{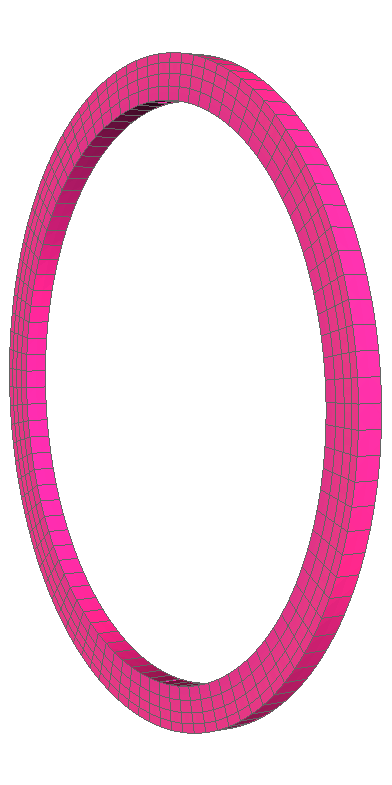}}
\put( 4.5, 0.0){60$\,$mmHg}
\put( 6.5, 0.5){\includegraphics[height=4.2cm]{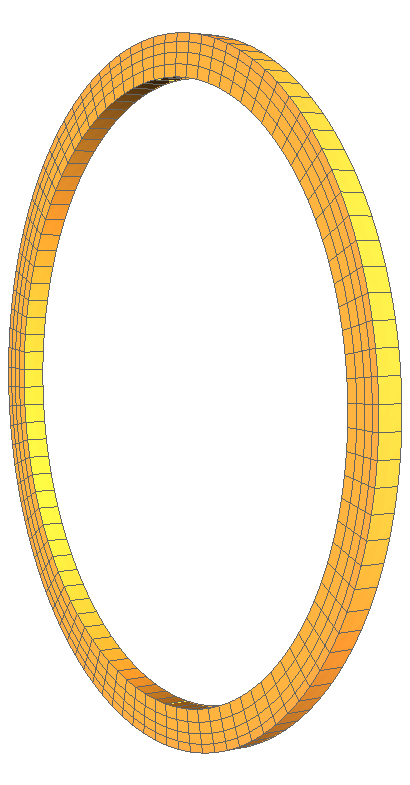}}
\put( 6.9, 0.0){80$\,$mmHg}
\put( 9.0, 0.5){\includegraphics[height=4.2cm]{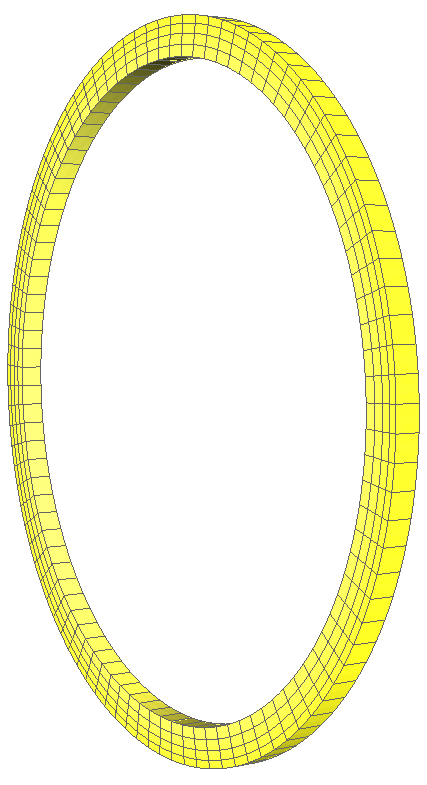}}
\put( 9.4, 0.0){100$\,$mmHg}
\put( 11.5, 0.5){\includegraphics[height=4.2cm]{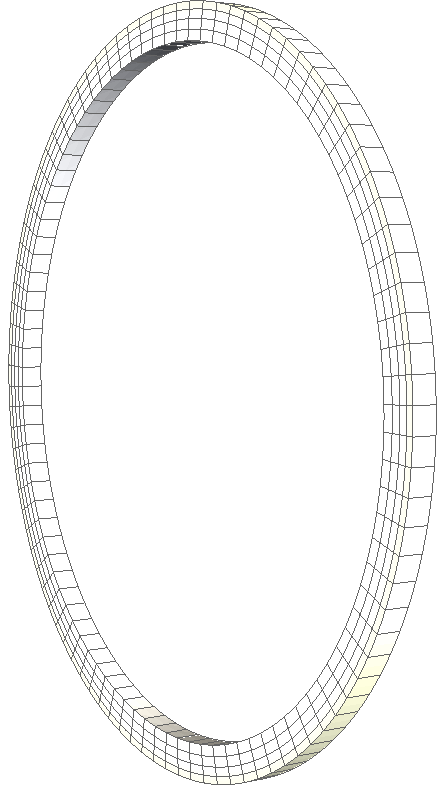}}
\put( 11.9, 0.0){120$\,$mmHg}
\put( 14.2, 0.5){\includegraphics[height=4.2cm]{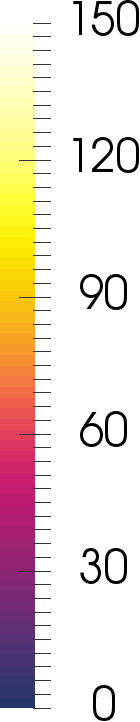}}
\put( 15.2, 0.3){\rotatebox{90}{Circ. Cauchy Stress in kPa}}
\put(0.0,0.6){(b)}

\end{picture}
\caption{
(a)~Comparison of simulation results for the passive response of the arterial wall (solid line) with experimental data from \cite{JohElyTakWalWalCol:2009:csv} (dots) showing an excellent agreement of the model response with the experimental data.
(b)~Comparison of circumferential Cauchy stresses of the passive material response at different times of the pressure profile showing almost homogeneous stress distributions in line with the homeostasis hypothesis.
However, the quantitative values of the stresses reach unrealistically high values, which should not be expected for physiological conditions.
This is caused by the application of a high intravascular pressure on a muscular artery while not including a realistic model for the active response.
\label{fig:validation}}
\end{figure}
Based on this setup, the optimization procedure is performed and leads to the results shown in Fig.~\ref{fig:opt1} and~\ref{fig:validation}, see the standard case in Table~\ref{tab:num}.
Fig.~\ref{fig:opt1}a and~b illustrate the distributions of the Cauchy stresses~$\sigma^{\text{(ax)}}$ and~$\sigma^{\text{(cir)}}$ in axial and circumferential direction, respectively, and the driving forces~$\phi^{\text{(2)}}$ and~$\phi^{\text{(3)}}$ before (dashed lines) and after (solid lines) the growth process.
For growth in the direction of the second eigenvector of the Cauchy stress, which is tagged here with~(ax), the stress and the driving force reach the predefined value of~0$\,$kPa.
Due to the aim to reduce the volumetric change of the reference geometry (see~$z_2$), the driving force for the growth in direction of the third eigenvector of the Cauchy stress (tagged with~(rad)) was optimized to 139.71$\,$kPa which was reached over the entire wall thickness (see Fig.~\ref{fig:opt1}b).
Consequently, the growth model achieves its purpose to equalize the driving force of the model over the entire wall and minimizes the absolute value of the gradient of the stresses.
Nonetheless, it should be noted that the gradient of the stresses over the wall in circumferential direction is not zero (see Fig.~\ref{fig:opt1}b).
This circumstance is related to the driving force~$\phi^{(3)}$ for~$\bF_{\text{g}}^{(3)}$ which is defined as the trace of the Cauchy stresses and involves the stresses in radial direction, which may be considered with caution.
The radial stresses are compelled to show a gradient over the wall thickness based on the boundary value problem which defines an intravascular pressure of~120$\,$mmHg at the inner side of the wall and a pressure of~0$\,$mmHg at the outer side.
Accordingly, the circumferential stresses have to show a gradient with opposed sign as soon as all driving forces~$\phi^{(a)}$ of the growth model reached a homogenization over the entire wall.
From a quantitative point of view, the results for the circumferential stresses could be further improved by excluding the radial stresses from the driving force~$\phi^{(3)}$.
This adjustment will be considered in the simulations with active response in Section~\ref{sec:act}.\\
As additional overview, Fig.~\ref{fig:opt1}c shows the final state of the growth factors~$\vartheta^{(2)}$ and~$\vartheta^{(3)}$.
Since the growth in axial direction is not restricted by the optimization procedure, the growth factor~$\vartheta^{(2)}$ reaches values between~1.35 at the inner side and~1.3 at the outer side of the arterial wall.
For the growth in radial direction, the volumetric change is minimized by the objective function~$z_2$.
Correspondingly, the growth factor is ranging from~1.3 at the inner side of the wall to~0.7 at the outer side of the wall and is, consequently, close to 1.0 in the middle of the arterial wall.
The inner and outer radii $r_{\text{i,} \, \text{ref}}$, and $r_{\text{o,} \, \text{ref}}$ of the arterial ring in the reference configuration are optimized to~82.91$\, \mu$m and~103.11$\, \mu$m.
The ratio between wall thickness and outer radius in the resulting geometry at an intravascular pressure of~15$\,$mmHg (simulation 2(b)) is~0.209997 which is close to the set target value of 0.21.
One further result is shown in Fig.~\ref{fig:validation}a which illustrates the outer diameter of the arterial wall for an increasing intravascular pressure for simulation~2(c) and the experimental data.
Equal to the values in the experiment, the intravascular pressure~$p_k$ was set to values from~0$\,$mmHg to~120$\,$mmHg.
The diagram shows that the optimization indeed accurately identified material parameters for the resulting geometry to have the model match the experimental data.
In Fig.~\ref{fig:validation}b, the circumferential Cauchy stress is illustrated as contour plots of the 3D arterial ring.
The illustration indicates that the stresses are nearly homogeneously distributed in all load scenarios.
Note that the stress values in the simulations of this section are unrealistically high, which can be explained by the fact that a muscular artery has been investigated without incorporating the active response of SMCs.
It can be expected that muscular arteries receive noticeable tissue damage during experiments where only the passive response is investigated.
Damage is not recorded during the experiments and would be expensive to consider in the material model.
However, from a simulation perspective, the combination of the optimization procedure and the growth model leads to convincing qualitative results.

\subsection{Convergence of Growth Over Time}
\label{subsec:mandel}
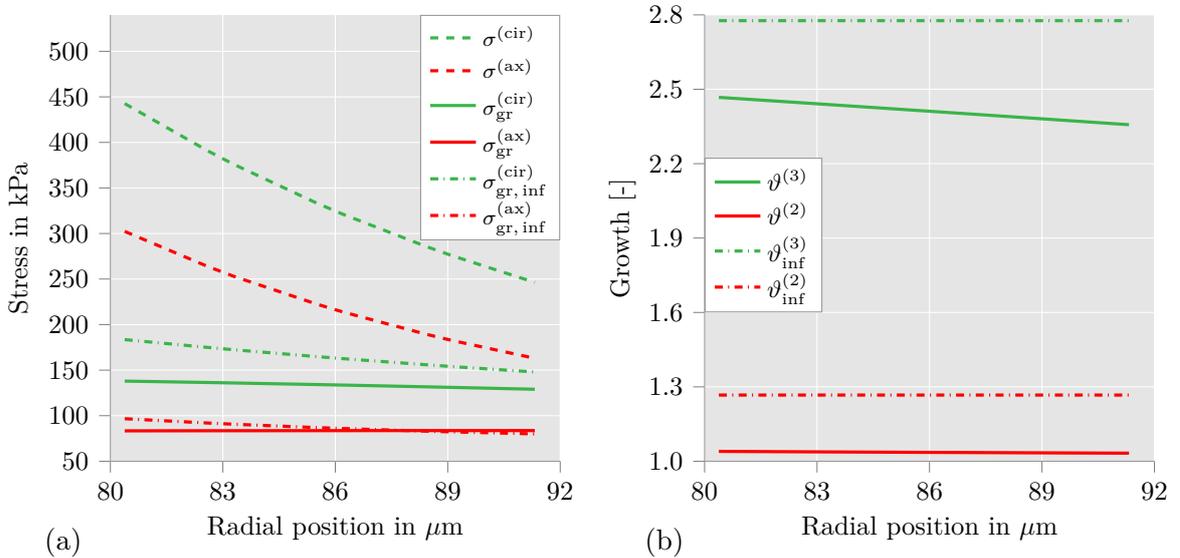
\begin{figure}[!b]
\unitlength1cm
\begin{picture}(15.5,7.0)
\put( 0.0, 0.0){
\put( 0.0, 0.0){
\begin{tikzpicture}
\begin{axis}[
width=7.5cm,
height=7.5cm,
xlabel={{Radial position in $\mu$m}},
xmin=80, xmax=92,
xtick={80, 83, 86, 89, 92},
extra x ticks={0.0},
extra x tick style={%
    grid=major,
},
ylabel={{Stress in kPa}},
ymin=50, ymax=540,
ytick={50, 100, 150, 200, 250, 300, 350, 400, 450, 500},
axis background/.style={fill=white!89.80392156862746!black},
axis line style={white!60.0!black},
axis x line*=bottom,
axis y line*=left,
tick align=outside,
tick pos=left,
scaled x ticks=false,
x grid style={white},
xmajorgrids,
x tick label style={
  font=\small,
  /pgf/number format/.cd,
  set decimal separator={.},
    fixed,
    fixed zerofill,
    precision=0,
  /tikz/.cd
},
xlabel style={at={(0.5,-0.1)},font=\small},
ylabel style={at={(-0.16,0.5)},font=\small},
y grid style={white},
ymajorgrids,
y tick label style={
  font=\small,
},
legend cell align={left},
legend style={at={(1.0,1.0)}, anchor=north east, draw=white!60.0!black, row sep=-0.02cm},
clip mode=individual,
]

\addplot [dashed, very thick,rub_green, mark options={solid, draw=\tcolorshade, line width=0.5pt}]
table {%
   80.3801403651576        442.599949604147 
   82.5892626974694        390.890697562682     
   83.2915818328207        375.914966512935     
   85.4998820736245        333.444927882327     
   86.2022027448999        321.148816818782      
   88.4113279896968        285.926255716476     
   89.1131703176982        275.675152841148    
   91.3219760376094        246.298863904747                  
};
\addlegendentry{{\scriptsize $\sigma^{\text{(cir)}}$}}

\addplot [dashed, very thick,red, mark options={solid, draw=\tcolorshade, line width=0.5pt}]
table {%
   80.3801403651576        302.311414832200
   82.5892626974694        263.916222097726     
   83.2915818328207        252.974335972548      
   85.4998820736245        222.537106290117     
   86.2022027448999        213.879787922209     
   88.4113279896968        189.535209607264     
   89.1131703176982        182.570973778058    
   91.3219760376094        163.000864432658                  
};
\addlegendentry{{\scriptsize $\sigma^{\text{(ax)}}$}}

\addplot [very thick,rub_green, mark options={solid, draw=\tcolorshade, line width=0.5pt}]
table {%
   80.3801403651576        137.967219330533 
   82.5892626974694        136.444947785858     
   83.2915818328207        135.905928811982      
   85.4998820736245        134.186916279872     
   86.2022027448999        133.608541521314   
   88.4113279896968        131.713510010693     
   89.1131703176982        131.074411804517     
   91.3219760376094        129.019960945424              
};
\addlegendentry{{\scriptsize $\sigma^{\text{(cir)}}_{\text{gr}}$}}

\addplot [very thick,red, mark options={solid, draw=\tcolorshade, line width=0.5pt}]
table {%
   80.3801403651576        83.3585948707313 
   82.5892626974694        83.4839037483962     
   83.2915818328207        83.5001510638008      
   85.4998820736245        83.5864454767338     
   86.2022027448999        83.6061240453832    
   88.4113279896968        83.6569800945717     
   89.1131703176982        83.6584758663785     
   91.3219760376094        83.6687114763577             
};
\addlegendentry{{\scriptsize $\sigma^{\text{(ax)}}_{\text{gr}}$}}

\addplot [dash dot,very thick,rub_green, mark options={solid, draw=\tcolorshade, line width=0.5pt}]
table {%
   80.3801403651576        183.525329051120 
   82.5892626974694        174.966800444247     
   83.2915818328207        172.375757536499     
   85.4998820736245        164.875200121412     
   86.2022027448999        162.623938689330    
   88.4113279896968        155.962162537288     
   89.1131703176982        153.935617600173      
   91.3219760376094        147.949991999836                     
};
\addlegendentry{{\scriptsize $\sigma^{\text{(cir)}}_{\text{gr,} \, \text{inf}}$}}

\addplot [dash dot,very thick,red, mark options={solid, draw=\tcolorshade, line width=0.5pt}]
table {%
   80.3801403651576        96.6819225636145   
   82.5892626974694        91.9853283631197     
   83.2915818328207        90.6240995471235   
   85.4998820736245        86.9482997002986     
   86.2022027448999        85.9097749656023     
   88.4113279896968        83.0471580114311     
   89.1131703176982        82.2279625345226    
   91.3219760376094        80.0164343895053                 
};
\addlegendentry{{\scriptsize $\sigma^{\text{(ax)}}_{\text{gr,} \, \text{inf}}$}}

\end{axis}
\end{tikzpicture}}
\put( 0.6, 0.0){(a)}
\put( 7.9, 0.0){
\begin{tikzpicture}
\begin{axis}[
width=7.5cm,
height=7.5cm,
xlabel={{Radial position in $\mu$m}},
xmin=80, xmax=92,
xtick={80, 83, 86, 89, 92},
extra x ticks={0.0},
extra x tick style={%
    grid=major,
},
ylabel={{Growth [-]}},
ymin=1.0, ymax=2.8,
ytick={1.0, 1.3, 1.6, 1.9, 2.2, 2.5, 2.8},
axis background/.style={fill=white!89.80392156862746!black},
axis line style={white!60.0!black},
axis x line*=bottom,
axis y line*=left,
tick align=outside,
tick pos=left,
scaled x ticks=false,
x grid style={white},
xmajorgrids,
x tick label style={
  font=\small,
  /pgf/number format/.cd,
  set decimal separator={.},
    fixed,
    fixed zerofill,
    precision=0,
  /tikz/.cd
},
y tick label style={
  font=\small,
  /pgf/number format/.cd,
  set decimal separator={.},
    fixed,
    fixed zerofill,
    precision=1,
  /tikz/.cd
},
xlabel style={at={(0.5,-0.1)},font=\small},
ylabel style={at={(-0.13,0.5)},font=\small},
y grid style={white},
ymajorgrids,
y tick label style={
  font=\small,
},
legend cell align={left},
legend style={at={(0.0,0.3333)}, anchor=south west, draw=white!60.0!black, row sep=-0.04cm},
clip mode=individual,
]

\addplot [very thick,rub_green, mark options={solid, draw=\tcolorshade, line width=0.5pt}]
table {%
   80.3801403651576        2.46731016108230
   82.5892626974694        2.44552975791739
   83.2915818328207        2.43847165444789
   85.4998820736245        2.41635915068612
   86.2022027448999        2.40926997195594
   88.4113279896968        2.38687912237059
   89.1131703176982        2.37967264328189
   91.3219760376094        2.35701997898356
};
\addlegendentry{{\footnotesize $\vartheta^{\text{(3)}}$}}

\addplot [very thick,red, mark options={solid, draw=\tcolorshade, line width=0.5pt}]
table {%
   80.3801403651576        1.03978689128300 
   82.5892626974694        1.03809792971610
   83.2915818328207        1.03758222217848          
   85.4998820736245        1.03605880197578     
   86.2022027448999        1.03559759239881    
   88.4113279896968        1.03421965201789     
   89.1131703176982        1.03379978979941     
   91.3219760376094        1.03254847186476                  
};
\addlegendentry{{\footnotesize $\vartheta^{\text{(2)}}$}}

\addplot [dash dot, very thick,rub_green, mark options={solid, draw=\tcolorshade, line width=0.5pt}]
table {%
   80.3801403651576        2.77662278200000 
   82.5892626974694        2.77662278199999     
   83.2915818328207        2.77662278199999     
   85.4998820736245        2.77662278199999     
   86.2022027448999        2.77662278199999    
   88.4113279896968        2.77662278199999     
   89.1131703176982        2.77662278199999     
   91.3219760376094        2.77662278199999       
};
\addlegendentry{{\footnotesize $\vartheta^{\text{(3)}}_{\text{inf}}$}}

\addplot [dash dot, very thick,red, mark options={solid, draw=\tcolorshade, line width=0.5pt}]
table {%
   80.3801403651576        1.26689864399997  
   82.5892626974694        1.26689864399997     
   83.2915818328207        1.26689864399997      
   85.4998820736245        1.26689864399996     
   86.2022027448999        1.26689864399996      
   88.4113279896968        1.26689864399996     
   89.1131703176982        1.26689864399996   
   91.3219760376094        1.26689864399996                       
};
\addlegendentry{{\footnotesize $\vartheta^{\text{(2)}}_{\text{inf}}$}}

\end{axis}
\end{tikzpicture}}
\put( 8.5, 0.0){(b)}
}
\end{picture}
\caption{Distribution of 
(a)~Cauchy stresses~$\sigma^{\text{(ax)}}$ and~$\sigma^{\text{(cir)}}$, and
(b)~growth values~$\vartheta^{(2)}$ and~$\vartheta^{(3)}$
in circumferential (green) and axial (red) direction over the wall thickness for the passive material response with application of growth model B which is based on the elastic part of the Mandel stress~$\BSigma_{\text{e}}$ (see~\cite{ZahBal:2018:acg}).
As for model A, the artery is loaded with an intravascular pressure of~120$\,$mmHg.
Dashed lines show the results before growth, solid lines show the results after the optimal time for growth, dashed-dotted lines show results after almost infinite time of growth (representing the saturated state).
As can be seen, the gradients and the values of the stresses are significantly decreased for the optimal growth time.
However, the growth increases again after this specifically chosen time instance and quite larger stress values are found in the saturated state.
\label{fig:opt_mandel}}
\end{figure}
\begin{figure}[!b]
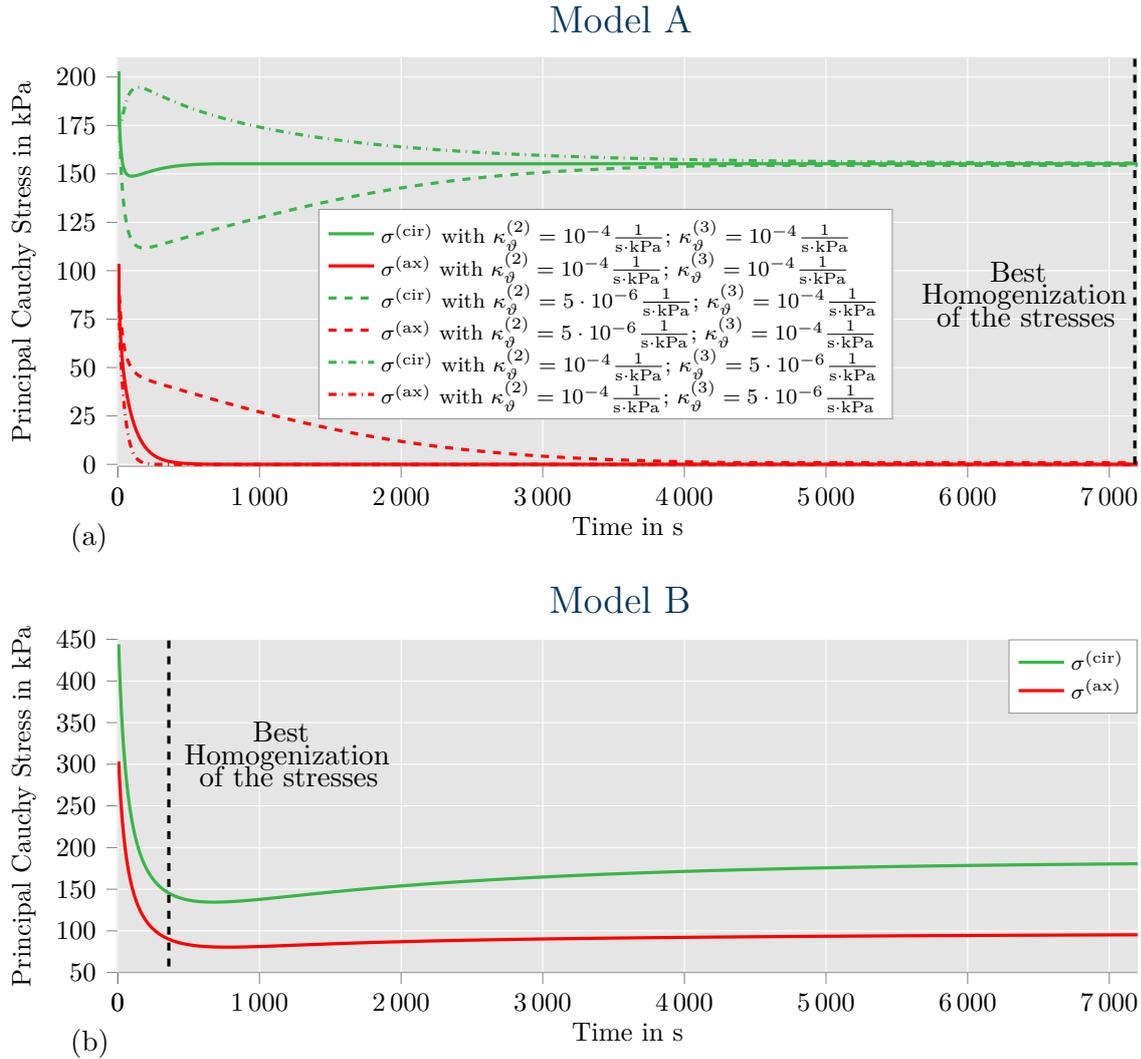

\unitlength1cm
\begin{picture}(15.5,13.8)
\put(0.0,6.7){
\put(7.2,6.8){\color{rub_blue} \Large Model A}
\put(0.0,0.0){\input{\figpath/Progress_Stress}}
\put(13.0,3.5){\color{black} Best}
\put(12.1,3.2){\color{black} Homogenization}
\put(12.3,2.9){\color{black} of the stresses}
\put(0.9,0.0){(a)}
}

\put(0.0,0.0){
\put(7.2,5.8){\color{rub_blue} \Large Model B}
\put(0.0,0.0){\input{\figpath/Progress_Stress_mandel}}
\put(3.3,4.1){\color{black} Best}
\put(2.4,3.8){\color{black} Homogenization}
\put(2.6,3.5){\color{black} of the stresses}
\put(0.9,0.0){(b)}
}
\end{picture}
\caption{
Comparison of the evolution of the stress in axial and circumferential direction at the most inner Gauss point (marked in Figure~\ref{fig:mesh}).
(a)~Varying values of the growth velocity factors~$\kappa_\vartheta^{(2)}$ and~$\kappa_\vartheta^{(3)}$ in the new growth model (model A).
The graphs show that variations of the growth velocity factors have no impact on the final value of the stresses.
(b)~For the previous version of the growth model (model B) which is dependent on the elastic part of the Mandel stress~$\BSigma_{\text{e}}$ as driving force and evolution equations aiming to saturate in defined growth states.
An almost homogeneous stress distribution is reached at 360s run time of the growth process.
But this growth state is not stable, growth actually converges to a state of non-homogeneous stress distributions.
Accordingly, a combination with additional active processes such as smooth muscle contraction is difficult if not impossible.
The proposed formulation (model A) does not show this instability.
\label{fig:progress}}
\end{figure}
In Section~\ref{subsec:sim}, it was shown that the new growth model can be applied to generate homogenized stress distributions in the arterial wall.
However, other models were already able to show similar results.
One example are growth models that depend on the elastic part of the Mandel stress~$\BSigma_{\text{e}} = \bC_e \bS_e$ as driving force and that take into account evolution equations which aim to converge to growth intensities, not stress intensities (see e.g. the previous version of our growth model~\cite{ZahBal:2018:acg} and~\cite{Zah:2020:mog}).
A comparison of the proposed growth model with the one from~\cite{ZahBal:2018:acg} will show that the previously established evolution equations can only generate suitably homogenized stress distributions when the growth process is stopped at a specific growth state, which differs from the state at infinite time.
Actually, an unstable growth may even be observed which not saturates in a specific grown state.
Since it is of course difficult to estimate this kind of optimal time instance, especially for varying load scenarios or when linking the formulation with an active SMC model, this non-time-converging response is problematic for practical application.
This will be shown to not be the case for the new growth model.
Note that from now on we denote the new proposed growth model as ``model A'' and the previous model~\cite{ZahBal:2018:acg} as ``model B'' to avoid misunderstandings.
The major difference between model A and model B is the definition of the evolution equations for the growth factor, see Eq.~(\ref{eq:groFac}).
A brief explanation of these equations for model B are shown in the Appendix~\ref{sec:mandel}.\\
To enable an objective display of the differences between both models, we executed a similar optimization as described in Section~\ref{subsec:opt_pro} for model B.
However, based on the definition of the evolution equations in model B, the growth factors are only able to increase (not descrease).
The volumetric change has thus to be positive after the application of model B.
In consequence, a restriction of the volumetric change from reference configuration to the resulting geometry after growth is not possible in the same manner.
Therefore, the optimization for objective function~$z_1$ is not applied for model B.
In addition, there is no specific target value set for the stresses~$\sigma^{\text{(ax)}}$ and~$\sigma^{\text{(cir)}}$, but only a minimization of the absolute value of the gradient for the stresses over the wall thickness.
To reach this objective, the parameters~$\vartheta^+_{(2)}$, $\vartheta^+_{(3)}$, $\kappa^+_{\vartheta, \, (2)}$ and~$\kappa^+_{\vartheta, \, (3)}$ were added as optimization parameters to the optimization procedure.\\
The results for the stress distribution of the optimized simulation are shown in Fig.~\ref{fig:opt_mandel}a.
The corresponding growth factors are displayed in Fig.~\ref{fig:opt_mandel}b.
It can be seen that the stresses~$\sigma^{\text{(ax)}}$ and~$\sigma^{\text{(cir)}}$ are nearly equal over the entire wall after an optimal growth time (solid lines).
At this state of the growth process, the material parameters~$\alpha_1$, $\alpha_4$ and~$\alpha_5$ are optimized to fit the experimental data (see Fig.~\ref{fig:validation_mandel} in the Appendix~\ref{sec:mandel}) and the ratio between the wall thickness and the outer radius is~0.20995 (compare objective function~$z_2$).
However, the growth process is stopped manually at this optimal run-time.
This manual stop was accomplished by checking the evolution of the stress distribution over the wall thickness during the simulation.
If the growth model is not manually stopped, the growth process continues until the evolution equations of the growth factors are zero.
Based on the definition of the evolution equations in model B, this circumstance is not reached when the growth factors are equal to their maximal values~$\vartheta^+_{(2)}$ and~$\vartheta^+_{(3)}$ which occurs after a longer run-time.
The corresponding results are illustrated by the dashed-dotted lines in Fig.~\ref{fig:opt_mandel}.
As can be seen, the circumferential Cauchy stresses on the inner side of the artery are approximately~40$\,$kPa larger than the circumferential Cauchy stresses at the outer side of the artery which yields noticeably worse results for the stress distribution. \\
As an additional comparison between model A and model B, Fig.~\ref{fig:progress} illustrates the evolution of the axial and circumferential stresses ($\sigma^{\text{(ax)}}$ and~$\sigma^{\text{(cir)}}$) for the Gauss point at the inner side of the arterial wall (compare Fig.~\ref{fig:mesh}b).
In Fig.~\ref{fig:progress}a, the evolution is shown for different values of the growth velocity factors~$\kappa_\vartheta^{(2)}$ and~$\kappa_\vartheta^{(3)}$ for model A.
The values~$\kappa_\vartheta^{(2)} = 10^{-4} (\text{s} \cdot \text{kPa})^{-1}$ and~$\kappa_\vartheta^{(3)} = 10^{-4} (\text{s} \cdot \text{kPa})^{-1}$ (solid lines) were used in the optimization in Section~\ref{subsec:sim}.
A reduction of~$\kappa_\vartheta^{(2)}$ (dashed lines) or~$\kappa_\vartheta^{(3)}$ (dashed-dotted lines) by a factor of~20 results in a considerable change of the evolution of the stresses.
However, the final results for the stresses are independent from the ratios between the growth velocity factors.
In comparison, the evolution of the stresses for model B confirms in Fig.~\ref{fig:progress}b what was already indicated by the changes of the stress distributions in Fig.~\ref{fig:opt_mandel}: the optimal homogenization of stresses is reached after 360$\,$s run-time of the growth process, but still changes as long as the growth process is not manually stopped.
In consequence, a qualitatively accurate coupling of model B with another active material process such as the active SMC response is difficult to achieve.

\subsection{Efficiency of the Optimization Procedure}
\label{sec:num}
There are mainly two aspects which have to be defined to obtain a characteristic structural problem of an artery when the optimization procedure with the new growth model is applied.
In context of the optimization shown in Section~\ref{subsec:sim}, these aspects are: the fiber direction of the collagen fibers (and SMCs) described by the angle~$\beta$ and the in vivo prestretch leading to residual stresses in axial direction which are linked to the convergence value~$\phi^{(2)}_{\text{con}}$ of the driving force~$\phi^{(2)}$ in axial direction.
To demonstrate the performance of the optimization procedure in combination with the new growth model, further optimizations were performed which include a variation of~$\beta$ or~$\phi^{(2)}_{\text{con}}$ in the simulations.
Furthermore, the boundary conditions of the artery were varied in additional optimizations by applying different values for the axial stretch~$\lambda_{\text{ax}}$ of the artery at the start of the simulation.
Note that this axial stretch value~$\lambda_{\text{ax}}$ represents an externally applied axial deformation induced by e.g. moving parts of the body which surround the artery.
It should not be confused with the in vivo axial prestretch of an artery which contributes to the residual stress state induced by growth which prevents supra-physiological stresses when externally stretching the artery.
The results for the optimized parameters and the final value of the objective functions~$z_1$, $z_2$ and~$z_3$ are listed in Table~\ref{tab:num}.
\begin{table}[h]
\caption{Results for the optimization parameters and the corresponding values of the objective functions~$z_1$, $z_2$ and~$z_3$ for variations of the angle of the collagen fibers~$\beta$, the axial stretch~$\lambda_{\text{ax}}$ or the convergence value of the growth factor~$\phi^{(2)}_{\text{con}}$.
The standard case refers to the scenario discussed in Section~\ref{subsec:sim} where the axial stretch is~$\lambda_{\text{ax}}=1.0$. \label{tab:num}}
\small
\centering
\resizebox{\textwidth}{!}{%
\begin{tabular}{|r|cccccc|ccc|}
\hline
 & \multicolumn{6}{|c|}{Optimization Parameters} & \multicolumn{3}{|c|}{Objective Function} \\
\hline
Variation & $\alpha_1$ & $\alpha_4$ & $\alpha_5$ & $r_{\text{o,} \, \text{ref}}$ & $r_{\text{i,} \, \text{ref}}$ & $w_{\text{min}}$ & $z_1$ & $z_2$ & $z_3$ \\
\hline
Standard Case & 2.79$\,$kPa & 9.90$\,$kPa & 2.99 & 103.11$\, \mu$m & 82.91$\, \mu$m & 0.77 & 0.00645 & 2.74e-06 & 0.00494 \\
$\beta = 10^\circ-30^\circ$ & 3.08$\,$kPa & 10.48$\,$kPa & 3.08 & 102.44$\, \mu$m & 81.84$\, \mu$m & 0.783 & 0.00868 & 8.51e-06 & 0.00658 \\
$\beta = 10^\circ-40^\circ$ & 2.83$\,$kPa & 19.87$\,$kPa & 2.54 & 108.44$\, \mu$m & 87.54$\, \mu$m & 0.857 & 0.01371 & 3.47e-05 & 0.00632 \\
$\beta = 30^\circ$ & 3.62$\,$kPa & 14.51$\,$kPa & 3.13 & 96.95$\, \mu$m & 77.50$\, \mu$m & 0.776 & 0.00535 & 8.75e-05 & 0.00460 \\
$\beta = 45^\circ$ & 3.76$\,$kPa & 39.86$\,$kPa & 2.76 & 88.64$\, \mu$m & 69.99$\, \mu$m & 0.915 & 0.00529 & 9.53e-06 & 0.00476 \\
$\lambda_{\text{ax}}=1.1$ & 2.79$\,$kPa & 11.46$\,$kPa & 3.10 & 102.98$\, \mu$m & 83.24$\, \mu$m & 0.799 & 0.00552 & 9.31e-05 & 0.00652 \\
$\lambda_{\text{ax}}=1.2$ & 2.38$\,$kPa & 14.89$\,$kPa & 3.00 & 105.17$\, \mu$m & 85.19$\, \mu$m & 0.895 & 0.00583 & 7.94e-05 & 0.00614 \\
$\lambda_{\text{ax}}=1.3$ & 1.75$\,$kPa & 19.83$\,$kPa & 2.92 & 107.22$\, \mu$m & 87.48$\, \mu$m & 0.986 & 0.00607 & 7.18e-06 & 0.00584 \\
$\lambda_{\text{ax}}=1.4$ & 1.56$\,$kPa & 25.077$\,$kPa & 2.85 & 108.39$\, \mu$m & 89.36$\, \mu$m & 1.079 & 0.00619 & 9.54e-05 &  0.00557 \\
$\phi^{(2)}_{\text{con}}=10 \, $kPa & 1.28$\,$kPa & 14.02$\,$kPa & 2.56 & 107.04$\, \mu$m & 87.04$\, \mu$m & 0.852 & 0.00634 & 8.84e-05 & 0.00540 \\
$\phi^{(2)}_{\text{con}}=20 \, $kPa & 1.52$\,$kPa & 10.71$\,$kPa & 2.52 & 105.82$\, \mu$m & 84.50$\, \mu$m & 0.937 & 0.00663 & 3.54e-05 & 0.00554 \\
$\phi^{(2)}_{\text{con}}=30 \, $kPa & 1.56$\,$kPa & 8.61$\,$kPa & 2.54 & 103.83$\, \mu$m & 82.29$\, \mu$m & 0.959 & 0.00639 & 2.59e-05 & 0.00581 \\
$\phi^{(2)}_{\text{con}}=40 \, $kPa & 3.11$\,$kPa & 5.47$\,$kPa & 2.66 & 101.01$\, \mu$m & 78.53$\, \mu$m & 0.984 & 0.00607 & 7.11e-05 & 0.00637 \\
$\phi^{(2)}_{\text{con}}=50 \, $kPa & 4.93$\,$kPa & 3.56$\,$kPa & 2.75 & 101.09$\, \mu$m & 77.17$\, \mu$m & 1.009 & 0.00572 & 0.000232 & 0.00656 \\
\hline
\end{tabular}
}
\end{table}
Note that the notation $\beta = 10^\circ - 30^\circ$ indicates a fiber angle which is linearly distributed from~
$10^\circ$ at the inner side of the wall to~$30^\circ$ at the outer side.
As can be seen in Table~\ref{tab:num}, the optimization is able to generate comparable values for the objective function, i.e. for almost all cases the same order of magnitude, nearly independent from the variations of the simulations.
The case with~$\phi^{(2)}_{\text{con}}=50 \,$kPa could be seen as an outlier for the fitting of the objective function~$z_2$ with~0.000232.
However, $z_2$ is used for the fitting of the resulting geometry in the grown state and the corresponding ratio between wall thickness and outer radius is still 0.2098 with 0.21 as target value.
Additional outliers may be identified in the cases with~$\beta = 10^\circ - 30^\circ$ and~$\beta = 10^\circ - 40^\circ$.
Here, the objective function~$z_1$, which is responsible for the reduction of the volumetric change during the growth process, is noticeably larger than in the other optimizations.
However, this can be explained by the necessary change of the growth factor~$\vartheta^{(3)}$ to adjust the stresses appropriately.\\
%
\begin{figure}[!tbhp]
\unitlength1cm
\begin{picture}(15.5,21.0)
\put( 3.5, 20.5){\color{rub_blue}\LARGE Variation of Collagen Fiber Angle $\beta$}

\put( 1.4, 11.0){
\put( 0.5, 8.5){\color{rub_blue} Circumferential Stress}
\put( -1.5, 0.0){
\begin{tikzpicture}
\begin{axis}[
width=6.0cm,
height=8.5cm,
xlabel={{Radial position [-]}},
xmin=0, xmax=1,
xtick={0.2, 0.4, 0.6, 0.8, 1.0},
extra x ticks={0.0},
extra x tick style={%
    grid=major,
},
ylabel={{Stress in kPa}},
ymin=50, ymax=225,
ytick={50, 75, 100, 125, 150, 175, 200, 225},
axis background/.style={fill=white!89.80392156862746!black},
axis line style={white!60.0!black},
axis x line*=bottom,
axis y line*=left,
tick align=outside,
tick pos=left,
scaled x ticks=false,
x grid style={white},
xmajorgrids,
x tick label style={
  font=\small,
  /pgf/number format/.cd,
  set decimal separator={.},
    fixed,
    fixed zerofill,
    precision=1,
  /tikz/.cd
},
xlabel style={at={(0.5,-0.08)},font=\small},
ylabel style={at={(-0.20,0.5)},font=\small},
y grid style={white},
ymajorgrids,
y tick label style={
  font=\small,
},
legend cell align={left},
legend style={at={(1.0,1.0)}, anchor=north west, draw=white!60.0!black, row sep=0.155cm},
clip mode=individual,
]

\addplot [dash dot,very thick,dia1, mark options={solid, draw=\tcolorshade, line width=0.5pt}]
table {%
0.0     180.870068687687
0.2018322547784532     154.332450807604
0.265971938705891     146.877743108143
0.46777759361303484     126.392358363511
0.5319381975319625     120.590305559627
0.7338473588066857     104.458379329313
0.7980529157464226     99.9026713042858
1.0     86.7212284807056            
};
\addlegendentry{{\small $\sigma^{\text{(cir)}}$}}

\addplot [dash dot,very thick,dia2, mark options={solid, draw=\tcolorshade, line width=0.5pt}]
table {%
0.0     184.617534660731
0.20186735446189283     154.253063089455
0.26604175578101197     145.594419285704
0.4679512271855926     121.254735324981
0.5322242377214047     114.314973461478
0.7339295714345633     94.7116831924245
0.7979854963117159     89.0077633895703
1.0     73.1226036958475    
};
\addlegendentry{{\small $\sigma^{\text{(cir)}}$}}

\addplot [dash dot,very thick,dia3, mark options={solid, draw=\tcolorshade, line width=0.5pt}]
table {%
0.0     182.576900727298
0.20189633443257785     148.872705068854
0.2660744781762834     139.063394996955
0.4678472045497876     110.781530084882
0.5320132276644746     102.591354143074
0.7340607372922476     79.4867656070555
0.7983365585965556     72.9207340551501
1.0     55.0619475307678
};
\addlegendentry{{\small $\sigma^{\text{(cir)}}$}}

\addplot [dash dot,very thick,dia4, mark options={solid, draw=\tcolorshade, line width=0.5pt}]
table {%
0.0     188.749283219399
0.2018623932538243     148.034565534594
0.2660448017440008     137.463483407448
0.4679174260446089     110.491341003529
0.5320767812940534     103.372839508338
0.7339215929671891     84.992922113721
0.7981060015400226     80.0648087150745
1.0     67.1968693647595    
};
\addlegendentry{{\small $\sigma^{\text{(cir)}}$}}

\addplot [dash dot,very thick,dia5, mark options={solid, draw=\tcolorshade, line width=0.5pt}]
table {%
0.0     144.127405750541
0.20185868858615783     113.0873014243
0.2660250203273116     105.055558522734
0.46793077834421604     84.6974119653351
0.5320979386088899     79.3317100663612
0.733971007631767     65.555879606872
0.7981413114132941     61.8617490428641
1.0     52.2855093694315    
};
\addlegendentry{{\small $\sigma^{\text{(cir)}}$}}

\addplot [very thick,dia1, mark options={solid, draw=\tcolorshade, line width=0.5pt}]
table {%
0.0     154.232280890971
0.2018322547784532     150.462185522285
0.265971938705891     149.354813134089
0.46777759361303484     146.154187340912
0.5319381975319625     145.230021148618
0.7338473588066857     142.652394956701
0.7980529157464226     141.977089058914
1.0     139.544945172281     
};
\addlegendentry{{\small $\sigma^{\text{(cir)}}_{\text{gr}}$}}

\addplot [very thick,dia2, mark options={solid, draw=\tcolorshade, line width=0.5pt}]
table {%
0.0     154.610806600533
0.20186735446189283     150.220289290048
0.26604175578101197     148.922584970058
0.4679512271855926     145.301592490265
0.5322242377214047     144.315429443396
0.7339295714345633     141.677806650774
0.7979854963117159     141.046485959549
1.0     139.398664760707
};
\addlegendentry{{\small $\sigma^{\text{(cir)}}_{\text{gr}}$}}

\addplot [very thick,dia3, mark options={solid, draw=\tcolorshade, line width=0.5pt}]
table {%
0.0     154.954113775573
0.20189633443257785     149.841617165783
0.2660744781762834     148.358391527506
0.4678472045497876     144.433279055611
0.5320132276644746     143.533590511825
0.7340607372922476     141.236781338618
0.7983365585965556     141.243723623886
1.0     140.149017799191
};
\addlegendentry{{\small $\sigma^{\text{(cir)}}_{\text{gr}}$}}

\addplot [very thick,dia4, mark options={solid, draw=\tcolorshade, line width=0.5pt}]
table {%
0.0     153.139500857986
0.2018623932538243     149.553217206212
0.2660448017440008     148.491824089198
0.4679174260446089     145.41770380795
0.5320767812940534     144.517275793658
0.7339215929671891     141.906259904227
0.7981060015400226     141.142800865621
1.0     138.906989047782
};
\addlegendentry{{\small $\sigma^{\text{(cir)}}_{\text{gr}}$}}

\addplot [very thick,dia5, mark options={solid, draw=\tcolorshade, line width=0.5pt}]
table {%
0.0     156.064723459774
0.20185868858615783     152.453479559679
0.2660250203273116     151.395057840507
0.46793077834421604     148.325198857083
0.5320979386088899     147.425455848295
0.733971007631767     144.797027730488
0.7981413114132941     144.028965551549
1.0     141.802044500742  
};
\addlegendentry{{\small $\sigma^{\text{(cir)}}_{\text{gr}}$}}

\end{axis}
\end{tikzpicture}}
\put( -0.6, 0.0){(a)}
}
\put( -0.15, 11.0){
\put( 9.4, 8.5){\color{rub_blue} Axial Stress/Driving Force}
\put( 8.0, 0.0){
\begin{tikzpicture}
\begin{axis}[
width=6.0cm,
height=8.5cm,
xlabel={{Radial position [-]}},
xmin=0, xmax=1,
xtick={0.2, 0.4, 0.6, 0.8, 1.0},
extra x ticks={0.0},
extra x tick style={%
    grid=major,
},
ylabel={{Stress in kPa}},
ymin=-1, ymax=125,
ytick={0, 25, 50, 75, 100, 125},
axis background/.style={fill=white!89.80392156862746!black},
axis line style={white!60.0!black},
axis x line*=bottom,
axis y line*=left,
tick align=outside,
tick pos=left,
scaled x ticks=false,
x grid style={white},
xmajorgrids,
x tick label style={
  font=\small,
  /pgf/number format/.cd,
  set decimal separator={.},
    fixed,
    fixed zerofill,
    precision=1,
  /tikz/.cd
},
xlabel style={at={(0.5,-0.08)},font=\small},
ylabel style={at={(-0.16,0.5)},font=\small},
y grid style={white},
ymajorgrids,
y tick label style={
  font=\small,
},
legend cell align={left},
legend style={at={(1.0,1.0)}, anchor=north west, draw=white!60.0!black, row sep=0.155cm},
clip mode=individual,
]

\addplot [dash dot,very thick,dia1, mark options={solid, draw=\tcolorshade, line width=0.5pt}]
table {%
0.0     80.5650547069373
0.2018322547784532     67.453130469594
0.265971938705891     63.929855912861
0.46777759361303484     54.8220264429333
0.5319381975319625     52.3535462948171
0.7338473588066857     45.8848408783539
0.7980529157464226     44.1557064740896
1.0     39.2259650735931       
};
\addlegendentry{{\small $\sigma^{\text{(ax)}}$}}

\addplot [dash dot,very thick,dia2, mark options={solid, draw=\tcolorshade, line width=0.5pt}]
table {%
0.0     81.3822952615271
0.20186735446189283     68.3580420735939
0.26604175578101197     64.9333929000132
0.4679512271855926     56.0830983757801
0.5322242377214047     53.742086842897
0.7339295714345633     47.4914487783165
0.7979854963117159     45.6868330420008
1.0     40.9112772627295    
};
\addlegendentry{{\small $\sigma^{\text{(ax)}}$}}

\addplot [dash dot,very thick,dia3, mark options={solid, draw=\tcolorshade, line width=0.5pt}]
table {%
0.0     77.3666654184308
0.20189633443257785     65.5081821101221
0.2660744781762834     62.5086787931942
0.4678472045497876     54.6186386726452
0.5320132276644746     52.4610927195421
0.7340607372922476     46.4635325878042
0.7983365585965556     44.6689891113355
1.0     39.4694304269185  
};
\addlegendentry{{\small $\sigma^{\text{(ax)}}$}}

\addplot [dash dot,very thick,dia4, mark options={solid, draw=\tcolorshade, line width=0.5pt}]
table {%
0.0     119.791854276078
0.2018623932538243     91.3770192581195
0.2660448017440008     84.1382373773412
0.4679174260446089     66.1368756708047
0.5320767812940534     61.4680494768925
0.7339215929671891     49.7394642016564
0.7981060015400226     46.6433615152677
1.0     38.7895253926824 
};
\addlegendentry{{\small $\sigma^{\text{(ax)}}$}}

\addplot [dash dot,very thick,dia5, mark options={solid, draw=\tcolorshade, line width=0.5pt}]
table {%
0.0     118.443960449119
0.20185868858615783     92.541771249038
0.2660250203273116     85.8651297192042
0.46793077834421604     69.07750192832
0.5320979386088899     64.6644118388467
0.733971007631767     53.4237221881069
0.7981413114132941     50.4133202984929
1.0     42.6749103935717      
};
\addlegendentry{{\small $\sigma^{\text{(ax)}}$}}

\addplot [very thick,dia1, mark options={solid, draw=\tcolorshade, line width=0.5pt}]
table {%
0.0     0.886674393986105
0.2018322547784532     0.855921524578735
0.265971938705891     0.845521031414263
0.46777759361303484     0.809229374174834
0.5319381975319625     0.796450352051417
0.7338473588066857     0.749798168205074
0.7980529157464226     0.73047283482165
1.0     0.66750264997764   
};
\addlegendentry{{\small $\sigma^{\text{(ax)}}_{\text{gr}}$}}

\addplot [very thick,dia2, mark options={solid, draw=\tcolorshade, line width=0.5pt}]
table {%
0.0     0.001416210417206796
0.20186735446189283     0.03605316853226532
0.26604175578101197     0.04938272223139405
0.4679512271855926     0.09994194330538621
0.5322242377214047     0.117994641611699
0.7339295714345633     0.172961869020756
0.7979854963117159     0.186637013387338
1.0     0.186713235086944
};
\addlegendentry{{\small $\sigma^{\text{(ax)}}_{\text{gr}}$}}

\addplot [very thick,dia3, mark options={solid, draw=\tcolorshade, line width=0.5pt}]
table {%
0.0     -0.02315487731129714
0.20189633443257785     0.008034747581780011
0.2660744781762834     0.018027445926405
0.4678472045497876     0.05297904737695634
0.5320132276644746     0.06144854830423449
0.7340607372922476     0.06645085768979611
0.7983365585965556     0.05595520927107164
1.0     0.01093880332317648
};
\addlegendentry{{\small $\sigma^{\text{(ax)}}_{\text{gr}}$}}

\addplot [very thick,dia4, mark options={solid, draw=\tcolorshade, line width=0.5pt}]
table {%
0.0     0.828653392130198
0.2018623932538243     0.765168441579624
0.2660448017440008     0.74452757301863
0.4679174260446089     0.677454056869344
0.5320767812940534     0.655275458632103
0.7339215929671891     0.582447525241564
0.7981060015400226     0.557942613636603
1.0     0.477275337506762
};
\addlegendentry{{\small $\sigma^{\text{(ax)}}_{\text{gr}}$}}

\addplot [very thick,dia5, mark options={solid, draw=\tcolorshade, line width=0.5pt}]
table {%
0.0     0.869217648556065
0.20185868858615783     0.779362565133843
0.2660250203273116     0.751182359312385
0.46793077834421604     0.66367890269241
0.5320979386088899     0.6361203729537
0.733971007631767     0.550916433685124
0.7981413114132941     0.524143981849545
1.0     0.441672767248011 
};
\addlegendentry{{\small $\sigma^{\text{(ax)}}_{\text{gr}}$}}

\end{axis}
\end{tikzpicture}}
\put( 8.9, 0.0){(b)}
}

\put( 1.4, 1.6){
\put( 0.65, 8.5){\color{rub_blue} Radial Driving Force}
\put( -1.5, 0.0){
\begin{tikzpicture}
\begin{axis}[
width=6.0cm,
height=8.5cm,
xlabel={{Radial position [-]}},
xmin=0, xmax=1,
xtick={0.2, 0.4, 0.6, 0.8, 1.0},
extra x ticks={0.0},
extra x tick style={%
    grid=major,
},
ylabel={{Driving Force in kPa}},
ymin=75, ymax=300,
ytick={75, 100, 125, 150, 175, 200, 225, 250, 275, 300},
axis background/.style={fill=white!89.80392156862746!black},
axis line style={white!60.0!black},
axis x line*=bottom,
axis y line*=left,
tick align=outside,
tick pos=left,
scaled x ticks=false,
x grid style={white},
xmajorgrids,
x tick label style={
  font=\small,
  /pgf/number format/.cd,
  set decimal separator={.},
    fixed,
    fixed zerofill,
    precision=1,
  /tikz/.cd
},
xlabel style={at={(0.5,-0.08)},font=\small},
ylabel style={at={(-0.20,0.5)},font=\small},
y grid style={white},
ymajorgrids,
y tick label style={
  font=\small,
},
legend cell align={left},
legend style={at={(1.0,1.0)}, anchor=north west, draw=white!60.0!black, row sep=0.155cm},
clip mode=individual,
]

\addplot [dash dot,very thick,dia1, mark options={solid, draw=\tcolorshade, line width=0.5pt}]
table {%
0.0     246.951805215933
0.2018322547784532     211.284965274232
0.265971938705891     201.372957404071
0.46777759361303484     174.981854936285
0.5319381975319625     167.57386847484
0.7338473588066857     147.514400862012
0.7980529157464226     141.957562260272
1.0     125.666792434802         
};
\addlegendentry{{\small $\phi^{(3)}$}}

\addplot [dash dot,very thick,dia2, mark options={solid, draw=\tcolorshade, line width=0.5pt}]
table {%
0.0     251.615232581595
0.20186735446189283     212.400862823729
0.26604175578101197     201.443865943057
0.4679512271855926     171.487761903947
0.5322242377214047     163.106540749528
0.7339295714345633     139.73774719502
0.7979854963117159     132.833541107961
1.0     114.048374395307
};
\addlegendentry{{\small $\phi^{(3)}$}}

\addplot [dash dot,very thick,dia3, mark options={solid, draw=\tcolorshade, line width=0.5pt}]
table {%
0.0     245.691548118373
0.20189633443257785     204.570141973313
0.2660744781762834     192.997269891588
0.4678472045497876     160.116142816652
0.5320132276644746     150.631252403765
0.7340607372922476     123.890465713617
0.7983365585965556     116.122050704859
1.0     94.7513815768138
};
\addlegendentry{{\small $\phi^{(3)}$}}

\addplot [dash dot,very thick,dia4, mark options={solid, draw=\tcolorshade, line width=0.5pt}]
table {%
0.0     294.778158760689
0.2018623932538243     229.917797906638
0.2660448017440008     213.152657313204
0.4679174260446089     171.278434527645
0.5320767812940534     160.254105496514
0.7339215929671891     132.456237441472
0.7981060015400226     125.004501907106
1.0     106.033535064227   
};
\addlegendentry{{\small $\phi^{(3)}$}}

\addplot [dash dot,very thick,dia5, mark options={solid, draw=\tcolorshade, line width=0.5pt}]
table {%
0.0     249.854748289301
0.20185868858615783     196.895955171474
0.2660250203273116     183.154263283109
0.46793077834421604     148.88137155986
0.5320979386088899     139.80917455571
0.733971007631767     116.91506991202
0.7981413114132941     110.7350790289
1.0     95.0340971721978    
};
\addlegendentry{{\small $\phi^{(3)}$}}

\addplot [very thick,dia1, mark options={solid, draw=\tcolorshade, line width=0.5pt}]
table {%
0.0     139.914269001022
0.2018322547784532     139.934969074565
0.265971938705891     139.947934061599
0.46777759361303484     140.016878941658
0.5319381975319625     140.047749987445
0.7338473588066857     140.1770583062
0.7980529157464226     140.229185678851
1.0     140.403129431109     
};
\addlegendentry{{\small $\phi^{(3)}_{\text{gr}}$}}

\addplot [very thick,dia2, mark options={solid, draw=\tcolorshade, line width=0.5pt}]
table {%
0.0     139.456065129223
0.20186735446189283     139.295951978488
0.26604175578101197     139.243618064421
0.4679512271855926     139.128860000786
0.5322242377214047     139.119481156625
0.7339295714345633     139.235212814535
0.7979854963117159     139.315159641443
1.0     139.588335672304
};
\addlegendentry{{\small $\phi^{(3)}_{\text{gr}}$}}

\addplot [very thick,dia3, mark options={solid, draw=\tcolorshade, line width=0.5pt}]
table {%
0.0     140.010803661351
0.20189633443257785     139.708470841755
0.2660744781762834     139.588104857042
0.4678472045497876     139.317599972602
0.5320132276644746     139.289795921717
0.7340607372922476     139.478036431297
0.7983365585965556     139.559060367384
1.0     139.607159669632
};
\addlegendentry{{\small $\phi^{(3)}_{\text{gr}}$}}

\addplot [very thick,dia4, mark options={solid, draw=\tcolorshade, line width=0.5pt}]
table {%
0.0     138.835535565107
0.2018623932538243     138.928835675968
0.2660448017440008     138.958999965151
0.4679174260446089     139.051373816236
0.5320767812940534     139.078991064698
0.7339215929671891     139.15398698181
0.7981060015400226     139.172636135411
1.0     139.207308894447
};
\addlegendentry{{\small $\phi^{(3)}_{\text{gr}}$}}

\addplot [very thick,dia5, mark options={solid, draw=\tcolorshade, line width=0.5pt}]
table {%
0.0     141.743205291842
0.20185868858615783     141.824198303089
0.2660250203273116     141.847659538689
0.46793077834421604     141.910201389656
0.5320979386088899     141.925424370917
0.733971007631767     141.954679354231
0.7981413114132941     141.956763982444
1.0     141.939199501467
};
\addlegendentry{{\small $\phi^{(3)}_{\text{gr}}$}}

\end{axis}
\end{tikzpicture}}
\put( -0.6, 0.0){(c)}
}
\put( -0.15, 1.6){
\put( 10.4, 8.5){\color{rub_blue} Growth Factors}
\put( 8.0, 0.0){
\begin{tikzpicture}
\begin{axis}[
width=6.0cm,
height=8.5cm,
xlabel={{Radial position [-]}},
xmin=0, xmax=1,
xtick={0.2, 0.4, 0.6, 0.8, 1.0},
extra x ticks={0.0},
extra x tick style={%
    grid=major,
},
ylabel={{Growth [-]}},
ymin=0.28, ymax=1.8,
ytick={0.4, 0.6, 0.8, 1.0, 1.2, 1.4, 1.6, 1.8},
axis background/.style={fill=white!89.80392156862746!black},
axis line style={white!60.0!black},
axis x line*=bottom,
axis y line*=left,
tick align=outside,
tick pos=left,
scaled x ticks=false,
x grid style={white},
xmajorgrids,
x tick label style={
  font=\small,
  /pgf/number format/.cd,
  set decimal separator={.},
    fixed,
    fixed zerofill,
    precision=1,
  /tikz/.cd
},
y tick label style={
  font=\small,
  /pgf/number format/.cd,
  set decimal separator={.},
    fixed,
    fixed zerofill,
    precision=1,
  /tikz/.cd
},
xlabel style={at={(0.5,-0.08)},font=\small},
ylabel style={at={(-0.16,0.5)},font=\small},
y grid style={white},
ymajorgrids,
y tick label style={
  font=\small,
  /pgf/number format/.cd,
  set decimal separator={.},
    fixed,
    fixed zerofill,
    precision=1,
  /tikz/.cd
},
legend cell align={left},
legend style={at={(1.0,1.0)}, anchor=north west, draw=white!60.0!black, row sep=0.185cm},
clip mode=individual,
]

\addplot [dash dot,very thick,dia1, mark options={solid, draw=\tcolorshade, line width=0.5pt}]
table {%
0.0     1.27721546219187
0.2018322547784532     1.27350402278278
0.265971938705891     1.2724523908156
0.46777759361303484     1.26930753426469
0.5319381975319625     1.26840656364303
0.7338473588066857     1.26560894277588
0.7980529157464226     1.26453124892761
1.0     1.26160577175881             
};
\addlegendentry{{\small $\vartheta^{\text{(2)}}$}}

\addplot [dash dot,very thick,dia2, mark options={solid, draw=\tcolorshade, line width=0.5pt}]
table {%
0.0     1.29137256663983
0.20186735446189283     1.29280800466338
0.26604175578101197     1.29368743821311
0.4679512271855926     1.29715893513834
0.5322242377214047     1.29847744887792
0.7339295714345633     1.30327733032786
0.7979854963117159     1.30485531293343
1.0     1.31043432831045       
};
\addlegendentry{{\small $\vartheta^{\text{(2)}}$}}

\addplot [dash dot,very thick,dia3, mark options={solid, draw=\tcolorshade, line width=0.5pt}]
table {%
0.0     1.29517307669974
0.20189633443257785     1.30331737559382
0.2660744781762834     1.30677657469757
0.4678472045497876     1.31942671936356
0.5320132276644746     1.32389295539957
0.7340607372922476     1.34048497454682
0.7983365585965556     1.34542104316412
1.0     1.36641107140593 
};
\addlegendentry{{\small $\vartheta^{\text{(2)}}$}}

\addplot [dash dot,very thick,dia4, mark options={solid, draw=\tcolorshade, line width=0.5pt}]
table {%
0.0     1.47747815912577
0.2018623932538243     1.46188105002548
0.2660448017440008     1.4570476365901
0.4679174260446089     1.44213726822855
0.5320767812940534     1.43747433916347
0.7339215929671891     1.42293817046283
0.7981060015400226     1.41832936955783
1.0     1.40384784665868      
};
\addlegendentry{{\small $\vartheta^{\text{(2)}}$}}

\addplot [dash dot,very thick,dia5, mark options={solid, draw=\tcolorshade, line width=0.5pt}]
table {%
0.0     1.73977430550095
0.20185868858615783     1.7168948819579
0.2660250203273116     1.70985195889818
0.46793077834421604     1.68814494709063
0.5320979386088899     1.68136753006516
0.733971007631767     1.66030503281976
0.7981413114132941     1.65366621659252
1.0     1.63285445170621        
};
\addlegendentry{{\small $\vartheta^{\text{(2)}}$}}

\addplot [very thick,dia1, mark options={solid, draw=\tcolorshade, line width=0.5pt}]
table {%
0.0     1.22008158928416
0.2018322547784532     1.14698525639413
0.265971938705891     1.1220088027794
0.46777759361303484     1.03736408921315
0.5319381975319625     1.00900618320375
0.7338473588066857     0.914847557330942
0.7980529157464226     0.882624432176735
1.0     0.779674716994965       
};
\addlegendentry{{\small $\vartheta^{\text{(3)}}$}}

\addplot [very thick,dia2, mark options={solid, draw=\tcolorshade, line width=0.5pt}]
table {%
0.0     1.33888417976984
0.20186735446189283     1.22823968337881
0.26604175578101197     1.18742173491886
0.4679512271855926     1.04035809705514
0.5322242377214047     0.988420810403893
0.7339295714345633     0.813838525309442
0.7979854963117159     0.755461475057203
1.0     0.57090333464397
};
\addlegendentry{{\small $\vartheta^{\text{(3)}}$}}

\addplot [very thick,dia3, mark options={solid, draw=\tcolorshade, line width=0.5pt}]
table {%
0.0     1.45928802276757
0.20189633443257785     1.30012314134548
0.2660744781762834     1.23635699149374
0.4678472045497876     0.996977837669909
0.5320132276644746     0.910928844054575
0.7340607372922476     0.630623065056527
0.7983365585965556     0.540368391249314
1.0     0.293724949764173
};
\addlegendentry{{\small $\vartheta^{\text{(3)}}$}}

\addplot [very thick,dia4, mark options={solid, draw=\tcolorshade, line width=0.5pt}]
table {%
0.0     1.23466783579435
0.2018623932538243     1.15414293622257
0.2660448017440008     1.12848277662408
0.4679174260446089     1.04769699791577
0.5320767812940534     1.02204773401692
0.7339215929671891     0.941536827662953
0.7981060015400226     0.915979973706469
1.0     0.836138907041739
};
\addlegendentry{{\small $\vartheta^{\text{(3)}}$}}

\addplot [very thick,dia5, mark options={solid, draw=\tcolorshade, line width=0.5pt}]
table {%
0.0     1.23264408409567
0.20185868858615783     1.14453465186794
0.2660250203273116     1.11682202081335
0.46793077834421604     1.03016618460354
0.5320979386088899     1.00285237813511
0.733971007631767     0.917774789455893
0.7981413114132941     0.891039230104046
1.0     0.808082918859776    
};
\addlegendentry{{\small $\vartheta^{\text{(3)}}$}}

\end{axis}
\end{tikzpicture}}
\put( 8.9, 0.0){(d)}
}

\put(1.0,0.0){\includegraphics[width=13.5cm]{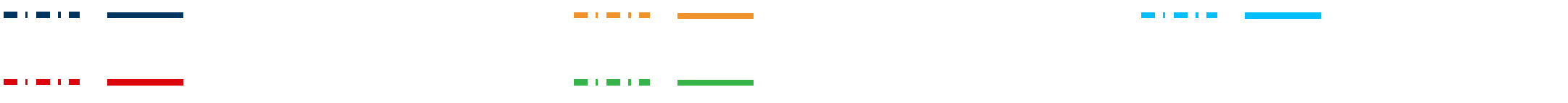}}

\put(2.8,0.68){\small with $\beta=10^{\circ}-20^{\circ}$}
\put(2.8,0.09){\small with $\beta=10^{\circ}-30^{\circ}$}
\put(7.68,0.68){\small with $\beta=10^{\circ}-40^{\circ}$}
\put(7.68,0.09){\small with $\beta=30^{\circ}$}
\put(12.55,0.68){\small with $\beta=45^{\circ}$}

\linethickness{0.4mm}
\put(0.9,1.13){\color{dia_grau_hell}\line(1,0){13.72}}
\put(14.6,1.13){\color{dia_grau_hell}\line(0,-1){1.25}}
\put(14.6,-0.1){\color{dia_grau_hell}\line(-1,0){13.72}}
\put(0.9,-0.1){\color{dia_grau_hell}\line(0,1){1.25}}

\end{picture}
\caption{Comparison of results from optimizations with different angles $\beta$ for the collagen fibers.
Distribution of
(a)~circumferential Cauchy stress $\sigma^{\text{(cir)}}$,
(b)~axial stress/~driving force $\sigma^{\text{(ax)}}$/~$\phi^{(2)}$,
(c)~radial driving force $\phi^{(3)}$ which directly influences the stresses in circumferential direction, and
(d)~growth factors $\vartheta^{(2)}$ (axial) and $\vartheta^{(3)}$ (radial).
Note that $\beta = 10^{\circ}-20^{\circ}$ describes fiber angles which are distributed over the wall thickness with $\beta = 10^{\circ}$ at the inner side and $\beta = 20^{\circ}$ at the outer side of the arterial wall.
Since the optimized values for $r_{\text{i,} \, \text{ref}}$ and $r_{\text{o,} \, \text{ref}}$ are not equal for the considered variation of the fiber angle, the radial positions are illustrated as normalized values.
Optimizations show convincing results with well homogenized stresses independent from the choice of the fiber angle~$\beta$, however some small gradient is still visible in the distributions of circumferential stresses.
\label{fig:variation_beta}}
\end{figure}
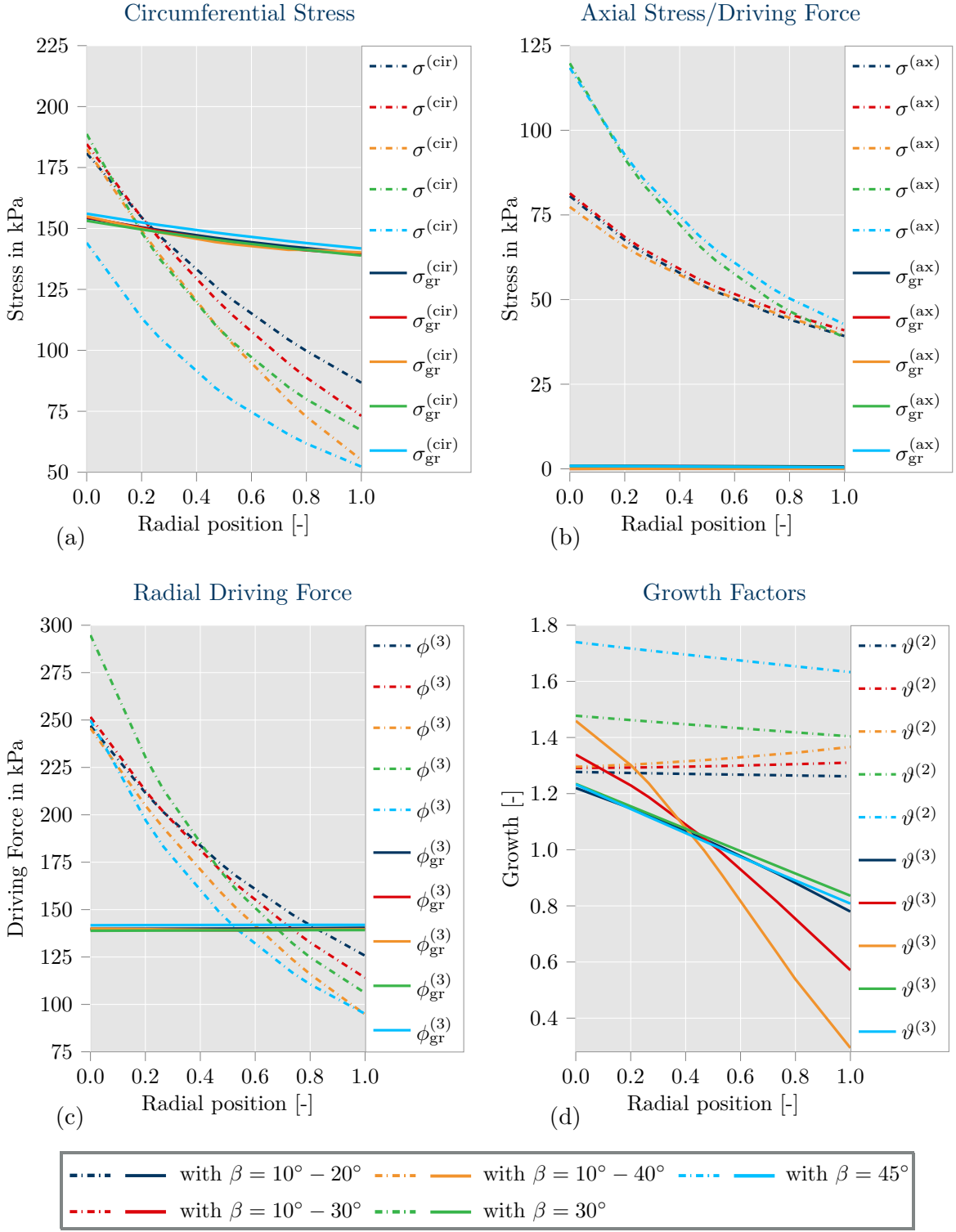
To obtain a better interpretation of these results, the distribution of the resulting Cauchy stresses~$\sigma^{\text{(ax)}}$ and~$\sigma^{\text{(cir)}}$, the driving forces~$\phi^{(2)}$ and~$\phi^{(3)}$, as well as the growth factors~$\vartheta^{(2)}$ and~$\vartheta^{(3)}$ are illustrated in Fig.~\ref{fig:variation_beta} for the variation of the fiber angle~$\beta$.
The same illustrations can be found for the variations of the axial stretch~$\lambda_{\text{ax}}$ and the convergence value~$\phi^{(2)}_{\text{con}}$ in Fig.~\ref{fig:variation_lambda} and Fig.~\ref{fig:variation_phi} of Appendix~\ref{sec:opt2}.
Note that the optimized values for the radii~$r_{\text{i,} \, \text{ref}}$ and~$r_{\text{o,} \, \text{ref}}$ are slightly different for every optimization and, hence, we depict the distributions over the normalized radial positions through the wall thickness where 0.0 represents the inner and 1.0 the outer side of the wall.
As can be seen in Fig.~\ref{fig:variation_beta}d, the growth values~$\vartheta^{(3)}$ for~$\beta = 10^\circ - 30^\circ$ and~$\beta = 10^\circ - 40^\circ$ have a larger difference to the initial value of~1.0 than in other cases.
Accordingly, the optimizations with~$\beta = 10^\circ - 30^\circ$ and~$\beta = 10^\circ - 40^\circ$ have larger values for~$z_1$ based on the higher gradient of~$\vartheta^{(3)}$.
However, $z_2$ and~$z_3$ were not influenced by this circumstance.
Therefore, it can be concluded that the optimization procedure is actually quite effective independent from variations of input parameters.\\
Taking another look at the results for the optimization with different fiber angles~$\beta$ in Fig.~\ref{fig:variation_beta}, it is visible that the stresses in the grown state of the artery are nearly identical in every single case.
While this had to be expected for the Cauchy stresses in axial direction~$\sigma^{\text{(ax)}}_{\text{gr}}$ caused by the corresponding convergence value~$\phi^{(2)}_{\text{con}} = 0 \,$kPa, it is surprising that the stresses in circumferential direction~$\sigma^{\text{(cir)}}_{\text{gr}}$ adjusted nearly equally while the fiber angles are strongly different.
Note that the convergence value~$\phi^{(3)}_{\text{con}}$ is not directly set prior to the optimization, it is strongly influenced by the demand for matching the
experimental data.
Consequently, this indicates that the generation of a reliable structural problem with the application of the new growth model to obtain a homogeneous stress distribution is independent from a perfect choice of the fiber angles.
For large arteries, experiments can be performed to quantify the microstructure distribution adequately, see e.g.~\cite{SchReiSanPieHol:2012:qao}.
This is, however, rather difficult for muscular and thus smaller arteries.
Therefore, a combination of growth models with fiber reorientation algorithms can be applied to improve the description of the mechanical fields of the arterial wall, cf~\cite{ZahBal:2018:acg}.

\section{Kinematic Growth and Active Contraction}
\label{sec:act}
To obtain a realistic simulation of muscular arteries, models describing the active contractile behavior of SMCs need to be incorporated.
This can be directly achieved by adding an associated energy in the strain energy density function~(\ref{eq:stain-energy}).
Here, we basically replace the strain energy density by the complete expression introduced in~\cite{UhlBal:2023:cmm} to directly also describe the stretch-dependent chemo-mechanical processes in smooth muscle.
However, the combination of this active material response with any kind of growth model is unfortunately not straightforward, because they both depend on the current mechanical state and thus, affect one another.
Then, an unstable growth will cause problems as a final physiological state given as a stable interplay of SMC activation and growth-induced residual stresses can hardly be reached.
Therefore, the growth model proposed in this paper, which clearly saturates in a stable state (see Subsection~\ref{subsec:mandel}) is actually convenient to allow an overall stable combined process.
However, a minor adjustment of the growth model with regard to growth directions and driving forces is proposed in this Section which actually improves the results regarding the circumferential Cauchy stress distribution when combining growth with active SMC contraction.
Furthermore, it has a more convincing mechanism from a biological point of view which cures the critical aspects discussed in the remark in Section~\ref{subsec:grow}.

\subsection{Improved Growth Model}
As shown and discussed in Section~\ref{subsec:sim}, a certain gradient over the wall thickness remains for the circumferential stresses after the growth process is applied.
This gradient can be linked to the dependency of the driving force~$\phi^{(3)}$ on the third principal stress which is equal to the stress in radial direction when considering an artery idealized as a hollow cylinder.
The circumferential Cauchy stresses were found to be approximately~$10\%$ larger at the inner side of the vessel wall than at the outer side as long as only the passive material response is considered.
Unfortunately, this difference increases significantly when the active material response is included into the simulations, since the absolute values of the circumferential stresses decrease considerably while the maximal and minimal values of the radial stresses remain unchanged due to the boundary conditions of the simulation.
Corresponding results can be seen in Fig.~\ref{fig:opt_act_app} in Appendix~\ref{sec:act2}.\\
To improve the results of the growth model, the driving force~$\phi^{(3)}$ should not depend on the third principal stress.
In consequence, the definition of the driving force~$\phi^{(3)}$ is adjusted to include only the first and second principal stresses (circumferential and axial direction) for simulations with active material response.
As the load-bearing mechanism in arteries is dominated by transferring the internal pressure to tensile stresses in circumferential direction, adding up to the axial tensile stresses reacting to external axial displacements, these stresses basically constitute the stress state in the layered structure of the circumferential-axial plane.
Furthermore, in previous simulations of this paper, growth in the direction of the first principal stress (circumferential) was not considered.
However, it can be expected that proliferation of SMCs and protein synthesis extends the tissue in every direction inside the plane of the layers.
Hence, the growth process is three-dimensional as was already suggested in~\cite{HolSomAueRegOgd:2007:lst} and is further supported by data from more recent publications, but not isotropic.
In~\cite{JadRazAntDoaAdaPipKam:2021:com}, the differences of the geometry of the human popliteal and superficial femoral arteries of donors between the age of 13 to 92 were investigated.
The measurements show that the wall thickness of the arteries increases significantly with age of the donors.
This progress corresponds to an increase of the outer radius of the artery which underlines the significance of growth in radial direction.
Additionally, there is also a smaller increase of the inner radius identified which may indicate that a certain amount of growth in circumferential direction occurs.
Considering that the growth process of tissue in axial direction is proven by the decrease of in vivo axial prestretch with age, consequently, a realistic model has to include growth in all directions.
Hence, the Eqs.~(\ref{eq:groFac}) and~(\ref{eq:drivingForce}) for the evolution of the growth factors~$\vartheta^{(a)}$ of the kinematic growth model are changed in this Section to
\eb
\label{eq:groFac2}
\dot{\vartheta}^{(a)} = \kappa_\vartheta^{(a)} \left(\phi^{(a)} - \phi^{(a)}_{\text{con}} \right) \quad \text{with} \quad a = 1,2,3 \, ,
\ee
\eb
\phi^{(1/2)}(\Bsigma) = \Bsigma : \left( \bn^{(1/2)} \otimes \bn^{(1/2)} \right) \qquad \text{and} \qquad
\phi^{(3)}(\Bsigma) = \phi^{(1)} + \phi^{(2)} \, .
\ee
Note that the form of the evolution equation is identical to Eq.~(\ref{eq:groFac}), it is, however, also applied to describe growth in the direction of the first principle Cauchy stress.
The modified driving force~$\phi^{(3)}$ does not contain components of the third (typically negative) principal Cauchy stresses anymore.
Instead, it is defined to let the evolution of growth adjust to the sum of the first and second principal Cauchy stresses~$\Bsigma_{\uproman{1}}$ and~$\Bsigma_{\uproman{2}}$.
Accordingly, in the optimization procedure for parameter identification, the convergence value~$\phi^{(3)}_{\text{con}}$ is defined as
\eb
\phi^{(3)}_{\text{con}} = w_{\text{min}} \bar{\sigma}_{\uproman{1}/ \uproman{2}} \qquad \text{with} \qquad \bar{\sigma}_{\uproman{1}/ \uproman{2}} = \frac{1}{n_{\text{gp}}}\sum_{g=1}^{n_{\text{gp}}} \left( \Bsigma_{\uproman{1}, \, g} + \Bsigma_{\uproman{2}, \, g} \right) \, .
\ee
Furthermore, the optimization to receive a reliable structural problem for the arterial wall is expanded here to fit the in vivo axial prestretch of 1.1 measured in~\cite{BelKunMon:2013:baf}.
For this purpose, the convergence value~$\phi^{(2)}_{\text{con}}$ is added as optimization parameter.
An additional simulation is executed in the optimization to determine the axial prestretch after growth of the artery.
The boundary conditions and material behavior of the artery are formulated to be comparable to the mechanical conditions of an artery after dissection from the body.
One end of the artery is held in axial direction and the other end can freely move.
During the simulation, there is no active response considered and no load is applied equal to measurements of axial prestretches in experiments.
Over a time frame of one second, the growth factors at every Gauss point of the artery are linearly increased from 1.0 to the final values of~$\vartheta^{(a)}$ at the end of simulation 2(a).
To evaluate the precision of the convergence value~$\phi^{(2)}_{\text{con}}$ to receive the correct axial prestretch~$\lambda_{\text{ax,pre}}$, the objective function $z$ is extended by
\eb
z_4 = 0.1 \sqrt{ \left( \frac{1}{\lambda_{\text{ax}}} - 1.1 \right)^2 } \, ,
\ee
where~$\lambda_{\text{ax}}$ is the axial stretch of the artery in the simulations after the growth factors~$\vartheta^{(a)}$ are applied.
Note that the axial prestretch is~$\lambda_{\text{ax, pre}} = \lambda_{\text{ax}}^{-1}$ and the convergence value~$\phi^{(1)}_{\text{con}}$ can be calculated from the optimization parameters for~$\phi^{(2)}_{\text{con}}$ and~$\phi^{(3)}_{\text{con}}$ as~$\phi^{(1)}_{\text{con}} = \phi^{(3)}_{\text{con}} - \phi^{(2)}_{\text{con}}$.

\subsection{Optimization Results Incorporating the Active Material Response}

The optimization, which was described in Section~\ref{subsec:opt_pro} and expanded in the previous Subsection, is now applied to an arterial ring in which the active material model is included.
The detailed description of this material model can be found in~\cite{UhlBal:2023:cmm}.
For the fiber directions of collagen and SMCs, the fiber angle~$\beta = 10^{\circ}-20^{\circ}$ is used for both fiber types.
Since the smooth muscle model is time- and stretch-dependent, the simulations~2(a) and~2(c) (see Fig.~\ref{fig:optimization}) have to be adjusted accordingly.
For the fitting of the material parameters (simulation 2(c)) to experimental data in~\cite{JohElyTakWalWalCol:2009:csv}, we apply the same procedure as described in~\cite{UhlBal:2023:cmm}:
An intravascular pressure of~10$\,$mmHg is set on the resulting geometry over a time of~600$\,$s to generate the initial contractile state of the arterial wall which represents the state of the artery before the experiment is performed.
This part of the simulation is executed after the final values of the growth factors~$\vartheta^{(a)}$ from the end of simulation~2(a) are reached.
Afterwards, the intravascular pressure is increased stepwise to 20, 40, 60, 80, 100 and~120$\,$mmHg where every pressure level is applied for~300$\,$s.
Over a time frame of~1800$\,$s, the evolution of the outer diameter of the arterial ring is recorded.
This simulation protocol is used for the fully active case and the suppressed version which is influenced by the Rho kinase inhibitor Y27632.
To enable a suitable fit of the material response, further optimization parameters are included for~1(c) of the optimization (see Fig.~\ref{fig:optimization}).
These optimization parameters are part of the active material response.
In accordance to the original paper of the SMC model, the same parameters are used as chosen for the fitting of the material response there (see Table~3 in~\cite{UhlBal:2023:cmm}).
The remaining parameters of the active material model (which are not included as optimization parameters) are equal to the values of Table~2 in~\cite{UhlBal:2023:cmm}.
For the growth process in simulation~2(a), the same simulation protocol is applied at the beginning of the simulation as described above for simulation~2(c): An intravascular pressure of~10$\,$mmHg is applied over a time of~600$\,$s to generate the initial contractile state and, afterwards, the intravascular pressure is increased stepwise to reach a value of~120$\,$mmHg after~1800$\,$s.
This procedure leads to a referential state of the mechanical fields at an intravascular pressure of~120$\,$mmHg without growth.
Subsequently, the convergence values~$\phi^{(a)}_{\text{con}}$ are calculated and the growth model is activated.
To regulate the interaction between active response and growth process, the growth velocity factors are chosen to be~$\kappa_\vartheta^{(1)} = 10^{-5} (\text{s} \cdot \text{kPa})^{-1}$, $\kappa_\vartheta^{(2)} = 10^{-4} (\text{s} \cdot \text{kPa})^{-1}$ and~$\kappa_\vartheta^{(3)} = 10^{-4} (\text{s} \cdot \text{kPa})^{-1}$.
As a consequence, the contraction of SMCs reacts significantly faster on growth-induced changes of the stretch in fiber direction of the SMCs than the growth factors change over time.
Accordingly, the active response is considered to react rather quickly to changes of the mechanical fields and the growth factors adjust comparably slow.
Thereby, a small interaction between both processes is realized and helps the overall convergence of the saturation process.
Note that the growth in the direction of the first and third principal stress (circumferential and radial) are primarily responsible for the adaptation of the stresses in circumferential direction.
Simulations with a large value for the growth velocity factor~$\kappa_\vartheta^{(1)}$ have shown that the convergence values cannot be reached, but infinite growth in circumferential direction occurs.
Consequently, the ratio between~$\kappa_\vartheta^{(1)}$ and~$\kappa_\vartheta^{(3)}$ has to be chosen adequately to avoid infinite growth.
When~$\kappa_\vartheta^{(3)}$ is set to a value of~$10^{-4} (\text{s} \cdot \text{kPa})^{-1}$, test simulations indicated that infinite growth can be avoided for growth velocities of~$\kappa_\vartheta^{(1)} = 2 \, 10^{-5} (\text{s} \cdot \text{kPa})^{-1}$ and smaller.
However, a value of~$\kappa_\vartheta^{(1)} = 10^{-5} (\text{s} \cdot \text{kPa})^{-1}$ resulted in shortest run-times to reach almost homogeneous stress distributions.
The infinite growth process in these simulations can be explained by the increase of the inner radius~$r_{\text{i}}$ of the arterial wall caused by the growth in circumferential direction.
The increase of~$r_{\text{i}}$ results in an increase of the inner surface area.
Since a static value of~120$\,$mmHg is applied for the intravascular pressure, the increasing surface area generates a rise of the load on the arterial wall.
In consequence, the circumferential stresses can reach values larger than the convergence value~$\phi^{(1)}_{\text{con}}$ over the entire wall which results in an infinite growth in circumferential direction.
This issue should not be expected in simulations with realistic fluid-solid interactions, where the blood pressure will also depend on deformations and processes of the artery.
Since an increase of the inner radius~$r_{\text{i}}$ results in an increase of the lumen, the pressure from the blood on the arterial wall would decrease there. \\
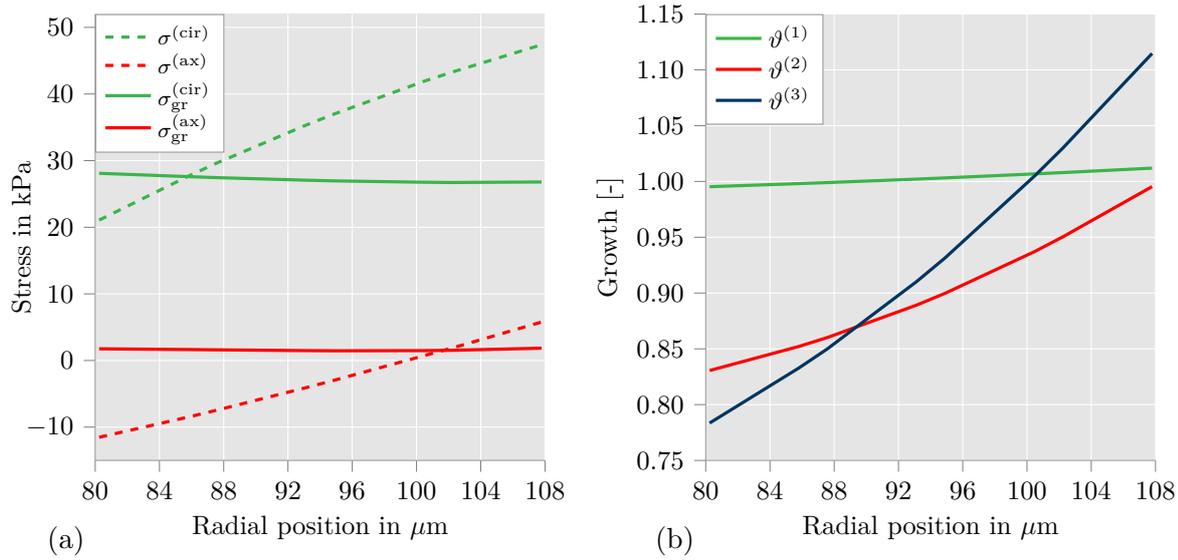
\begin{figure}[!t]
\unitlength1cm
\begin{picture}(15.5,7.0)
\put(-0.2, 0.0){
\put( 0.0, 0.0){
\begin{tikzpicture}
\begin{axis}[
width=7.5cm,
height=7.5cm,
xlabel={{Radial position in $\mu$m}},
xmin=80, xmax=108,
xtick={80, 84, 88, 92, 96, 100, 104, 108},
extra x ticks={0.0},
extra x tick style={%
    grid=major,
},
ylabel={{Stress in kPa}},
ymin=-15, ymax=52,
ytick={-10, 0, 10, 20, 30, 40, 50},
axis background/.style={fill=white!89.80392156862746!black},
axis line style={white!60.0!black},
axis x line*=bottom,
axis y line*=left,
tick align=outside,
tick pos=left,
scaled x ticks=false,
x grid style={white},
xmajorgrids,
x tick label style={
  font=\small,
  /pgf/number format/.cd,
  set decimal separator={.},
    fixed,
    fixed zerofill,
    precision=0,
  /tikz/.cd
},
xlabel style={at={(0.5,-0.1)},font=\small},
ylabel style={at={(-0.12,0.5)},font=\small},
y grid style={white},
ymajorgrids,
y tick label style={
  font=\small,
},
legend cell align={left},
legend style={at={(0.0,1.0)}, anchor=north west, draw=white!60.0!black, row sep=-0.08cm},
clip mode=individual,
]

\addplot [dashed, very thick,rub_green, mark options={solid, draw=\tcolorshade, line width=0.5pt}]
table {%
80.243924515629     21.0688048843223
85.8041348168634     27.683581217793
87.5719028651069     29.6386738472557
93.1322925046738     35.3384618819925
94.8995677607057     37.0183575217653
100.461893176436     41.9076500337753
102.228242365196     43.2957286837323
107.786012787448     47.3933598001205        
};
\addlegendentry{{\scriptsize $\sigma^{\text{(cir)}}$}}

\addplot [dashed, very thick,red, mark options={solid, draw=\tcolorshade, line width=0.5pt}]
table {%
80.243924515629     -11.5268420301784
85.8041348168634     -8.4840595204565
87.5719028651069     -7.45517983250857
93.1322925046738     -4.07347385093439
94.8995677607057     -2.95025172377457
100.461893176436     0.716030336683086
102.228242365196     1.88213816725043
107.786012787448     5.77328910046487                      
};
\addlegendentry{{\scriptsize $\sigma^{\text{(ax)}}$}}

\addplot [very thick,rub_green, mark options={solid, draw=\tcolorshade, line width=0.5pt}]
table {%
80.243924515629     28.0971109768719
85.8041348168634     27.5971089114211
87.5719028651069     27.4477405371594
93.1322925046738     27.0635287023635
94.8995677607057     26.9590509495604
100.461893176436     26.7562518367845
102.228242365196     26.717323828348
107.786012787448     26.7952114102482                            
};
\addlegendentry{{\scriptsize $\sigma^{\text{(cir)}}_{\text{gr}}$}}

\addplot [very thick,red, mark options={solid, draw=\tcolorshade, line width=0.5pt}]
table {%
80.243924515629     1.75788265739627
85.8041348168634     1.64544085716977
87.5719028651069     1.600244845151
93.1322925046738     1.48422441345863
94.8995677607057     1.45793605485706
100.461893176436     1.48354887006467
102.228242365196     1.53244278473824
107.786012787448     1.8498414988394                       
};
\addlegendentry{{\scriptsize $\sigma^{\text{(ax)}}_{\text{gr}}$}}

\end{axis}
\end{tikzpicture}}
\put( 0.6, 0.0){(a)}

\put( 7.7, 0.0){
\begin{tikzpicture}
\begin{axis}[
width=7.5cm,
height=7.5cm,
xlabel={{Radial position in $\mu$m}},
xmin=80, xmax=108,
xtick={80, 84, 88, 92, 96, 100, 104, 108},
extra x ticks={0.0},
extra x tick style={%
    grid=major,
},
ylabel={{Growth [-]}},
ymin=0.75, ymax=1.15,
ytick={0.75, 0.8, 0.85, 0.9, 0.95, 1.0, 1.05, 1.1, 1.15},
axis background/.style={fill=white!89.80392156862746!black},
axis line style={white!60.0!black},
axis x line*=bottom,
axis y line*=left,
tick align=outside,
tick pos=left,
scaled x ticks=false,
x grid style={white},
xmajorgrids,
x tick label style={
  font=\small,
  /pgf/number format/.cd,
  set decimal separator={.},
    fixed,
    fixed zerofill,
    precision=0,
  /tikz/.cd
},
y tick label style={
  font=\small,
  /pgf/number format/.cd,
  set decimal separator={.},
    fixed,
    fixed zerofill,
    precision=2,
  /tikz/.cd
},
xlabel style={at={(0.5,-0.1)},font=\small},
ylabel style={at={(-0.16,0.5)},font=\small},
y grid style={white},
ymajorgrids,
y tick label style={
  font=\small,
},
legend cell align={left},
legend style={at={(0.0,1.0)}, anchor=north west, draw=white!60.0!black, row sep=-0.08cm},
clip mode=individual,
]

\addplot [very thick,rub_green, mark options={solid, draw=\tcolorshade, line width=0.5pt}]
table {%
80.243924515629     0.99529104588038
85.8041348168634     0.99807038140948
87.5719028651069     0.998982153987527
93.1322925046738     1.00210905930515
94.8995677607057     1.00317313327976
100.461893176436     1.00676718239952
102.228242365196     1.00795211273169
107.786012787448     1.01193006553505                                     
};
\addlegendentry{{\footnotesize $\vartheta^{\text{(1)}}$}}

\addplot [very thick,red, mark options={solid, draw=\tcolorshade, line width=0.5pt}]
table {%
80.243924515629     0.830573787653786
85.8041348168634     0.852168420484092
87.5719028651069     0.860119977944241
93.1322925046738     0.889228801127404
94.8995677607057     0.899762394572054
100.461893176436     0.93707228669379
102.228242365196     0.95031658800304
107.786012787448     0.995372145222222                             
};
\addlegendentry{{\footnotesize $\vartheta^{\text{(2)}}$}}

\addplot [very thick,rub_blue, mark options={solid, draw=\tcolorshade, line width=0.5pt}]
table {%
80.243924515629     0.783484246457587
85.8041348168634     0.832872234578891
87.5719028651069     0.849941517819503
93.1322925046738     0.910319394178904
94.8995677607057     0.931493727369665
100.461893176436     1.00474411068899
102.228242365196     1.02983771531988
107.786012787448     1.11467280057274                                  
};
\addlegendentry{{\footnotesize $\vartheta^{\text{(3)}}$}}

\end{axis}
\end{tikzpicture}}
\put( 8.6, 0.0){(b)}
}
\end{picture}
\caption{Distribution of 
(a)~Cauchy stresses and driving forces $\sigma^{\text{(cir)}}=\phi^{(1)}$ and $\sigma^{\text{(ax)}}=\phi^{(2)}$
(b)~growth factors~$\vartheta^{(1)}$, $\vartheta^{(2)}$ and $\vartheta^{(3)}$
in circumferential (green), axial (red) and radial (blue) direction over the wall thickness for the including the fully active material model for SMCs.
Artery is loaded with an intravascular pressure of 120mmHg. 
Dashed lines show results before growth, solid lines show results after growth. 
The proposed, improved growth model results in a well-homogenized stress distribution when smooth muscle contraction is activated.
\label{fig:opt_act}}
\end{figure}
The simulation results for the optimized structural problem are illustrated in Fig.~\ref{fig:opt_act} and~\ref{fig:validation_act}.
In Fig.~\ref{fig:opt_act}a, the distribution of the Cauchy stresses~$\sigma^{\text{(ax)}}$ and~$\sigma^{\text{(cir)}}$ in axial and circumferential direction before (dashed) and after (solid) the growth process can be seen.
Furthermore, Fig.~\ref{fig:opt_act}b shows the distribution of the growth factors~$\vartheta^{(a)}$.
Note that the stresses~$\sigma^{\text{(cir)}}$ and~$\sigma^{\text{(ax)}}$ are equal to the driving forces~$\phi^{(1)}$ and~$\phi^{(2)}$, respectively.
The convergence values for the driving forces in circumferential and axial direction were optimized to values of~$\phi^{(1)}_{\text{con}} = 27.46 \,$kPa and~$\phi^{(2)}_{\text{con}} = 2.88 \,$kPa.
Note that the Cauchy stresses would only reach the values of~$\phi^{(1)}_{\text{con}}$ and~$\phi^{(2)}_{\text{con}}$ after an infinite run time of the growth process.
To regulate the computational cost of the optimization procedure, the growth process was stopped automatically when the standard deviation of the stresses over the wall thickness were lower than a predefined tolerance value.
The corresponding circumferential stresses are~28.1$\,$kPa at the inner side of the wall and~26.8$\,$kPa at the outer side of the wall.
For the axial direction, the stresses reached values between~1.45$\,$kPa and~1.85$\,$kPa over the entire wall thickness which lead to an axial prestretch of~$\lambda_{\text{ax, pre}} = 1.105$.
The ratio between the wall thickness and outer radius is~0.216.
Consequently, it can be stated that the combination of active response with the proposed improved growth model lead to reasonably homogenized stresses which are in line with the homeostasis hypothesis.
It should be noted that the circumferential Cauchy stresses~$\sigma^{\text{(cir)}}$ have a higher value at the outer side of the artery than at the inner side of the artery before growth which is in contrast to the simulations with a purely passive material response.
This difference is caused by the contraction of the active material model which can lead to a smaller diameter of the arterial ring in the contracted state than in the reference configuration.
This influences also the stresses in axial direction $\sigma^{\text{(ax)}}$ which can reach negative values before growth, which further motivates the necessity for residual tensile stresses in axial direction.
When regarding the growth factor~$\vartheta^{(3)}$ in Fig.~\ref{fig:opt_act}b, it can be noticed that the gradient over the wall thickness is positive when active response is included.
In simulations with a purely passive material response, this gradient was negative (see Fig.~\ref{fig:opt1}).
These results underline the importance of the combination of growth and active response to receive reliable residual stresses for muscular arteries.
For such arteries, an estimation of the residual stresses cannot be achieved by executing simulations which calculate the growth process based on the purely passive material response.
This would actually make the mechanical fields even less realistic in comparison to not considering residual stresses at all.\\
\begin{figure}[!bt]
\unitlength1cm
\begin{picture}(15.5,8.0)
\put( 0.0,-0.4){
\put( 0.0, 0.0){\input{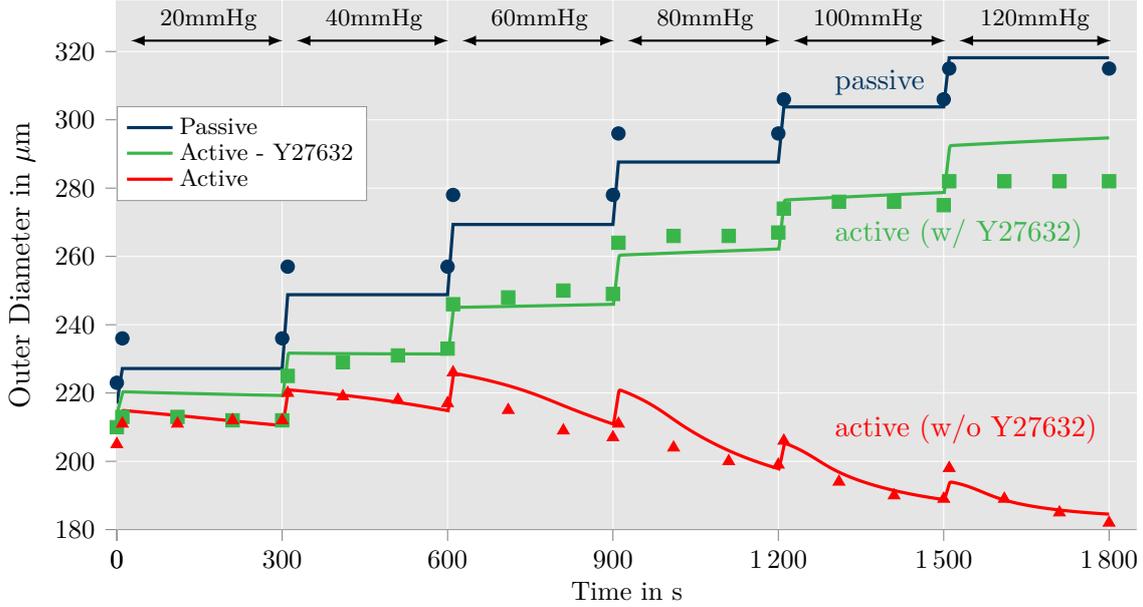}}

\put( 11.0,6.9){\color{rub_blue} passive}
\put( 11.0,4.9){\color{rub_green} active (w/ Y27632)}
\put( 11.0,2.3){\color{red} active (w/o Y27632)}

\put(-0.08,0.0){
\put(2.2,7.75){\color{black} \footnotesize 20mmHg}
\put(4.38,7.75){\color{black} \footnotesize 40mmHg}
\put(6.56,7.75){\color{black} \footnotesize 60mmHg}
\put(8.74,7.75){\color{black} \footnotesize 80mmHg}
\put(10.8,7.75){\color{black} \footnotesize 100mmHg}
\put(13.0,7.75){\color{black} \footnotesize 120mmHg}
}

\put(-0.08,-0.1){
\put(1.8,7.55){\begin{tikzpicture}[thick, scale=1.0, transform shape, color=black]
\draw [latex-latex](0,0) -- (2.0,0.0);
\useasboundingbox (0,-0.1) rectangle (2.0,0.1);
\end{tikzpicture}}
\put(3.98,7.55){\begin{tikzpicture}[thick, scale=1.0, transform shape, color=black]
\draw [latex-latex](0,0) -- (2.0,0.0);
\useasboundingbox (0,-0.1) rectangle (2.0,0.1);
\end{tikzpicture}}
\put(6.16,7.55){\begin{tikzpicture}[thick, scale=1.0, transform shape, color=black]
\draw [latex-latex](0,0) -- (2.0,0.0);
\useasboundingbox (0,-0.1) rectangle (2.0,0.1);
\end{tikzpicture}}
\put(8.34,7.55){\begin{tikzpicture}[thick, scale=1.0, transform shape, color=black]
\draw [latex-latex](0,0) -- (2.0,0.0);
\useasboundingbox (0,-0.1) rectangle (2.0,0.1);
\end{tikzpicture}}
\put(10.52,7.55){\begin{tikzpicture}[thick, scale=1.0, transform shape, color=black]
\draw [latex-latex](0,0) -- (2.0,0.0);
\useasboundingbox (0,-0.1) rectangle (2.0,0.1);
\end{tikzpicture}}
\put(12.70,7.55){\begin{tikzpicture}[thick, scale=1.0, transform shape, color=black]
\draw [latex-latex](0,0) -- (2.0,0.0);
\useasboundingbox (0,-0.1) rectangle (2.0,0.1);
\end{tikzpicture}}
}
}
\end{picture}
\caption{
Comparison of model response with experimental data from~\cite{JohElyTakWalWalCol:2009:csv} for three different setups: passive response, active response under influence of 1$\mu$M Rho kinase inhibitor Y27632 and fully active response (see material model in~\cite{UhlBal:2023:cmm}).
The results for the active model agree well with the experimental data.
\label{fig:validation_act}}
\end{figure}
The results for the fitting of the material response for the optimization with active response can be seen in Fig.~\ref{fig:validation_act} and the corresponding values of the optimization parameters in Table~\ref{tab:opt_act}.
The results for the two different active material graphs look comparable to the results without growth (see Fig.~3 in~\cite{UhlBal:2023:cmm}).
However, the passive material response shows higher differences between simulation and experiments than before.
As can be seen in equation~(\ref{eq:psi}), the passive material model for collagen fibers only contributes to the material response when the stretch in fiber direction is larger than~1.0.
For the artery in a contracted state, the stretch can reach values lower than~1.0.
In consequence, collagen fibers at these stretch values do not contribute to the growth process.
In addition, the gradient of the growth factors is not beneficial for the stress distribution when only the passive response is considered.
As a result, the optimized value of the material parameter~$\alpha_4$, which contributes to the description of the stiffness of the collagen fibers, is considerably smaller than in the optimizations of the previous Sections.
In contrast, the parameter~$\alpha_1$ increased which describes the material stiffness of elastin.
However, the material model for elastin is not sufficient to describe the non-linearity of the material.
One attempt to fix this issue could be the consideration of separated growth of SMCs and the passive material response.
In this regard, separated growth tensors~$\bF_{\text{g}}$ and growth factors~$\vartheta$ are calculated for each part of the material model.
This approach would combine the simplicity of kinematic growth with the idea of constrained mixture models where growth is regarded individually for every constituent of the arterial wall.
However, since the application of inhibitors anyhow leads to a somewhat artificial response, the fitting quality just for the passive behavior should not be overrated.\\
Summarizing, an excellent agreement with experiments is achieved for the fully active, i.e. realistic, response (red curve in Fig.~\ref{fig:validation_act}) while still showing almost homogeneous stress distributions through the vessel wall (Fig.~\ref{fig:opt_act}) thanks to the proposed growth model.
\begin{table}[bht]
\caption{Optimization parameters for inclusion of active response.\label{tab:opt_act}}
\begin{center}
\begin{tabular}{cccccc} 
\toprule
$\alpha_1$ & $\alpha_4$ & $\alpha_5$ & $r_{\text{o,} \, \text{ref}}$ & $r_{\text{i,} \, \text{ref}}$ & $w_{\text{min}}$ \\
\midrule
6.69$\,$kPa & 1.1$\,$kPa & 2.38 & 108.67$\, \mu$m & 79.36$\, \mu$m & 0.884\\ \bottomrule
\toprule
$\eta$ & $\gamma_1$ & $\dot{\bar{\lambda}}_{\text{c}, \, \text{max}}$ & $\dot{\bar{\lambda}}_{\text{c}, \, \text{min}}$ & $\dot{k}_{2/5, \, \text{max}}$ & $\dot{k}_{2/5, \, \text{min}}$ \\ \midrule
$0.1349\, \text{s}^{-1}$ & $0.3095\, \mu\text{M}$ & $0.0471\, \text{s}^{-1}$ & $-0.0471\, \text{s}^{-1}$ & $0.000569\, \text{s}^{-2}$ & $-0.001004\, \text{s}^{-2}$ \\ \bottomrule
\toprule
$\dot{\bar{\lambda}}_{\text{p}, \, \text{max}}$ & $\dot{\bar{\lambda}}_{\text{p}, \, \text{min}}$ & $\mu_\text{a}$ & $\kappa$ & $\beta_1$ & $k_{2/5, \, \text{start}}$\\ \midrule
$0.000094\, \text{s}^{-1}$ & $-0.000108\, \text{s}^{-1}$ & $29.04\, \text{kPa}$ & $104.09\, \text{kPa}$ & $0.00069\, \text{s}^{-1}$ & $1.9088\, \text{s}^{-1}$\\ \bottomrule
\end{tabular}
\end{center}
\end{table} 

\section{Conclusion}

A new kinematic growth model has been proposed which was motivated by the results of~\cite{Zah:2020:mog}.
It extends rather classical kinematic growth models, but in particular the one of~\cite{ZahBal:2018:acg}, by a new set of evolution equations and driving forces which depend on the principal values of Cauchy stresses.
As main difference to previous formulations, the evolution equations were formulated to let growth adjust to homeostatic stresses instead of growth intensities.
As a major advantage thereof, the model was shown to allow for stable growth states, i.e. growth that converges over time to saturated states, which becomes important in the context of connecting growth to further active processes like SMC contraction.
Furthermore, much more homogenized stress distributions were obtained that are rather independent from changes of fiber angle distribution, axial loading, and convergence values of principal stresses defining the homeostatic stress state.
In addition to the proposed model, a suitable optimization setup for the solution of inverse problems related to the identification of material parameters based on experiments performed on whole arterial segments was described.
The first variant of the proposed model was formulated to partly coincide with ideas from the literature, i.e. that growth in radial direction should be driven by an isotropic stress measure.
In line with~\cite{Zah:2020:mog}, this was combined with growth in axial direction, which could be motivated by mechanical considerations.
When only considering a purely passive mechanical response resulting from collagen and elastin, reasonable results were obtained.
However, when considering the layered structure of arterial walls possessing layers in the axial-circumferential plane, it was rationalized that not considering growth in circumferential direction appears somewhat unreasonable.
Therefore, an improved growth model was proposed which includes growth in the directions of all principal Cauchy stresses and a modified definition of the driving force related the third (negative) principal stress.
By applying this improved growth model in connection with the mechanical model from~\cite{UhlBal:2023:cmm} including the active response of SMCs and by updating the optimization scheme, convincing results were obtained.
These were characterized by (i) an accurate agreement with sophisticated experiments including the active response and (ii) almost perfectly homogeneous stress distributions.
Results showed that the proposed growth model can be suitably used to obtain realistic stresses even if an advanced material model for the SMC activation is used.

\section*{Acknowledgement}

Daniel Balzani acknowledges the Deutsche Forschungsgemeinschaft (DFG) for financial support within the Priority Program 2311 ``Robust Coupling of Continuum-biomechanical In Silico Models to Establish Active Biological System Models for Later Use in Clinical Applications – Co-design of Modelling, Numerics and Usability'', project ID 465228106, reference ID BA2823/18-1.
Furthermore, both authors appreciate fruitful discussions with Christian Cyron (University of Technology Hamburg) regarding the growth model.

\bibliographystyle{stylefiles/references_style}
\bibliography{references}

\begin{thebibliography}{58}
\providecommand{\natexlab}[1]{#1}
\providecommand{\url}[1]{\texttt{#1}}
\expandafter\ifx\csname urlstyle\endcsname\relax
  \providecommand{\doi}[1]{doi: #1}\else
  \providecommand{\doi}{doi: \begingroup \urlstyle{rm}\Url}\fi

\bibitem[Ambrosi et~al.(2019)Ambrosi, Ben~Amar, Cyron, DeSimone, Goriely,
  Humphrey, and Kuhl]{AmbAmaCyrDesGorHumKuh:2019:gar}
D.~Ambrosi, M.~Ben~Amar, C.~J. Cyron, A.~DeSimone, A.~Goriely, J.~D. Humphrey,
  and E.~Kuhl.
\newblock Growth and remodelling of living tissues: perspectives, challenges
  and opportunities.
\newblock \emph{J R Soc Interface}, 16:\penalty0 20190233, 2019.

\bibitem[Anttila et~al.(2019)Anttila, Balzani, Desyatova, Deegan, Mac{T}aggart,
  and Kamenskiy]{AntDeeBalMacKem:2019:cod}
E.~Anttila, D.~Balzani, A.~Desyatova, P.~Deegan, J.~Mac{T}aggart, and
  A.~Kamenskiy.
\newblock Mechanical damage characterization in human femoropopliteal arteries
  of different ages.
\newblock \emph{Acta Biomaterialia}, 90:\penalty0 225--240, 2019.
\newblock \doi{10.1016/j.actbio.2019.03.053}.

\bibitem[Bagalad et~al.(2017)Bagalad, Mohan~Kumar, and
  Puneeth]{BagMohPun:2017:mmo}
B.~S. Bagalad, K.~P. Mohan~Kumar, and H.~K. Puneeth.
\newblock Myofibroblasts: Master of disguise.
\newblock \emph{J Oral Maxillofac Pathol}, 21:\penalty0 462--463, 2017.

\bibitem[Ball(1977)]{Bal:1977:cca}
J.~M. Ball.
\newblock Convexity conditions and existence theorems in non-linear elasticity.
\newblock \emph{Archive for Rational Mechanics and Analysis}, 63:\penalty0
  337--403, 1977.

\bibitem[Balzani et~al.(2010)Balzani, Brands, Klawonn, Rheinbach, and
  Schr\"oder]{BalBraKlaRheSch:2010:otm}
D.~Balzani, D.~Brands, A.~Klawonn, O.~Rheinbach, and J.~Schr\"oder.
\newblock On the mechanical modeling of soft biological tissue and iterative
  parallel solution strategies.
\newblock \emph{Archive of Applied Mechanics}, 80:\penalty0 479--488, 2010.

\bibitem[Balzani et~al.(2015)Balzani, Gandhi, Tanaka, and
  Schr\"{o}der]{BalGanTanSch:2015:nco}
D.~Balzani, A.~Gandhi, M.~Tanaka, and J.~Schr\"{o}der.
\newblock Numerical calculation of thermo-mechanical problems at large strains
  based on complex step derivative approximation scheme for tangent stiffness
  matrices.
\newblock \emph{Computational Mechanics}, 55:\penalty0 861--871, 2015.
\newblock \doi{10.1007/S00466-015-1139-0}.

\bibitem[Bell et~al.(2013)Bell, Kunjir, and Monson]{BelKunMon:2013:baf}
E.~D. Bell, R.~S. Kunjir, and K.~L. Monson.
\newblock Biaxial and failure properties of passive rat middle cerebral
  arteries.
\newblock \emph{J Biomech}, 46:\penalty0 91--96, 2013.

\bibitem[Boehler(1987)]{Boe:1987:itt}
J.~P. Boehler.
\newblock \emph{Introduction to the Invariant Formulation of Anisotropic
  Constitutive Equations}, pages 13--30.
\newblock Springer Vienna, Vienna, 1987.

\bibitem[Bonetti et~al.(2021)Bonetti, Corti, Lerouge, Pompella, and
  Gaucher]{BonCorLerPomGau:2021:pmo}
J.~Bonetti, A.~Corti, L.~Lerouge, A.~Pompella, and C.~Gaucher.
\newblock Phenotypic modulation of macrophages and vascular smooth muscle cells
  in atherosclerosis—nitro-redox interconnections.
\newblock \emph{Antioxidants}, 10:\penalty0 516, 2021.

\bibitem[Braeu et~al.(2017)Braeu, Seitz, Aydin, and
  Cyron]{BraSeiAydCyr:2017:hcm}
F.~A. Braeu, A.~Seitz, R.~C. Aydin, and C.~J. Cyron.
\newblock Homogenized constrained mixture models for anisotropic volumetric
  growth and remodeling.
\newblock \emph{Biomech. Model. Mechan.}, 16:\penalty0 889, 2017.

\bibitem[Comellas et~al.(2018)Comellas, Carriero, Giorgi, Pereira, and
  Shefelbine]{ComCarGioPerShe:2018:mti}
E.~Comellas, A.~Carriero, M.~Giorgi, A.~Pereira, and S.~Shefelbine.
\newblock Chapter 2 - modeling the influence of mechanics on biological
  growth.
\newblock In M.~Cerrolaza, S.~J. Shefelbine, and D.~Garzón-Alvarado, editors,
  \emph{Numerical Methods and Advanced Simulation in Biomechanics and
  Biological Processes}, pages 17--35. Academic Press, 2018.

\bibitem[Cyron and Humphrey(2017)]{CyrHum:2017:gar}
C.~Cyron and J.~Humphrey.
\newblock Growth and remodeling of load-bearing biological soft tissues.
\newblock \emph{Meccanica}, 52:\penalty0 645–--664, 2017.

\bibitem[Cyron et~al.(2016)Cyron, Aydin, and Humphrey]{CyrAydHum:2016:ahc}
C.~J. Cyron, R.~C. Aydin, and J.~D. Humphrey.
\newblock A homogenized constrained mixture (and mechanical analog) model for
  growth and remodeling of soft tissue.
\newblock \emph{Biomech. Model. Mechan.}, 15:\penalty0 1389, 2016.

\bibitem[Cyron et~al.(2017)Cyron, Wilson, and Humphrey]{CyrWilHum:2017:cff}
C.~J. Cyron, J.~S. Wilson, and J.~D. Humphrey.
\newblock Chapter 4 - constitutive formulations for soft tissue growth and
  remodeling.
\newblock In Y.~Payan and J.~Ohayon, editors, \emph{Biomechanics of Living
  Organs}, volume~1 of \emph{Translational Epigenetics}, pages 79--100.
  Academic Press, Oxford, 2017.

\bibitem[Eichinger et~al.(2021)Eichinger, Haeusel, Paukner, Aydin, Humphrey,
  and Cyron]{EicHaePauAydHumCyr:2021:mhi}
J.~F. Eichinger, L.~J. Haeusel, D.~Paukner, R.~C. Aydin, J.~D. Humphrey, and
  C.~J. Cyron.
\newblock Mechanical homeostasis in tissue equivalents: a review.
\newblock \emph{Biomech Model Mechanobiol}, 20:\penalty0 833–850, 2021.

\bibitem[Famaey et~al.(2018)Famaey, Vastmans, Fehervary, Maes, Vanderveken,
  Rega, Mousavi, and Avril]{FamVasFehMaeVanRegMouAvr:2018:nso}
N.~Famaey, J.~Vastmans, H.~Fehervary, L.~Maes, E.~Vanderveken, F.~Rega, S.~J.
  Mousavi, and S.~Avril.
\newblock Numerical simulation of arterial remodeling in pulmonary autografts.
\newblock \emph{Z Angew Meth Mech}, 98:\penalty0 2239--2257, 2018.

\bibitem[Finlay et~al.(1995)Finlay, McCullough, and Canham]{FinMccCan:1995:tdc}
H.~M. Finlay, L.~McCullough, and P.~B. Canham.
\newblock Three-dimensional collagen organization of human brain arteries at
  different transmural pressures.
\newblock \emph{J Vasc Res}, 32:\penalty0 301--312, 1995.

\bibitem[Gannon et~al.(2008)Gannon, Vanlandingham, Jernigan, Grifoni, Hamilton,
  and Drummond]{GanVanJerGriHamDru:2008:ipi}
K.~P. Gannon, L.~G. Vanlandingham, N.~L. Jernigan, S.~C. Grifoni, G.~Hamilton,
  and H.~A. Drummond.
\newblock Impaired pressure-induced constriction in mouse middle cerebral
  arteries of asic2 knockout mice.
\newblock \emph{Am J Physiol Heart Circ Physiol}, 294:\penalty0 H1793--H1803,
  2008.

\bibitem[Göktepe et~al.(2010)Göktepe, Abilez, and Kuhl]{GokAbiKuh:2010:aga}
S.~Göktepe, O.~J. Abilez, and E.~Kuhl.
\newblock A generic approach towards finite growth with examples of athlete’s
  heart, cardiac dilation, and cardiac wall thickening.
\newblock \emph{Journal of the Mechanics and Physics of Solids}, 58:\penalty0
  1661--1680, 2010.

\bibitem[Hayash et~al.(2001)Hayash, Stergiopulos, Meister, Greenwald, and
  Rachev]{HaySteMeiGreRac:2001:tit}
K.~Hayash, N.~Stergiopulos, J.-J. Meister, S.~E. Greenwald, and A.~Rachev.
\newblock Techniques in the determination of the mechanical properties and
  constitutive laws of arterial walls.
\newblock In C.~Leondes, editor, \emph{Biomechanical Systems: Techniques and
  Applications, Volume II: Cardiovascular Techniques}, chapter~6, pages 1--61.
  {CRC Press}, 2001.

\bibitem[He and Ku(1996)]{HeKu:1996:pfi}
X.~He and D.~Ku.
\newblock Pulsatile flow in the human left coronary artery bifurcation: average
  conditions.
\newblock \emph{J. Biomech. Eng.}, 118:\penalty0 74--–82, 1996.
\newblock \doi{10.1115/1.2795948}.

\bibitem[Himpel et~al.(2005)Himpel, Kuhl, Menzel, and
  Steinmann]{HimKuhMenSte:2005:cmo}
G.~Himpel, E.~Kuhl, A.~Menzel, and P.~Steinmann.
\newblock Computational modelling of isotropic multiplicative growth.
\newblock \emph{Comput. Model. Eng. Sci.}, 8:\penalty0 119--134, 2005.

\bibitem[Hinz(2015)]{Hin:2015:tem}
B.~Hinz.
\newblock The extracellular matrix and transforming growth factor-$\beta$1:
  Tale of a strained relationship.
\newblock \emph{Matrix Biol}, 47:\penalty0 54--65, 2015.

\bibitem[Holzapfel et~al.(2007)Holzapfel, Sommer, Auer, Regitnig, and
  Ogden]{HolSomAueRegOgd:2007:lst}
G.~A. Holzapfel, G.~Sommer, M.~Auer, P.~Regitnig, and R.~W. Ogden.
\newblock Layer-specific 3d residual deformations of human aortas with
  non-atherosclerotic intimal thickening.
\newblock \emph{Ann Biomed Eng}, 35:\penalty0 530–545, 2007.

\bibitem[Horny et~al.(2011)Horny, Adamek, Gultova, Zitny, Vesely, Chlup, and
  Konvickova]{HorAdaGulZitVesChlKon:2011:cba}
L.~Horny, T.~Adamek, E.~Gultova, R.~Zitny, J.~Vesely, H.~Chlup, and
  S.~Konvickova.
\newblock Correlations between age, prestrain, diameter and atherosclerosis in
  the male abdominal aorta.
\newblock \emph{J Mech Behav Biomed}, 4:\penalty0 2128--2132, 2011.

\bibitem[Humphrey and Rajagopal(2002)]{HumRaj:2002:acm}
J.~D. Humphrey and K.~R. Rajagopal.
\newblock A constrained mixture model for growth and remodeling of soft
  tissues.
\newblock \emph{Math Mod Meth Appl S}, 12:\penalty0 407, 2002.

\bibitem[Imatani and Maugin(2002)]{ImaMau:2002:acm}
S.~Imatani and G.~A. Maugin.
\newblock A constitutive model for material growth and its application to
  three-dimensional finite element analysis.
\newblock \emph{Mechanics Research Communications}, 29\penalty0 (6):\penalty0
  477--483, 2002.
\newblock ISSN 0093-6413.

\bibitem[Jadidi et~al.(2021{\natexlab{a}})Jadidi, Razian, Anttila, Doan,
  Adamson, Pipinos, and Kamenskiy]{JadRazAntDoaAdaPipKam:2021:com}
M.~Jadidi, S.~A. Razian, E.~Anttila, T.~Doan, J.~Adamson, M.~Pipinos, and
  A.~Kamenskiy.
\newblock Comparison of morphometric, structural, mechanical, and physiologic
  characteristics of human superficial femoral and popliteal arteries.
\newblock \emph{Acta Biomater}, 121:\penalty0 431--443, 2021{\natexlab{a}}.

\bibitem[Jadidi et~al.(2021{\natexlab{b}})Jadidi, Razian, Habibnezhad, and
  Kamenskiy]{JadRazHabAntKam:2021:msa}
M.~Jadidi, S.~A. Razian, E.~Habibnezhad, M.and~Anttila, and A.~Kamenskiy.
\newblock Mechanical, structural, and physiologic differences in human elastic
  and muscular arteries of different ages: Comparison of the descending
  thoracic aorta to the superficial femoral artery.
\newblock \emph{Acta Biomater}, 119:\penalty0 268--283, 2021{\natexlab{b}}.

\bibitem[Johnson et~al.(2009)Johnson, El-Yazbi, Takeya, Walsh, Walsh, and
  Cole]{JohElyTakWalWalCol:2009:csv}
R.~P. Johnson, A.~F. El-Yazbi, K.~Takeya, E.~J. Walsh, M.~P. Walsh, and W.~C.
  Cole.
\newblock Ca2+ sensitization via phosphorylation of myosin phosphatase
  targeting subunit at threonine-855 by rho kinase contributes to the arterial
  myogenic response.
\newblock \emph{J Physiol}, 587.11:\penalty0 2537--2553, 2009.

\bibitem[Kelly et~al.(1989)Kelly, Hayward, Avolio, and
  O’Rourke]{KelHayAvoOro:1989:ndo}
R.~Kelly, C.~Hayward, A.~Avolio, and M.~O’Rourke.
\newblock Noninvasive determination of age-related changes in the human
  arterial pulse.
\newblock \emph{Circulation}, 80:\penalty0 1652--1659, 1989.

\bibitem[Kuhl(2014)]{Kuh:2014:gma}
E.~Kuhl.
\newblock Growing matter: a review of growth in living systems.
\newblock \emph{J Mech Behav Biomed Mater}, 29:\penalty0 529--543, 2014.

\bibitem[Kuhl and Holzapfel(2007)]{KuhHol:2007:acm}
E.~Kuhl and G.~A. Holzapfel.
\newblock A continuum model for remodeling in living structures.
\newblock \emph{J Mater Sci}, 42:\penalty0 8811–--8823, 2007.

\bibitem[Kuhl et~al.(2007)Kuhl, Maas, Himpel, and
  Menzel]{KuhMaaHimMen:2007:cmo}
E.~Kuhl, R.~Maas, G.~Himpel, and A.~Menzel.
\newblock Computational modeling of arterial wall growth.
\newblock \emph{Biomech Model Mechanobiol}, 6:\penalty0 321--331, 2007.

\bibitem[Lamm et~al.(2022)Lamm, Holthusen, Brepols, Jockenh\"ovel, and
  Reese]{LamHolBreJocRee:2022:ama}
L.~Lamm, H.~Holthusen, T.~Brepols, S.~Jockenh\"ovel, and S.~Reese.
\newblock A macroscopic approach for stress-driven anisotropic growth in
  bioengineered soft tissues.
\newblock \emph{Biomechanics and Modeling in Mechanobiology}, 21:\penalty0
  627--645, 2022.

\bibitem[Latorre and Humphrey(2018)]{LatHum:2018:ame}
M.~Latorre and J.~D. Humphrey.
\newblock A mechanobiologically equilibrated constrained mixture model for
  growth and remodeling of soft tissues.
\newblock \emph{Z Angew Math Mech}, 98(12):\penalty0 2048--2071, 2018.

\bibitem[Lee et~al.(2015)Lee, Genet, Acevedo-Bolton, Ordovas, Guccinone, and
  Kuhl]{LeeGenAceOrdGucKuhl:2015:acm}
L.~C. Lee, M.~Genet, G.~Acevedo-Bolton, K.~Ordovas, J.~M. Guccinone, and
  E.~Kuhl.
\newblock A computational model that predicts reverse growth in response to
  mechanical unloading.
\newblock \emph{Biomech Model Mechanobiol}, 14:\penalty0 217–--229, 2015.

\bibitem[Lubarda and Hoger(2002)]{LubHog:2002:otm}
V.~A. Lubarda and A.~Hoger.
\newblock On the mechanics of solids with a growing mass.
\newblock \emph{IJSS}, 39:\penalty0 4627--4664, 2002.

\bibitem[Menzel and Kuhl(2012)]{MenKuh:2012:fig}
A.~Menzel and E.~Kuhl.
\newblock Frontiers in growth and remodeling.
\newblock \emph{Mech Res Commun}, 42:\penalty0 1--14, 2012.

\bibitem[Mousavi and Avril(2017)]{MouAvr:2017:pss}
S.~J. Mousavi and S.~Avril.
\newblock Patient-specific stress analyses in the ascending thoracic aorta
  using a finite-element implementation of the constrained mixture theory.
\newblock \emph{Biomech Model Mechanobiol}, 16:\penalty0 1765–--1777, 2017.

\bibitem[Parasuraman and Raveendran(2012)]{ParRav:2012:moi}
S.~Parasuraman and R.~Raveendran.
\newblock Measurement of invasive blood pressure in rats.
\newblock \emph{J Pharmacol Pharmacother}, 3\penalty0 (2):\penalty0
  172–--177, 2012.

\bibitem[Rodriguez et~al.(1994)Rodriguez, Hoger, and
  McCulloch]{RodHogMcc:1994:sdf}
E.~K. Rodriguez, A.~Hoger, and A.~D. McCulloch.
\newblock Stress-dependent finite growth in soft elastic tissues.
\newblock \emph{J. Biomech.}, 27:\penalty0 455--467, 1994.

\bibitem[Saez(2016)]{Sae:2016:ott}
P.~Saez.
\newblock On the theories and numerics of continuum models for adaptation
  processes in biological tissues.
\newblock \emph{Arch Computat Methods Eng}, 23:\penalty0 301--322, 2016.

\bibitem[Schmidt and Balzani(2016)]{SchBal:2016:riv}
T.~Schmidt and D.~Balzani.
\newblock Relaxed incremental variational approach for the modeling of
  damage-induced stress hysteresis in arterial walls.
\newblock \emph{Journal of the Mechanical Behavior of Biomedical Materials},
  58:\penalty0 149--162, 2016.
\newblock \doi{10.1016/j.jmbbm.2015.08.005}.

\bibitem[Schriefl et~al.(2012)Schriefl, Reinisch, Sankaran, Pierce, and
  Holzapfel]{SchReiSanPieHol:2012:qao}
A.~J. Schriefl, A.~J. Reinisch, S.~Sankaran, D.~M. Pierce, and G.~A. Holzapfel.
\newblock Quantitative assessment of collagen fibre orientations from
  two-dimensional images of soft biological tissues.
\newblock \emph{J R Soc Interface}, 9:\penalty0 3081--3093, 2012.

\bibitem[Schr\"oder and Neff(2003)]{SchNef:2003:ifo}
J.~Schr\"oder and P.~Neff.
\newblock Invariant formulation of hyperelastic transverse isotropy based on
  polyconvex free energy functions.
\newblock \emph{International Journal of Solids and Structures}, 40:\penalty0
  401--445, 2003.

\bibitem[Schulze-Bauer et~al.(2003)Schulze-Bauer, M\"orth, and
  Holzapfel]{SchMorHol:2003:pbm}
C.~A.~J. Schulze-Bauer, C.~M\"orth, and G.~A. Holzapfel.
\newblock Passive biaxial mechanical response of aged human iliac arteries.
\newblock \emph{J Biomech Eng}, 125:\penalty0 395--406, 2003.

\bibitem[Sommer et~al.(2010)Sommer, Regitnig, K\"oltringer, and
  Holzapfel]{SomRegKolHol:2010:bmp}
G.~Sommer, P.~Regitnig, L.~K\"oltringer, and G.~A. Holzapfel.
\newblock Biaxial mechanical properties of intact and layer-dissected human
  carotid arteries at physiological and supraphysiological loadings.
\newblock \emph{Am J Physiol Heart Circ Physiol}, 298:\penalty0 H898--H912,
  2010.

\bibitem[Sáez et~al.(2014)Sáez, Peña, Martínez, and
  Kuhl]{SaePenMarKuh:2014:cmo}
P.~Sáez, E.~Peña, M.~Martínez, and E.~Kuhl.
\newblock Computational modeling of hypertensive growth in the human carotid
  artery.
\newblock \emph{Comput Mech}, 53:\penalty0 1183--1196, 2014.

\bibitem[Taber and Humphrey(2001)]{TabHum:2001:smg}
L.~A. Taber and J.~D. Humphrey.
\newblock Stress-modulated growth, residual stress, and vascular heterogeneity.
\newblock \emph{J Biomech Eng}, 123:\penalty0 528, 2001.

\bibitem[Tanaka et~al.(2016)Tanaka, Balzani, and
  Schr\"oder]{TanBalSch:2016:ioi}
M.~Tanaka, D.~Balzani, and J.~Schr\"oder.
\newblock Implementation of incremental variational formulations based on the
  numerical calculation of derivatives using hyper dual numbers.
\newblock \emph{Computer Method in Applied Mechanics and Engineering},
  301:\penalty0 216--240, 2016.

\bibitem[Uhlmann and Balzani(2023)]{UhlBal:2023:cmm}
K.~Uhlmann and D.~Balzani.
\newblock Chemo-mechanical modeling of smooth muscle cell activation for the
  simulation of arterial walls under changing blood pressure.
\newblock \emph{Biomech Model Mechanobiol}, 22:\penalty0 1049--1065, 2023.

\bibitem[Valentín and Humphrey(2009)]{ValHum:2009:eof}
A.~Valentín and J.~D. Humphrey.
\newblock Evaluation of fundamental hypotheses underlying constrained mixture
  models of arterial growth and remodelling.
\newblock \emph{Phil Trans R Soc A}, 367:\penalty0 3585–--3606, 2009.

\bibitem[Wang et~al.(2021)Wang, Balzani, Vedula, Uhlmann, and
  Varnik]{WanBalVedUhlVar:2021:otp}
H.~Wang, D.~Balzani, V.~Vedula, K.~Uhlmann, and F.~Varnik.
\newblock On the potential self-amplification of aneurysms due to tissue
  degradation and blood flow revealed from {FSI} simulations.
\newblock \emph{Frontiers in Physiology}, 12:\penalty0 785780, 2021.
\newblock \doi{10.3389/fphys.2021.785780}.

\bibitem[Wang et~al.(2022)Wang, Uhlmann, Vedula, Balzani, and
  Varnik]{WanUhlVedBalVar:2022:fsi}
H.~Wang, K.~Uhlmann, V.~Vedula, D.~Balzani, and F.~Varnik.
\newblock Fluid-structure interaction of tissue degradation and its effects on
  intra-aneurysm hemodynamics.
\newblock \emph{Biomechanics and Modeling in Mechanobiology}, 21:\penalty0
  671--683, 2022.
\newblock \doi{10.1007/s10237-022-01556-7}.

\bibitem[Yap et~al.(2021)Yap, Mieremet, de~Vries, Micha, and
  de~Waard]{YapMieVriMicWaa:2021:sso}
C.~Yap, A.~Mieremet, C.~J.~M. de~Vries, D.~Micha, and V.~de~Waard.
\newblock Six shades of vascular smooth muscle cells illuminated by klf4.
\newblock \emph{Arterioscler Thromb Vasc Biol}, 41:\penalty0 2693--2707, 2021.

\bibitem[Zahn(2020)]{Zah:2020:mog}
A.~Zahn.
\newblock \emph{Modeling of growth and fiber reorientation in soft biological
  tissues}.
\newblock PhD thesis, Ruhr Universit{\"a}t Bochum, 2020.

\bibitem[Zahn and Balzani(2018)]{ZahBal:2018:acg}
A.~Zahn and D.~Balzani.
\newblock A combined growth and remodeling framework for the approximation of
  residual stresses in arterial walls.
\newblock \emph{Z Angew Math Mech}, 98:\penalty0 2072--2100, 2018.

\end{thebibliography}
\clearpage
\begin{appendix}
\section{Recapitulation of Competitive Growth Model (Model B)}
\label{sec:mandel}
\begin{figure}[!b]
\unitlength1cm
\begin{picture}(15.5,5.4)
\put(0.0,0.0){
\begin{tikzpicture}
\begin{axis}[
width=15.0cm,
height=7.0cm,
xlabel={{Pressure in mmHg}},
xmin=0.0, xmax=125,
xtick={0, 20, 40, 60, 80, 100, 120},
extra x ticks={0.0},
extra x tick style={%
    grid=major,
},
ylabel={{Outer Diameter in $\mu$m}},
ymin=200, ymax=325,
ytick={200, 220, 240, 260, 280, 300, 320},
axis background/.style={fill=white!89.80392156862746!black},
axis line style={white!60.0!black},
axis x line*=bottom,
axis y line*=left,
tick align=outside,
tick pos=left,
scaled x ticks=false,
x grid style={white},
xmajorgrids,
x tick label style={
  font=\small,
  /pgf/number format/.cd,
    fixed,
    fixed zerofill,
    precision=0,
  /tikz/.cd
},
xlabel style={at={(0.5,-0.10)},font=\small},
ylabel style={at={(-0.07,0.5)},font=\small},
y grid style={white},
ymajorgrids,
y tick label style={
  font=\small,
},
legend cell align={left},
legend style={at={(0.0,1.0)}, font=\small, anchor=north west, draw=white!60.0!black, row sep=-0.08cm},
clip mode=individual,
]

\addplot [very thick, rub_blue, mark options={solid, draw=\tcolorshade, line width=0.5pt}]
table {%
0   200.906248167913
10  222.145741265094
20	237.047881794649 
40	259.635095565804 
60	277.368419251967 
80  292.152958029544 
100 304.948629503116  
120 316.292817881301
};
\addlegendentry{{\footnotesize Passive Response - Simulation}}

\addplot [only marks, color=rub_blue, mark=otimes*, mark size=2.5,]
table {%
10  223
20	236
40	257
60	278
80	296
100	306
120	315
};
\addlegendentry{{\footnotesize Passive Response - Experiment}}

\end{axis}
\end{tikzpicture}}
\end{picture}
\caption{
Comparison of simulation results for the passive response of the arterial wall (solid line) with experimental data from~\cite{JohElyTakWalWalCol:2009:csv} (dots) for the consideration of the competitive growth model (model B) in the optimization.
The simulation results agree well with the experimental data.
\label{fig:validation_mandel}}
\end{figure}
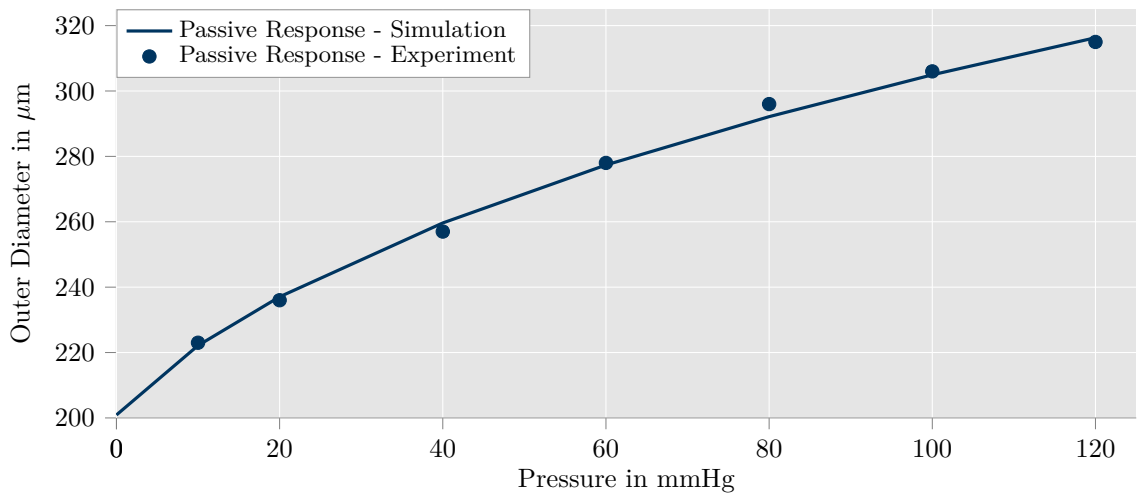

The previous version of the growth model (model B) includes a different set of evolution equations for the growth factors~$\vartheta^{(a)}$.
These evolution equations for model B are defined as
\eb
\label{eq:groFac_B}
\dot{\vartheta}^{(a)} = \kappa^+_{\vartheta,(a)} \left[ \frac{\vartheta^+_{(a)} - \vartheta^{(a)}}{\vartheta^+_{(a)} - 1} \right]^{m^+_{\vartheta,(a)}} \phi^{(a)} \, ,
\ee
where~$\vartheta^{(a)}$ is the growth factor, $\vartheta^+_{(a)}$ constitutes the predefined maximal value for the growth factor, $\kappa^+_{\vartheta,(a)}$ is the growth velocity factor and~$\phi^{(a)}$ is the driving force.
The exponent~$m^+_{\vartheta,(a)}$ determines the degree of nonlinearity of the growth process.
However, this value is set to a value of~1.0 in simulations presented in this paper.
The elastic part of the Mandel stress~$\BSigma_{\text{e}} = \bC_e \bS_e$ is used as foundation for the definition of driving forces~$\phi^{(a)}$.
As described in Section~\ref{sec:model} for model A, only growth in the direction of the first and second eigenvectors of~$\BSigma_{\text{e}}$ is applied here.
Since an idealized geometry (hollow cylinder) is used for the simulations, these eigenvectors point into axial and radial direction and, consequently, are equal to the eigenvectors of the Cauchy stress tensor~$\Bsigma$.
Comparable to model A, the driving forces in model B are defined as
\eb
\phi^{(2)}(\BSigma_{\text{e}}) = \BSigma_{\text{e}} : \left( \bn^{(2)} \otimes \bn^{(2)} \right) \qquad \text{and} \qquad
\phi^{(3)}(\BSigma_{\text{e}}) = \BSigma_{\text{e}} : \bI \, .
\ee
More details on investigations and variations of this growth model can be found in~\cite{ZahBal:2018:acg} and~\cite{Zah:2020:mog}.
The final values of the optimization parameters are listed in Table~\ref{tab:mandel}.
Furthermore, a comparison between the final simulation results with the experimental data from~\cite{JohElyTakWalWalCol:2009:csv} is illustrated in Fig.~\ref{fig:validation_mandel}.
\begin{table}[thb]
\caption{Optimized values of parameters for model B \label{tab:mandel}}
\small
\begin{center}
\begin{tabular}{rccccccc} \toprule
  Parameter &   $\alpha_1$ &   $\alpha_4$ &   $\alpha_5$ 
&   $\vartheta^+_{(2)}$ &   $\vartheta^+_{(3)}$ &   $\kappa^+_{\vartheta,(2)}$ &   $\kappa^+_{\vartheta,(3)}$ \\ \midrule
  Value &   2.97$\,$kPa &   4.77$\,$kPa &   2.55$\,$kPa 
&   1.27 &   2.78 &   4.95$\cdot 10^{-7} \, \text{s}^{-1}$ &   4.35$\cdot 10^{-4} \, \text{s}^{-1}$ \\ \bottomrule
\end{tabular}
\end{center}
\end{table}
\clearpage

\section{Efficiency of the Optimization Procedure for Varied Axial Stretch and Convergence Values}
\label{sec:opt2}
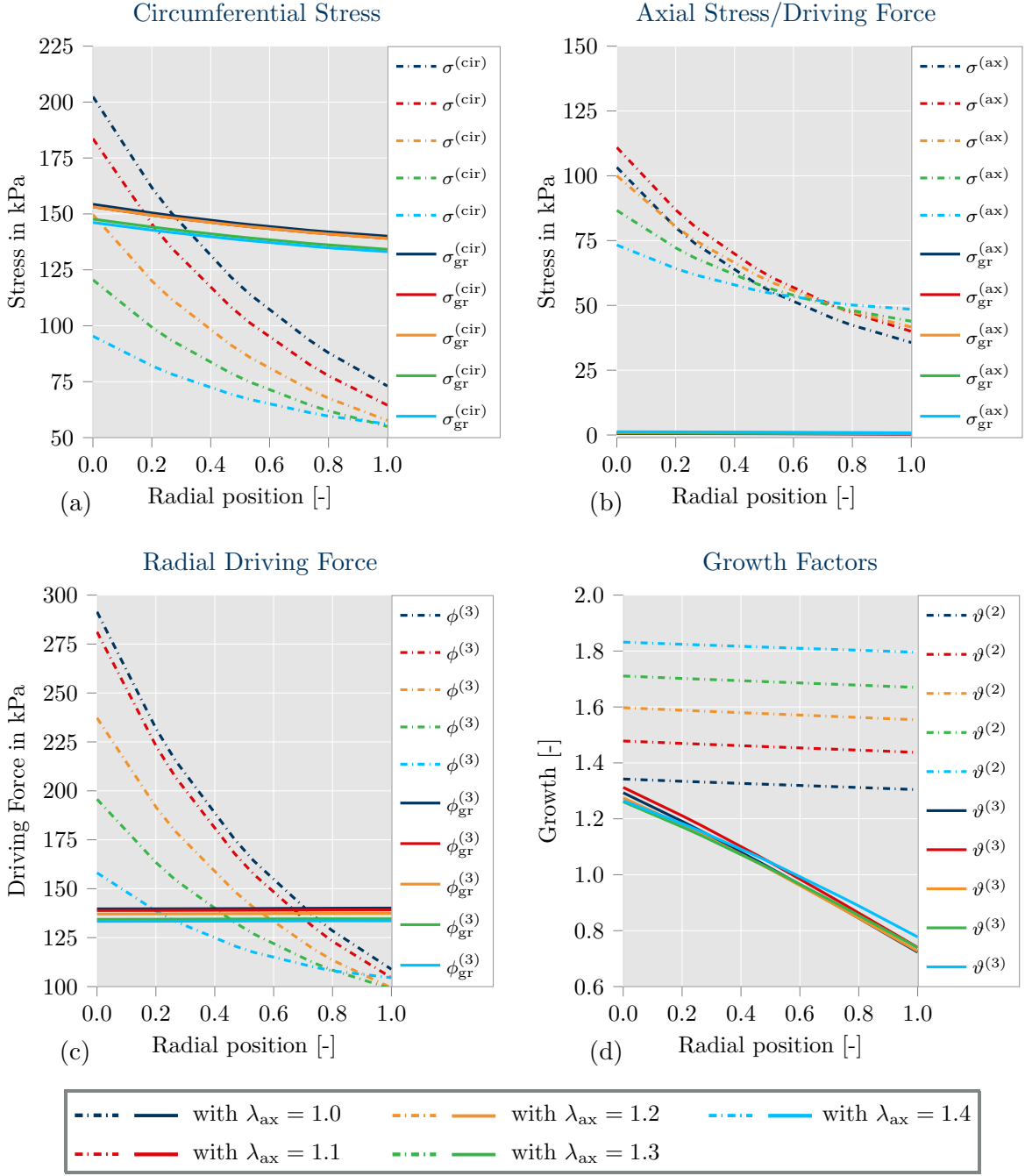
\begin{figure}[!h]
\unitlength1cm
\begin{picture}(15.5,18.5)
\put( 4.0, 18.1){\color{rub_blue}\LARGE Variation of Axial Stretch $\lambda_{\text{ax}}$}

\put( 1.4, 9.8){
\put( 0.5, 7.5){\color{rub_blue} Circumferential Stress}
\put( -1.5, 0.0){
\begin{tikzpicture}
\begin{axis}[
width=6.0cm,
height=7.5cm,
xlabel={{Radial position [-]}},
xmin=0, xmax=1,
xtick={0.2, 0.4, 0.6, 0.8, 1.0},
extra x ticks={0.0},
extra x tick style={%
    grid=major,
},
ylabel={{Stress in kPa}},
ymin=50, ymax=225,
ytick={50, 75, 100, 125, 150, 175, 200, 225},
axis background/.style={fill=white!89.80392156862746!black},
axis line style={white!60.0!black},
axis x line*=bottom,
axis y line*=left,
tick align=outside,
tick pos=left,
scaled x ticks=false,
x grid style={white},
xmajorgrids,
x tick label style={
  font=\small,
  /pgf/number format/.cd,
  set decimal separator={.},
    fixed,
    fixed zerofill,
    precision=1,
  /tikz/.cd
},
xlabel style={at={(0.5,-0.1)},font=\small},
ylabel style={at={(-0.20,0.5)},font=\small},
y grid style={white},
ymajorgrids,
y tick label style={
  font=\small,
},
legend cell align={left},
legend style={at={(1.0,1.0)}, anchor=north west, draw=white!60.0!black, row sep=0.107cm},
clip mode=individual,
]

\addplot [dash dot,very thick,dia1, mark options={solid, draw=\tcolorshade, line width=0.5pt}]
table {%
0.0     202.40645100786
0.20185596580163762     161.440541541506
0.26601054957955494     150.520485037003
0.4678399874118748     121.89662425841
0.5320028392220363     114.150243808358
0.7338491076444759     93.6750860178975
0.7980690667767155     88.0700034081744
1.0     73.0619601368322
};
\addlegendentry{{\scriptsize $\sigma^{\text{(cir)}}$}}

\addplot [dash dot,very thick,dia2, mark options={solid, draw=\tcolorshade, line width=0.5pt}]
table {%
0.0     183.635902681099
0.20188333742746575     145.124897329308
0.2660462743523159     134.929920900045
0.4679023555948115     108.454081593913
0.532093855771953     101.378972913256
0.7339673659647336     82.8011249107702
0.7981525585732263     77.7803462304167
1.0     64.5029126309624
};
\addlegendentry{{\scriptsize $\sigma^{\text{(cir)}}$}}

\addplot [dash dot,very thick,dia3, mark options={solid, draw=\tcolorshade, line width=0.5pt}]
table {%
0.0     149.651097666578
0.20185997498365138     119.826070345636
0.2660299566628939     111.944112084912
0.467890185399922     91.4466798987027
0.5320702441477826     85.9570661265304
0.7339854190801681     71.5908249751842
0.7981925671006774     67.7290806120584
1.0     57.6562304431193
};
\addlegendentry{{\scriptsize $\sigma^{\text{(cir)}}$}}

\addplot [dash dot,very thick,dia4, mark options={solid, draw=\tcolorshade, line width=0.5pt}]
table {%
0.0     120.529415966987
0.20183988927754767     99.1869967914573
0.26602049954314666     93.518369579147
0.46790487067223624     78.8339228229155
0.5320475928010207     74.9021472742934
0.7340559311293323     64.6986659435721
0.7982816272747595     62.0217440156419
1.0     54.9792623060118
};
\addlegendentry{{\scriptsize $\sigma^{\text{(cir)}}$}}

\addplot [dash dot,very thick,dia5, mark options={solid, draw=\tcolorshade, line width=0.5pt}]
table {%
0.0     95.3510224690786
0.20190041499246705     81.9968834165531
0.26608336293818485     78.4635328894961
0.4679375311265276     69.4423255679858
0.5321713590851198     67.0968684455471
0.7340911877736388     61.2319357263079
0.798116275977273     59.6460063020014
1.0     56.0603001999751 
};
\addlegendentry{{\scriptsize $\sigma^{\text{(cir)}}$}}

\addplot [very thick,dia1, mark options={solid, draw=\tcolorshade, line width=0.5pt}]
table {%
0.0     154.27907828825
0.20185596580163762     150.403584424015
0.26601054957955494     149.292326415394
0.4678399874118748     146.126692307899
0.5320028392220363     145.222986578734
0.7338491076444759     142.655159096935
0.7980690667767155     141.968686690109
1.0     140.04126924289   
};
\addlegendentry{{\scriptsize $\sigma^{\text{(cir)}}_{\text{gr}}$}}

\addplot [very thick,dia2, mark options={solid, draw=\tcolorshade, line width=0.5pt}]
table {%
0.0     153.175620481738
0.20188333742746575     149.394553190739
0.2660462743523159     148.293997919158
0.4679023555948115     145.171117378712
0.532093855771953     144.280941094617
0.7339673659647336     141.767832172469
0.7981525585732263     141.069504976431
1.0     139.067926954296
};
\addlegendentry{{\scriptsize $\sigma^{\text{(cir)}}_{\text{gr}}$}}

\addplot [very thick,dia3, mark options={solid, draw=\tcolorshade, line width=0.5pt}]
table {%
0.0     153.175620481738
0.20188333742746575     149.394553190739
0.2660462743523159     148.293997919158
0.4679023555948115     145.171117378712
0.532093855771953     144.280941094617
0.7339673659647336     141.767832172469
0.7981525585732263     141.069504976431
1.0     139.067926954296
};
\addlegendentry{{\scriptsize $\sigma^{\text{(cir)}}_{\text{gr}}$}}

\addplot [very thick,dia4, mark options={solid, draw=\tcolorshade, line width=0.5pt}]
table {%
0.0     147.654090970992
0.20183988927754767     144.126417448865
0.26602049954314666     143.072356526991
0.46790487067223624     140.110152255873
0.5320475928010207     139.254583767845
0.7340559311293323     136.799563690718
0.7982816272747595     136.160332349163
1.0     134.137617834624
};
\addlegendentry{{\scriptsize $\sigma^{\text{(cir)}}_{\text{gr}}$}}

\addplot [very thick,dia5, mark options={solid, draw=\tcolorshade, line width=0.5pt}]
table {%
0.0     146.159360878169
0.20190041499246705     142.738027401845
0.26608336293818485     141.73066059368
0.4679375311265276     138.854261874758
0.5321713590851198     138.024094423486
0.7340911877736388     135.606852386526
0.798116275977273     134.859956216278
1.0     133.115983822947
};
\addlegendentry{{\scriptsize $\sigma^{\text{(cir)}}_{\text{gr}}$}}

\end{axis}
\end{tikzpicture}}
\put( -0.6, 0.1){(a)}
}
\put( -0.15, 9.8){
\put( 9.6, 7.5){\color{rub_blue} Axial Stress/Driving Force}
\put( 8.0, 0.0){
\begin{tikzpicture}
\begin{axis}[
width=6.0cm,
height=7.5cm,
xlabel={{Radial position [-]}},
xmin=0, xmax=1,
xtick={0.2, 0.4, 0.6, 0.8, 1.0},
extra x ticks={0.0},
extra x tick style={%
    grid=major,
},
ylabel={{Stress in kPa}},
ymin=-1, ymax=150,
ytick={0, 25, 50, 75, 100, 125, 150},
axis background/.style={fill=white!89.80392156862746!black},
axis line style={white!60.0!black},
axis x line*=bottom,
axis y line*=left,
tick align=outside,
tick pos=left,
scaled x ticks=false,
x grid style={white},
xmajorgrids,
x tick label style={
  font=\small,
  /pgf/number format/.cd,
  set decimal separator={.},
    fixed,
    fixed zerofill,
    precision=1,
  /tikz/.cd
},
xlabel style={at={(0.5,-0.1)},font=\small},
ylabel style={at={(-0.18,0.5)},font=\small},
y grid style={white},
ymajorgrids,
y tick label style={
  font=\small,
},
legend cell align={left},
legend style={at={(1.0,1.0)}, anchor=north west, draw=white!60.0!black, row sep=0.107cm},
clip mode=individual,
]

\addplot [dash dot,very thick,dia1, mark options={solid, draw=\tcolorshade, line width=0.5pt}]
table {%
0.0     103.236855288882
0.20185596580163762     79.8553238549863
0.26601054957955494     73.8498172481616
0.4678399874118748     58.8400947914352
0.5320028392220363     54.9093138611257
0.7338491076444759     45.0199488101938
0.7980690667767155     42.3931521932223
1.0     35.6907499445511   
};
\addlegendentry{{\scriptsize $\sigma^{\text{(ax)}}$}}

\addplot [dash dot,very thick,dia2, mark options={solid, draw=\tcolorshade, line width=0.5pt}]
table {%
0.0     110.932966770892
0.20188333742746575     86.6428245891546
0.2660462743523159     80.3655833529257
0.4679023555948115     64.577726975795
0.532093855771953     60.4506708665146
0.7339673659647336     49.9378968810748
0.7981525585732263     47.1468097596048
1.0     39.9742886007532  
};
\addlegendentry{{\scriptsize $\sigma^{\text{(ax)}}$}}

\addplot [dash dot,very thick,dia3, mark options={solid, draw=\tcolorshade, line width=0.5pt}]
table {%
0.0     99.986144286641
0.20185997498365138     80.193750919895
0.2660299566628939     75.0612001656074
0.467890185399922     62.0404985677935
0.5320702441477826     58.6062695373835
0.7339854190801681     49.8446802532889
0.7981925671006774     47.5212140278972
1.0     41.6412372925384  
};
\addlegendentry{{\scriptsize $\sigma^{\text{(ax)}}$}}

\addplot [dash dot,very thick,dia4, mark options={solid, draw=\tcolorshade, line width=0.5pt}]
table {%
0.0     86.6402595658191
0.20183988927754767     72.0752158423464
0.26602049954314666     68.2632315008248
0.46790487067223624     58.627528807919
0.5320475928010207     56.0848763755909
0.7340559311293323     49.6758350460716
0.7982816272747595     48.0445124766944
1.0     43.87304367346
};
\addlegendentry{{\scriptsize $\sigma^{\text{(ax)}}$}}

\addplot [dash dot,very thick,dia5, mark options={solid, draw=\tcolorshade, line width=0.5pt}]
table {%
0.0     73.288625508398
0.20190041499246705     64.1507964600045
0.26608336293818485     61.7810794798313
0.4679375311265276     55.9458739808553
0.5321713590851198     54.4814698427986
0.7340911877736388     51.0482641209107
0.798116275977273     50.1500835413695
1.0     48.4964042049077    
};
\addlegendentry{{\scriptsize $\sigma^{\text{(ax)}}$}}

\addplot [very thick,dia1, mark options={solid, draw=\tcolorshade, line width=0.5pt}]
table {%
0.0     0.600906337242439
0.20185596580163762     0.560369807131971
0.26601054957955494     0.54646822529981
0.4678399874118748     0.499989156165868
0.5320028392220363     0.483738167402359
0.7338491076444759     0.429107801045305
0.7980690667767155     0.409195906721596
1.0     0.343531836128363  
};
\addlegendentry{{\scriptsize $\sigma^{\text{(ax)}}_{\text{gr}}$}}

\addplot [very thick,dia2, mark options={solid, draw=\tcolorshade, line width=0.5pt}]
table {%
0.0     0.689753112460743
0.20188333742746575     0.639103543206801
0.2660462743523159     0.622084905493117
0.4679023555948115     0.5660740838617
0.532093855771953     0.546966230547244
0.7339673659647336     0.484313854190701
0.7981525585732263     0.462727704465388
1.0     0.394268292913936
};
\addlegendentry{{\scriptsize $\sigma^{\text{(ax)}}_{\text{gr}}$}}

\addplot [very thick,dia3, mark options={solid, draw=\tcolorshade, line width=0.5pt}]
table {%
0.0     0.857812616376287
0.20185997498365138     0.795200987983325
0.2660299566628939     0.774324358993749
0.467890185399922     0.706891220405241
0.5320702441477826     0.68415696681524
0.7339854190801681     0.611278476927487
0.7981925671006774     0.587449362630482
1.0     0.510675503575287
};
\addlegendentry{{\scriptsize $\sigma^{\text{(ax)}}_{\text{gr}}$}}

\addplot [very thick,dia4, mark options={solid, draw=\tcolorshade, line width=0.5pt}]
table {%
0.0     1.05906869013482
0.20183988927754767     0.998086823956611
0.26602049954314666     0.977807448988643
0.46790487067223624     0.907583591045434
0.5320475928010207     0.882977128908204
0.7340559311293323     0.79832015449809
0.7982816272747595     0.766858304138618
1.0     0.663094467046577
};
\addlegendentry{{\scriptsize $\sigma^{\text{(ax)}}_{\text{gr}}$}}

\addplot [very thick,dia5, mark options={solid, draw=\tcolorshade, line width=0.5pt}]
table {%
0.0     1.25484507923876
0.20190041499246705     1.19128525396482
0.26608336293818485     1.16954984549773
0.4679375311265276     1.09585150390089
0.5321713590851198     1.07003347008103
0.7340911877736388     0.982600957744211
0.798116275977273     0.953153846911421
1.0     0.841520217753854
};
\addlegendentry{{\scriptsize $\sigma^{\text{(ax)}}_{\text{gr}}$}}

\end{axis}
\end{tikzpicture}}
\put( 8.9, 0.1){(b)}
}

\put( 1.4, 1.5){
\put( 0.65, 7.5){\color{rub_blue} Radial Driving Force}
\put( -1.5, 0.0){
\begin{tikzpicture}
\begin{axis}[
width=6.0cm,
height=7.5cm,
xlabel={{Radial position [-]}},
xmin=0, xmax=1,
xtick={0.2, 0.4, 0.6, 0.8, 1.0},
extra x ticks={0.0},
extra x tick style={%
    grid=major,
},
ylabel={{Driving Force in kPa}},
ymin=100, ymax=300,
ytick={100, 125, 150, 175, 200, 225, 250, 275, 300},
axis background/.style={fill=white!89.80392156862746!black},
axis line style={white!60.0!black},
axis x line*=bottom,
axis y line*=left,
tick align=outside,
tick pos=left,
scaled x ticks=false,
x grid style={white},
xmajorgrids,
x tick label style={
  font=\small,
  /pgf/number format/.cd,
  set decimal separator={.},
    fixed,
    fixed zerofill,
    precision=1,
  /tikz/.cd
},
xlabel style={at={(0.5,-0.1)},font=\small},
ylabel style={at={(-0.20,0.5)},font=\small},
y grid style={white},
ymajorgrids,
y tick label style={
  font=\small,
},
legend cell align={left},
legend style={at={(1.0,1.0)}, anchor=north west, draw=white!60.0!black, row sep=0.107cm},
clip mode=individual,
]

\addplot [dash dot,very thick,dia1, mark options={solid, draw=\tcolorshade, line width=0.5pt}]
table {%
0.0     291.496759389624
0.20185596580163762     231.434579362786
0.26601054957955494     215.585513751923
0.4678399874118748     175.160266924958
0.5320028392220363     164.276712123709
0.7338491076444759     136.30979427123
0.7980690667767155     128.688233588243
1.0     108.836243729195    
};
\addlegendentry{{\scriptsize $\phi^{(3)}$}}

\addplot [dash dot,very thick,dia2, mark options={solid, draw=\tcolorshade, line width=0.5pt}]
table {%
0.0     281.236774816524
0.20188333742746575     222.560940774216
0.2660462743523159     207.109436665597
0.4679023555948115     167.853053035688
0.532093855771953     157.410969513668
0.7339673659647336     130.549280651239
0.7981525585732263     123.302534225769
1.0     104.514639245659   
};
\addlegendentry{{\scriptsize $\phi^{(3)}$}}

\addplot [dash dot,very thick,dia3, mark options={solid, draw=\tcolorshade, line width=0.5pt}]
table {%
0.0     237.281332563484
0.20185997498365138     191.392153660767
0.2660299566628939     179.332321456575
0.467890185399922     148.584157943331
0.5320702441477826     140.362395136694
0.7339854190801681     119.309324451002
0.7981925671006774     113.650410085557
1.0     99.3094920764675 
};
\addlegendentry{{\scriptsize $\phi^{(3)}$}}

\addplot [dash dot,very thick,dia4, mark options={solid, draw=\tcolorshade, line width=0.5pt}]
table {%
0.0     195.737548544163
0.20183988927754767     163.163003982338
0.26602049954314666     154.534728311835
0.46790487067223624     132.729094171281
0.5320475928010207     126.904738161856
0.7340559311293323     112.232073399595
0.7982816272747595     108.482767148595
1.0     98.8043711326594
};
\addlegendentry{{\scriptsize $\phi^{(3)}$}}

\addplot [dash dot,very thick,dia5, mark options={solid, draw=\tcolorshade, line width=0.5pt}]
table {%
0.0     158.123935584334
0.20190041499246705     138.461521041564
0.26608336293818485     133.315893255479
0.4679375311265276     120.713214902459
0.5321713590851198     117.521963758599
0.7340911877736388     110.086271992951
0.798116275977273     108.06194249301
1.0     104.508065991652 
};
\addlegendentry{{\scriptsize $\phi^{(3)}$}}

\addplot [very thick,dia1, mark options={solid, draw=\tcolorshade, line width=0.5pt}]
table {%
0.0     139.694154700653
0.20185596580163762     139.740866900666
0.26601054957955494     139.761556660596
0.4678399874118748     139.846164195451
0.5320028392220363     139.878047501233
0.7338491076444759     139.985661875538
0.7980690667767155     140.019506544876
1.0     140.104560412817 
};
\addlegendentry{{\scriptsize $\phi^{(3)}_{\text{gr}}$}}

\addplot [very thick,dia2, mark options={solid, draw=\tcolorshade, line width=0.5pt}]
table {%
0.0     138.851176443006
0.20188333742746575     138.899964700752
0.2660462743523159     138.920628604357
0.4679023555948115     139.001941965841
0.532093855771953     139.031715305177
0.7339673659647336     139.12905567704
0.7981525585732263     139.158581629094
1.0     139.229257427662
};
\addlegendentry{{\scriptsize $\phi^{(3)}_{\text{gr}}$}}

\addplot [very thick,dia3, mark options={solid, draw=\tcolorshade, line width=0.5pt}]
table {%
0.0     137.073862008718
0.20185997498365138     137.128055955688
0.2660299566628939     137.150039802975
0.467890185399922     137.232429258846
0.5320702441477826     137.261707033384
0.7339854190801681     137.353647551903
0.7981925671006774     137.379909345514
1.0     137.438796622365
};
\addlegendentry{{\scriptsize $\phi^{(3)}_{\text{gr}}$}}

\addplot [very thick,dia4, mark options={solid, draw=\tcolorshade, line width=0.5pt}]
table {%
0.0     134.32393929355
0.20183988927754767     134.399293151135
0.26602049954314666     134.425690261268
0.46790487067223624     134.511803648796
0.5320475928010207     134.538134282751
0.7340559311293323     134.603557347597
0.7982816272747595     134.614780310569
1.0     134.602538893061
};
\addlegendentry{{\scriptsize $\phi^{(3)}_{\text{gr}}$}}

\addplot [very thick,dia5, mark options={solid, draw=\tcolorshade, line width=0.5pt}]
table {%
0.0     133.325260634921
0.20190041499246705     133.388219933131
0.26608336293818485     133.410748881837
0.4679375311265276     133.484723571186
0.5321713590851198     133.507805705629
0.7340911877736388     133.566746123707
0.798116275977273     133.577818482367
1.0     133.571417404791
};
\addlegendentry{{\scriptsize $\phi^{(3)}_{\text{gr}}$}}

\end{axis}
\end{tikzpicture}}
\put( -0.6, 0.1){(c)}
}
\put( -0.15, 1.5){
\put( 10.6, 7.5){\color{rub_blue} Growth Factors}
\put( 8.0, 0.0){
\begin{tikzpicture}
\begin{axis}[
width=6.0cm,
height=7.5cm,
xlabel={{Radial position [-]}},
xmin=0, xmax=1,
xtick={0.2, 0.4, 0.6, 0.8, 1.0},
extra x ticks={0.0},
extra x tick style={%
    grid=major,
},
ylabel={{Growth [-]}},
ymin=0.6, ymax=2.0,
ytick={0.6, 0.8, 1.0, 1.2, 1.4, 1.6, 1.8, 2.0},
axis background/.style={fill=white!89.80392156862746!black},
axis line style={white!60.0!black},
axis x line*=bottom,
axis y line*=left,
tick align=outside,
tick pos=left,
scaled x ticks=false,
x grid style={white},
xmajorgrids,
x tick label style={
  font=\small,
  /pgf/number format/.cd,
  set decimal separator={.},
    fixed,
    fixed zerofill,
    precision=1,
  /tikz/.cd
},
y tick label style={
  font=\small,
  /pgf/number format/.cd,
  set decimal separator={.},
    fixed,
    fixed zerofill,
    precision=1,
  /tikz/.cd
},
xlabel style={at={(0.5,-0.1)},font=\small},
ylabel style={at={(-0.18,0.5)},font=\small},
y grid style={white},
ymajorgrids,
y tick label style={
  font=\small,
  /pgf/number format/.cd,
  set decimal separator={.},
    fixed,
    fixed zerofill,
    precision=1,
  /tikz/.cd
},
legend cell align={left},
legend style={at={(1.0,1.0)}, anchor=north west, draw=white!60.0!black, row sep=0.135cm},
clip mode=individual,
]

\addplot [dash dot,very thick,dia1, mark options={solid, draw=\tcolorshade, line width=0.5pt}]
table {%
0.0     1.34221314682493
0.20185596580163762     1.33409311861651
0.26601054957955494     1.33163242652029
0.4678399874118748     1.32409984631628
0.5320028392220363     1.32175700638126
0.7338491076444759     1.31442390757258
0.7980690667767155     1.31201472254329
1.0     1.30429587761818       
};
\addlegendentry{{\scriptsize $\vartheta^{\text{(2)}}$}}

\addplot [dash dot,very thick,dia2, mark options={solid, draw=\tcolorshade, line width=0.5pt}]
table {%
0.0     1.47822406993441
0.20188333742746575     1.46951675558053
0.2660462743523159     1.46687218383918
0.4679023555948115     1.45876517850188
0.532093855771953     1.45622762971518
0.7339673659647336     1.44823623869852
0.7981525585732263     1.44561759523677
1.0     1.43738428134055  
};
\addlegendentry{{\scriptsize $\vartheta^{\text{(2)}}$}}

\addplot [dash dot,very thick,dia3, mark options={solid, draw=\tcolorshade, line width=0.5pt}]
table {%
0.0     1.59745830635372
0.20185997498365138     1.58808814850625
0.2660299566628939     1.58525143240012
0.467890185399922     1.57659946437025
0.5320702441477826     1.57386735382828
0.7339854190801681     1.56527666123676
0.7981925671006774     1.56257225980862
1.0     1.5539195587638
};
\addlegendentry{{\scriptsize $\vartheta^{\text{(2)}}$}}

\addplot [dash dot,very thick,dia4, mark options={solid, draw=\tcolorshade, line width=0.5pt}]
table {%
0.0     1.71075180837408
0.20183988927754767     1.70187106589422
0.26602049954314666     1.69925665223087
0.46790487067223624     1.69112804846945
0.5320475928010207     1.68859444394796
0.7340559311293323     1.68062870702324
0.7982816272747595     1.67794034973483
1.0     1.66974799699489
};
\addlegendentry{{\scriptsize $\vartheta^{\text{(2)}}$}}

\addplot [dash dot,very thick,dia5, mark options={solid, draw=\tcolorshade, line width=0.5pt}]
table {%
0.0     1.83193584920091
0.20190041499246705     1.82403315708187
0.26608336293818485     1.821651190307
0.4679375311265276     1.81443610095374
0.5321713590851198     1.81218680467657
0.7340911877736388     1.80516244757796
0.798116275977273     1.8030383839446
1.0     1.79537655392332   
};
\addlegendentry{{\scriptsize $\vartheta^{\text{(2)}}$}}

\addplot [very thick,dia1, mark options={solid, draw=\tcolorshade, line width=0.5pt}]
table {%
0.0     1.29317556140684
0.20185596580163762     1.19008820160038
0.26601054957955494     1.15560633104063
0.4678399874118748     1.04228302466076
0.5320028392220363     1.00499923484429
0.7338491076444759     0.884867733351752
0.7980690667767155     0.845848260575519
1.0     0.723088692685733 
};
\addlegendentry{{\scriptsize $\vartheta^{\text{(3)}}$}}

\addplot [very thick,dia2, mark options={solid, draw=\tcolorshade, line width=0.5pt}]
table {%
0.0     1.31223940696923
0.20188333742746575     1.21007492897111
0.2660462743523159     1.1757490856785
0.4679023555948115     1.06248071761609
0.532093855771953     1.02500504849744
0.7339673659647336     0.903730364984265
0.7981525585732263     0.864198733589962
1.0     0.739942207394063
};
\addlegendentry{{\scriptsize $\vartheta^{\text{(3)}}$}}

\addplot [very thick,dia3, mark options={solid, draw=\tcolorshade, line width=0.5pt}]
table {%
0.0     1.2748957473517
0.20185997498365138     1.17751960187961
0.2660299566628939     1.14474382350848
0.467890185399922     1.03654441866894
0.5320702441477826     1.00054817626141
0.7339854190801681     0.883728350754103
0.7981925671006774     0.845997572594017
1.0     0.726587060819165
};
\addlegendentry{{\scriptsize $\vartheta^{\text{(3)}}$}}

\addplot [very thick,dia4, mark options={solid, draw=\tcolorshade, line width=0.5pt}]
table {%
0.0     1.26087052954619
0.20183988927754767     1.17010811710437
0.26602049954314666     1.13967226842139
0.46790487067223624     1.03762141124934
0.5320475928010207     1.00358679112872
0.7340559311293323     0.892292312134544
0.7982816272747595     0.855467146108193
1.0     0.739262158918281
};
\addlegendentry{{\scriptsize $\vartheta^{\text{(3)}}$}}

\addplot [very thick,dia5, mark options={solid, draw=\tcolorshade, line width=0.5pt}]
table {%
0.0     1.26440691715616
0.20190041499246705     1.18222322506243
0.26608336293818485     1.15409968707886
0.4679375311265276     1.05981424978114
0.5321713590851198     1.02804960065001
0.7340911877736388     0.923598696072711
0.798116275977273     0.88958769148269
1.0     0.777448734470607
};
\addlegendentry{{\scriptsize $\vartheta^{\text{(3)}}$}}

\end{axis}
\end{tikzpicture}}
\put( 8.9, 0.1){(d)}
}

\put(1.0,0.0){\includegraphics[width=13.5cm]{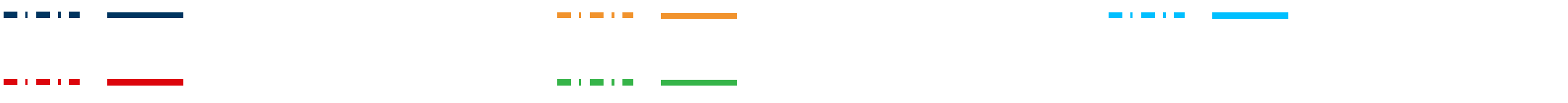}}

\put(2.8,0.68){\small with $\lambda_{\text{ax}}=1.0$}
\put(2.8,0.09){\small with $\lambda_{\text{ax}}=1.1$}
\put(7.58,0.68){\small with $\lambda_{\text{ax}}=1.2$}
\put(7.58,0.09){\small with $\lambda_{\text{ax}}=1.3$}
\put(12.25,0.68){\small with $\lambda_{\text{ax}}=1.4$}

\linethickness{0.4mm}
\put(0.9,1.13){\color{dia_grau_hell}\line(1,0){13.72}}
\put(14.6,1.13){\color{dia_grau_hell}\line(0,-1){1.25}}
\put(14.6,-0.1){\color{dia_grau_hell}\line(-1,0){13.72}}
\put(0.9,-0.1){\color{dia_grau_hell}\line(0,1){1.25}}

\end{picture}
\caption{Comparison of results from optimizations with different axial stretches $\lambda_{\text{ax}}$ as boundary condition. 
Distribution of
(a)~circumferential Cauchy stress $\sigma^{\text{(cir)}}$,
(b)~axial stress/~driving force $\sigma^{\text{(ax)}}$/~$\phi^{(2)}$,
(c)~radial driving force $\phi^{(3)}$ which directly influences the stresses in circumferential direction, and
(d)~growth factors $\vartheta^{(2)}$ (axial) and $\vartheta^{(3)}$ (radial).
Optimizations show convincing results independent from the value of the axial stretch $\lambda_{\text{ax}}$, however some small gradient is still visible in the distributions of circumferential stresses.
\label{fig:variation_lambda}}
\end{figure}
%
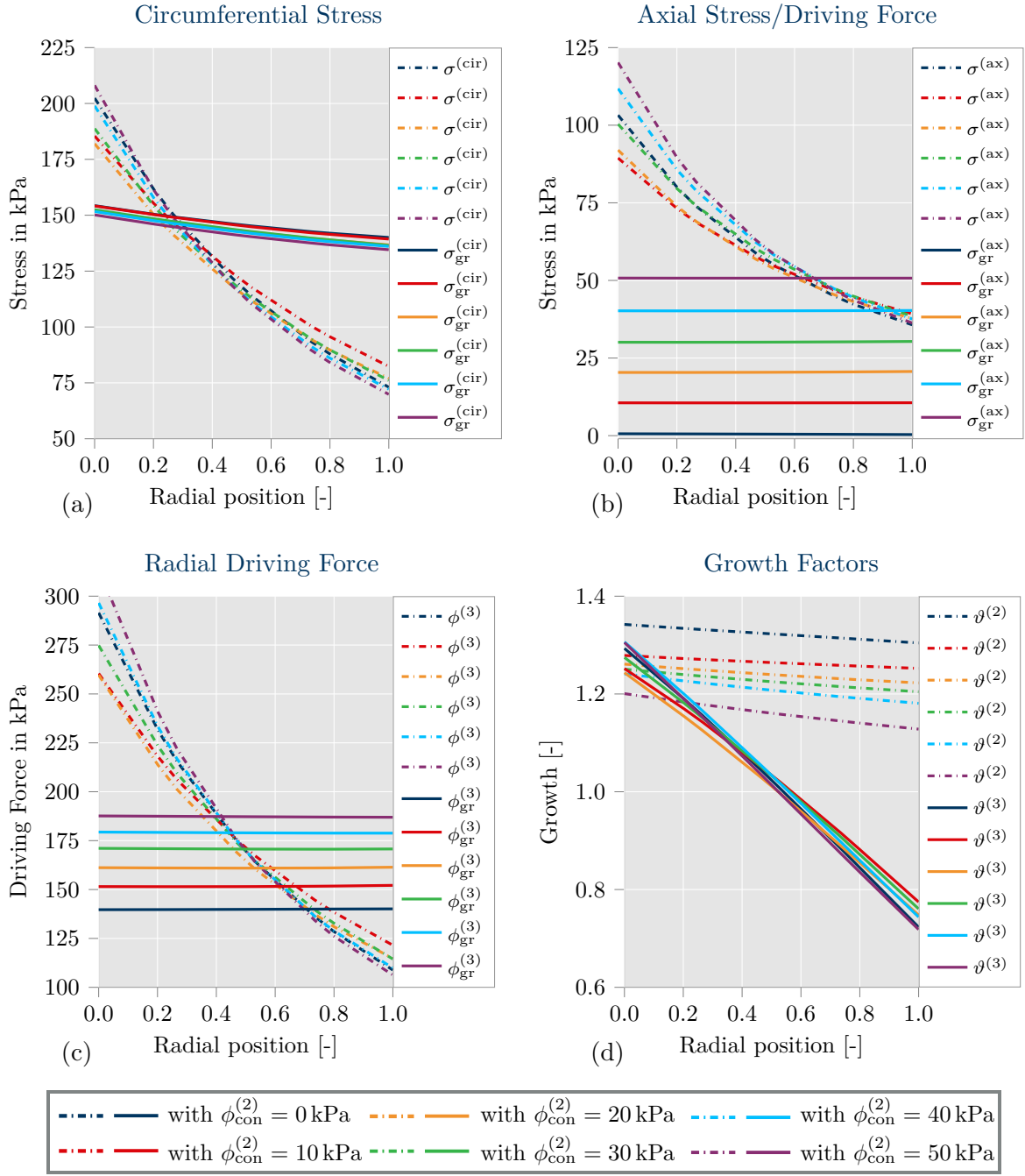
\begin{figure}[!h]
\unitlength1cm
\begin{picture}(15.5,18.5)
\put( 2.2, 18.1){\color{rub_blue}\LARGE Variation of Growth Convergence Value $\phi^{(2)}_{\text{con}}$}

\put( 1.4, 9.8){
\put( 0.5, 7.5){\color{rub_blue} Circumferential Stress}
\put( -1.5, 0.0){
\begin{tikzpicture}
\begin{axis}[
width=6.0cm,
height=7.5cm,
xlabel={{Radial position [-]}},
xmin=0, xmax=1,
xtick={0.2, 0.4, 0.6, 0.8, 1.0},
extra x ticks={0.0},
extra x tick style={%
    grid=major,
},
ylabel={{Stress in kPa}},
ymin=50, ymax=225,
ytick={50, 75, 100, 125, 150, 175, 200, 225},
axis background/.style={fill=white!89.80392156862746!black},
axis line style={white!60.0!black},
axis x line*=bottom,
axis y line*=left,
tick align=outside,
tick pos=left,
scaled x ticks=false,
x grid style={white},
xmajorgrids,
x tick label style={
  font=\small,
  /pgf/number format/.cd,
  set decimal separator={.},
    fixed,
    fixed zerofill,
    precision=1,
  /tikz/.cd
},
xlabel style={at={(0.5,-0.1)},font=\small},
ylabel style={at={(-0.20,0.5)},font=\small},
y grid style={white},
ymajorgrids,
y tick label style={
  font=\small,
},
legend cell align={left},
legend style={at={(1.0,1.0)}, anchor=north west, draw=white!60.0!black, row sep=0.000cm},
clip mode=individual,
]

\addplot [dash dot,very thick,dia1, mark options={solid, draw=\tcolorshade, line width=0.5pt}]
table {%
0.0     202.40645100786
0.20185596580163762     161.440541541506
0.26601054957955494     150.520485037003
0.4678399874118748     121.89662425841
0.5320028392220363     114.150243808358
0.7338491076444759     93.6750860178975
0.7980690667767155     88.0700034081744
1.0     73.0619601368322
};
\addlegendentry{{\scriptsize $\sigma^{\text{(cir)}}$}}

\addplot [dash dot,very thick,dia2, mark options={solid, draw=\tcolorshade, line width=0.5pt}]
table {%
0.0     185.445027125629
0.20188173903445528     155.166232865602
0.26605928202680235     146.794043585971
0.4679374876636629     124.16711479284
0.5320879598842401     117.845834288161
0.7339746338835688     100.581055319048
0.7980839091891314     95.7651614032747
1.0     82.3527824994002 
};
\addlegendentry{{\scriptsize $\sigma^{\text{(cir)}}$}}

\addplot [dash dot,very thick,dia3, mark options={solid, draw=\tcolorshade, line width=0.5pt}]
table {%
0.0     182.002279161995
0.20190324845516802     150.053080410713
0.2660875373703188     141.391918313869
0.46800322529752464     118.23408683608
0.5321739397598275     111.837313022468
0.7342675658521483     94.5888446485837
0.7983984910765181     89.8136363733975
1.0     77.0018699942771
};
\addlegendentry{{\scriptsize $\sigma^{\text{(cir)}}$}}

\addplot [dash dot,very thick,dia4, mark options={solid, draw=\tcolorshade, line width=0.5pt}]
table {%
0.0     188.775794253536
0.20186026294538864     154.010098988415
0.2660249927613069     144.632775943478
0.4679104598128649     119.827851105289
0.5320936936786088     113.05219746127
0.7339539566238789     94.848373601077
0.798081695581439     89.8629288565757
1.0     76.0891166111722
};
\addlegendentry{{\scriptsize $\sigma^{\text{(cir)}}$}}

\addplot [dash dot,very thick,dia5, mark options={solid, draw=\tcolorshade, line width=0.5pt}]
table {%
0.0     198.922419865058
0.20189714475417825     157.580039544747
0.2660788310235616     146.781949362716
0.46797588809359275     118.655715022751
0.5321551554811583     111.167270345864
0.7340370820244232     91.4749145733778
0.7982149706466702     86.1831517324374
1.0     72.1605670797898
};
\addlegendentry{{\scriptsize $\sigma^{\text{(cir)}}$}}

\addplot [dash dot,very thick,dia6, mark options={solid, draw=\tcolorshade, line width=0.5pt}]
table {%
0.0     208.108233117116
0.2018536709448879     161.122715503698
0.26600724520280783     149.056508060927
0.46787512663220865     118.444043482406
0.5320493156999452     110.42504892618
0.7338835068707057     89.823438013245
0.7980523442315836     84.2651275561462
1.0     69.8603259640899
};
\addlegendentry{{\scriptsize $\sigma^{\text{(cir)}}$}}

\addplot [very thick,dia1, mark options={solid, draw=\tcolorshade, line width=0.5pt}]
table {%
0.0     154.27907828825
0.20185596580163762     150.403584424015
0.26601054957955494     149.292326415394
0.4678399874118748     146.126692307899
0.5320028392220363     145.222986578734
0.7338491076444759     142.655159096935
0.7980690667767155     141.968686690109
1.0     140.04126924289   
};
\addlegendentry{{\scriptsize $\sigma^{\text{(cir)}}_{\text{gr}}$}}

\addplot [very thick,dia2, mark options={solid, draw=\tcolorshade, line width=0.5pt}]
table {%
0.0     154.155848332815
0.20188173903445528     150.277268675708
0.26605928202680235     149.116491775992
0.4679374876636629     145.806197825335
0.5320879598842401     144.844972761744
0.7339746338835688     142.163174270275
0.7980839091891314     141.458625470976
1.0     139.344434778231
};
\addlegendentry{{\scriptsize $\sigma^{\text{(cir)}}_{\text{gr}}$}}

\addplot [very thick,dia3, mark options={solid, draw=\tcolorshade, line width=0.5pt}]
table {%
0.0     152.24553930829
0.20190324845516802     148.184851319651
0.2660875373703188     146.987913984618
0.46800322529752464     143.533204859699
0.5321739397598275     142.529388444292
0.7342675658521483     139.689950762758
0.7983984910765181     138.904889863728
1.0     136.745720625405
};
\addlegendentry{{\scriptsize $\sigma^{\text{(cir)}}_{\text{gr}}$}}

\addplot [very thick,dia4, mark options={solid, draw=\tcolorshade, line width=0.5pt}]
table {%
0.0     152.457491212889
0.20186026294538864     148.411702673446
0.2660249927613069     147.207180138173
0.4679104598128649     143.717977975051
0.5320936936786088     142.703230995828
0.7339539566238789     139.786614124254
0.798081695581439     139.025310394901
1.0     136.507174155947
};
\addlegendentry{{\scriptsize $\sigma^{\text{(cir)}}_{\text{gr}}$}}

\addplot [very thick,dia5, mark options={solid, draw=\tcolorshade, line width=0.5pt}]
table {%
0.0     151.65710689128
0.20189714475417825     147.530180721727
0.2660788310235616     146.348836116623
0.46797588809359275     142.887522531577
0.5321551554811583     141.890516046464
0.7340370820244232     138.998148682295
0.7982149706466702     138.201975982155
1.0     136.034634271022
};
\addlegendentry{{\scriptsize $\sigma^{\text{(cir)}}_{\text{gr}}$}}

\addplot [very thick,dia6, mark options={solid, draw=\tcolorshade, line width=0.5pt}]
table {%
0.0     150.143092312616
0.2018536709448879     146.043097986768
0.26600724520280783     144.84468224808
0.46787512663220865     141.391152690378
0.5320493156999452     140.402399793941
0.7338835068707057     137.592721384218
0.7980523442315836     136.773522931886
1.0     134.577965021303
};
\addlegendentry{{\scriptsize $\sigma^{\text{(cir)}}_{\text{gr}}$}}

\end{axis}
\end{tikzpicture}}
\put( -0.6, 0.1){(a)}
}
\put( -0.15, 9.8){
\put( 9.6, 7.5){\color{rub_blue} Axial Stress/Driving Force}
\put( 8.0, 0.0){
\begin{tikzpicture}
\begin{axis}[
width=6.0cm,
height=7.5cm,
xlabel={{Radial position [-]}},
xmin=0, xmax=1,
xtick={0.2, 0.4, 0.6, 0.8, 1.0},
extra x ticks={0.0},
extra x tick style={%
    grid=major,
},
ylabel={{Stress in kPa}},
ymin=-1, ymax=125,
ytick={0, 25, 50, 75, 100, 125},
axis background/.style={fill=white!89.80392156862746!black},
axis line style={white!60.0!black},
axis x line*=bottom,
axis y line*=left,
tick align=outside,
tick pos=left,
scaled x ticks=false,
x grid style={white},
xmajorgrids,
x tick label style={
  font=\small,
  /pgf/number format/.cd,
  set decimal separator={.},
    fixed,
    fixed zerofill,
    precision=1,
  /tikz/.cd
},
xlabel style={at={(0.5,-0.1)},font=\small},
ylabel style={at={(-0.18,0.5)},font=\small},
y grid style={white},
ymajorgrids,
y tick label style={
  font=\small,
},
legend cell align={left},
legend style={at={(1.0,1.0)}, anchor=north west, draw=white!60.0!black, row sep=0.000cm},
clip mode=individual,
]

\addplot [dash dot,very thick,dia1, mark options={solid, draw=\tcolorshade, line width=0.5pt}]
table {%
0.0     103.236855288882
0.20185596580163762     79.8553238549863
0.26601054957955494     73.8498172481616
0.4678399874118748     58.8400947914352
0.5320028392220363     54.9093138611257
0.7338491076444759     45.0199488101938
0.7980690667767155     42.3931521932223
1.0     35.6907499445511
};
\addlegendentry{{\scriptsize $\sigma^{\text{(ax)}}$}}

\addplot [dash dot,very thick,dia2, mark options={solid, draw=\tcolorshade, line width=0.5pt}]
table {%
0.0     89.4203638144422
0.20188173903445528     73.0535809056314
0.26605928202680235     68.7019637123152
0.4679374876636629     57.5629477130253
0.5320879598842401     54.5668737094595
0.7339746338835688     46.8232871190263
0.7980839091891314     44.758735089593
1.0     39.237377202167
};
\addlegendentry{{\scriptsize $\sigma^{\text{(ax)}}$}}

\addplot [dash dot,very thick,dia3, mark options={solid, draw=\tcolorshade, line width=0.5pt}]
table {%
0.0     91.9908834360966
0.20190324845516802     73.7144091084264
0.2660875373703188     68.9571253956417
0.46800322529752464     56.8612877113191
0.5321739397598275     53.6420318549339
0.7342675658521483     45.4325632895545
0.7983984910765181     43.2432483482375
1.0     37.7160557386889
};
\addlegendentry{{\scriptsize $\sigma^{\text{(ax)}}$}}

\addplot [dash dot,very thick,dia4, mark options={solid, draw=\tcolorshade, line width=0.5pt}]
table {%
0.0     100.346513590995
0.20186026294538864     79.4349157025472
0.2660249927613069     73.9955804266314
0.4679104598128649     60.2801889044365
0.5320936936786088     56.6678157891642
0.7339539566238789     47.4097249667474
0.798081695581439     44.9788018803434
1.0     38.4729031815952
};
\addlegendentry{{\scriptsize $\sigma^{\text{(ax)}}$}}

\addplot [dash dot,very thick,dia5, mark options={solid, draw=\tcolorshade, line width=0.5pt}]
table {%
0.0     111.78808170041
0.20189714475417825     85.473721725696
0.2660788310235616     78.8544174605035
0.46797588809359275     62.2976899077958
0.5321551554811583     58.0449115775537
0.7340370820244232     47.3427067820719
0.7982149706466702     44.5735135522721
1.0     37.5656070180011
};
\addlegendentry{{\scriptsize $\sigma^{\text{(ax)}}$}}

\addplot [dash dot,very thick,dia6, mark options={solid, draw=\tcolorshade, line width=0.5pt}]
table {%
0.0     120.20037256348
0.2018536709448879     89.3373614532259
0.26600724520280783     81.6799011641565
0.46787512663220865     63.0623784867781
0.5320493156999452     58.3445229238453
0.7338835068707057     46.7831552050691
0.7980523442315836     43.7401966602047
1.0     36.2703326480961
};
\addlegendentry{{\scriptsize $\sigma^{\text{(ax)}}$}}

\addplot [very thick,dia1, mark options={solid, draw=\tcolorshade, line width=0.5pt}]
table {%
0.0     0.600906337242439
0.20185596580163762     0.560369807131971
0.26601054957955494     0.54646822529981
0.4678399874118748     0.499989156165868
0.5320028392220363     0.483738167402359
0.7338491076444759     0.429107801045305
0.7980690667767155     0.409195906721596
1.0     0.343531836128363 
};
\addlegendentry{{\scriptsize $\sigma^{\text{(ax)}}_{\text{gr}}$}}

\addplot [very thick,dia2, mark options={solid, draw=\tcolorshade, line width=0.5pt}]
table {%
0.0     10.5735051273896
0.20188173903445528     10.5579531207898
0.26605928202680235     10.5558917014967
0.4679374876636629     10.55617840364
0.5320879598842401     10.5589107593835
0.7339746338835688     10.5733022347488
0.7980839091891314     10.5775700518951
1.0     10.5995864147058
};
\addlegendentry{{\scriptsize $\sigma^{\text{(ax)}}_{\text{gr}}$}}

\addplot [very thick,dia3, mark options={solid, draw=\tcolorshade, line width=0.5pt}]
table {%
0.0     20.3592957766888
0.20190324845516802     20.3529672091243
0.2660875373703188     20.3583272041718
0.46800322529752464     20.3967016485191
0.5321739397598275     20.4159774551185
0.7342675658521483     20.4998373357394
0.7983984910765181     20.5336114179533
1.0     20.6670557377332
};
\addlegendentry{{\scriptsize $\sigma^{\text{(ax)}}_{\text{gr}}$}}

\addplot [very thick,dia4, mark options={solid, draw=\tcolorshade, line width=0.5pt}]
table {%
0.0     30.0887460053412
0.20186026294538864     30.0772584166402
0.2660249927613069     30.0803930918491
0.4679104598128649     30.1110774478006
0.5320936936786088     30.1275488888956
0.7339539566238789     30.2012589181098
0.798081695581439     30.2280272533226
1.0     30.3434471334284
};
\addlegendentry{{\scriptsize $\sigma^{\text{(ax)}}_{\text{gr}}$}}

\addplot [very thick,dia5, mark options={solid, draw=\tcolorshade, line width=0.5pt}]
table {%
0.0     40.2129419176433
0.20189714475417825     40.1932112026716
0.2660788310235616     40.192072535374
0.46797588809359275     40.2038193955766
0.5321551554811583     40.2117892850373
0.7340370820244232     40.2488801954717
0.7982149706466702     40.2632806862358
1.0     40.3107180828595
};
\addlegendentry{{\scriptsize $\sigma^{\text{(ax)}}_{\text{gr}}$}}

\addplot [very thick,dia6, mark options={solid, draw=\tcolorshade, line width=0.5pt}]
table {%
0.0     50.7570574695111
0.2018536709448879     50.7333076925876
0.26600724520280783     50.7293771021341
0.46787512663220865     50.7262446441154
0.5320493156999452     50.7273200980732
0.7338835068707057     50.7341611270372
0.7980523442315836     50.736192971583
1.0     50.7353254450391
};
\addlegendentry{{\scriptsize $\sigma^{\text{(ax)}}_{\text{gr}}$}}

\end{axis}
\end{tikzpicture}}
\put( 8.9, 0.1){(b)}
}

\put( 1.4, 1.5){
\put( 0.65, 7.5){\color{rub_blue} Radial Driving Force}
\put( -1.5, 0.0){
\begin{tikzpicture}
\begin{axis}[
width=6.0cm,
height=7.5cm,
xlabel={{Radial position [-]}},
xmin=0, xmax=1,
xtick={0.2, 0.4, 0.6, 0.8, 1.0},
extra x ticks={0.0},
extra x tick style={%
    grid=major,
},
ylabel={{Driving Force in kPa}},
ymin=100, ymax=300,
ytick={100, 125, 150, 175, 200, 225, 250, 275, 300},
axis background/.style={fill=white!89.80392156862746!black},
axis line style={white!60.0!black},
axis x line*=bottom,
axis y line*=left,
tick align=outside,
tick pos=left,
scaled x ticks=false,
x grid style={white},
xmajorgrids,
x tick label style={
  font=\small,
  /pgf/number format/.cd,
  set decimal separator={.},
    fixed,
    fixed zerofill,
    precision=1,
  /tikz/.cd
},
xlabel style={at={(0.5,-0.1)},font=\small},
ylabel style={at={(-0.20,0.5)},font=\small},
y grid style={white},
ymajorgrids,
y tick label style={
  font=\small,
},
legend cell align={left},
legend style={at={(1.0,1.0)}, anchor=north west, draw=white!60.0!black, row sep=0.000cm},
clip mode=individual,
]

\addplot [dash dot,very thick,dia1, mark options={solid, draw=\tcolorshade, line width=0.5pt}]
table {%
0.0     291.496759389624
0.20185596580163762     231.434579362786
0.26601054957955494     215.585513751923
0.4678399874118748     175.160266924958
0.5320028392220363     164.276712123709
0.7338491076444759     136.30979427123
0.7980690667767155     128.688233588243
1.0     108.836243729195 
};
\addlegendentry{{\scriptsize $\phi^{(3)}$}}

\addplot [dash dot,very thick,dia2, mark options={solid, draw=\tcolorshade, line width=0.5pt}]
table {%
0.0     260.45883134254
0.20188173903445528     217.90669415006
0.26605928202680235     206.24674565653
0.4679374876636629     175.685348572204
0.5320879598842401     167.206409453907
0.7339746338835688     144.706261965625
0.7980839091891314     138.524122241015
1.0     121.517031323698
};
\addlegendentry{{\scriptsize $\phi^{(3)}$}}

\addplot [dash dot,very thick,dia3, mark options={solid, draw=\tcolorshade, line width=0.5pt}]
table {%
0.0     259.709026166103
0.20190324845516802     213.584761506965
0.2660875373703188     201.242554757464
0.46800322529752464     169.168116979206
0.5321739397598275     160.374918968129
0.7342675658521483     137.422964334919
0.7983984910765181     131.115020470248
1.0     114.75055247695
};
\addlegendentry{{\scriptsize $\phi^{(3)}$}}

\addplot [dash dot,very thick,dia4, mark options={solid, draw=\tcolorshade, line width=0.5pt}]
table {%
0.0     274.825833716827
0.20186026294538864     223.331559777537
0.2660249927613069     209.581223817345
0.4679104598128649     174.269178568597
0.5320936936786088     164.711551080304
0.7339539566238789     139.687379000439
0.798081695581439     132.945550784821
1.0     114.450050481921
};
\addlegendentry{{\scriptsize $\phi^{(3)}$}}

\addplot [dash dot,very thick,dia5, mark options={solid, draw=\tcolorshade, line width=0.5pt}]
table {%
0.0     296.554064002098
0.20189714475417825     233.140657700582
0.2660788310235616     216.838600863218
0.46797588809359275     175.334526949714
0.5321551554811583     164.413370113847
0.7340370820244232     136.425497502448
0.7982149706466702     129.01102659761
1.0     109.882025574251
};
\addlegendentry{{\scriptsize $\phi^{(3)}$}}

\addplot [dash dot,very thick,dia6, mark options={solid, draw=\tcolorshade, line width=0.5pt}]
table {%
0.0     314.097242046096
0.2018536709448879     240.667068054526
0.26600724520280783     222.02156289818
0.46787512663220865     176.015322350171
0.5320493156999452     164.063648557093
0.7338835068707057     134.327278306055
0.7980523442315836     126.287304743582
1.0     106.261180384731
};
\addlegendentry{{\scriptsize $\phi^{(3)}$}}

\addplot [very thick,dia1, mark options={solid, draw=\tcolorshade, line width=0.5pt}]
table {%
0.0     139.694154700653
0.20185596580163762     139.740866900666
0.26601054957955494     139.761556660596
0.4678399874118748     139.846164195451
0.5320028392220363     139.878047501233
0.7338491076444759     139.985661875538
0.7980690667767155     140.019506544876
1.0     140.104560412817 
};
\addlegendentry{{\scriptsize $\phi^{(3)}_{\text{gr}}$}}

\addplot [very thick,dia2, mark options={solid, draw=\tcolorshade, line width=0.5pt}]
table {%
0.0     151.500193763472
0.20188173903445528     151.455702939295
0.26605928202680235     151.450148527158
0.4679374876636629     151.481033014502
0.5320879598842401     151.509226123688
0.7339746338835688     151.679823259094
0.7980839091891314     151.764883209599
1.0     152.131069536724
};
\addlegendentry{{\scriptsize $\phi^{(3)}_{\text{gr}}$}}

\addplot [very thick,dia3, mark options={solid, draw=\tcolorshade, line width=0.5pt}]
table {%
0.0     161.170564671137
0.20190324845516802     161.0408779957
0.2660875373703188     161.005384582616
0.46800322529752464     160.936032220642
0.5321739397598275     160.932751160621
0.7342675658521483     161.013734629793
0.7983984910765181     161.075117698078
1.0     161.400892998121
};
\addlegendentry{{\scriptsize $\phi^{(3)}_{\text{gr}}$}}

\addplot [very thick,dia4, mark options={solid, draw=\tcolorshade, line width=0.5pt}]
table {%
0.0     171.05682750986
0.20186026294538864     170.897261860542
0.2660249927613069     170.846814594179
0.4679104598128649     170.706154146142
0.5320936936786088     170.671369273813
0.7339539566238789     170.617359036641
0.798081695581439     170.624348138224
1.0     170.738163765021
};
\addlegendentry{{\scriptsize $\phi^{(3)}_{\text{gr}}$}}

\addplot [very thick,dia5, mark options={solid, draw=\tcolorshade, line width=0.5pt}]
table {%
0.0     179.369545010774
0.20189714475417825     179.214739117066
0.2660788310235616     179.163566715088
0.46797588809359275     179.010261287643
0.5321551554811583     178.966661800163
0.7340370820244232     178.862315500965
0.7982149706466702     178.842022706377
1.0     178.828129854793
};
\addlegendentry{{\scriptsize $\phi^{(3)}_{\text{gr}}$}}

\addplot [very thick,dia6, mark options={solid, draw=\tcolorshade, line width=0.5pt}]
table {%
0.0     187.629151379577
0.2018536709448879     187.484442001743
0.26600724520280783     187.43337147414
0.46787512663220865     187.269865800506
0.5320493156999452     187.218180495505
0.7338835068707057     187.074913145363
0.7980523442315836     187.035374233085
1.0     186.949906847262
};
\addlegendentry{{\scriptsize $\phi^{(3)}_{\text{gr}}$}}

\end{axis}
\end{tikzpicture}}
\put( -0.6, 0.1){(c)}
}
\put( -0.15, 1.5){
\put( 10.6, 7.5){\color{rub_blue} Growth Factors}
\put( 8.0, 0.0){
\begin{tikzpicture}
\begin{axis}[
width=6.0cm,
height=7.5cm,
xlabel={{Radial position [-]}},
xmin=0, xmax=1,
xtick={0.2, 0.4, 0.6, 0.8, 1.0},
extra x ticks={0.0},
extra x tick style={%
    grid=major,
},
ylabel={{Growth [-]}},
ymin=0.6, ymax=1.4,
ytick={0.6, 0.8, 1.0, 1.2, 1.4},
axis background/.style={fill=white!89.80392156862746!black},
axis line style={white!60.0!black},
axis x line*=bottom,
axis y line*=left,
tick align=outside,
tick pos=left,
scaled x ticks=false,
x grid style={white},
xmajorgrids,
x tick label style={
  font=\small,
  /pgf/number format/.cd,
  set decimal separator={.},
    fixed,
    fixed zerofill,
    precision=1,
  /tikz/.cd
},
y tick label style={
  font=\small,
  /pgf/number format/.cd,
  set decimal separator={.},
    fixed,
    fixed zerofill,
    precision=1,
  /tikz/.cd
},
xlabel style={at={(0.5,-0.1)},font=\small},
ylabel style={at={(-0.18,0.5)},font=\small},
y grid style={white},
ymajorgrids,
y tick label style={
  font=\small,
  /pgf/number format/.cd,
  set decimal separator={.},
    fixed,
    fixed zerofill,
    precision=1,
  /tikz/.cd
},
legend cell align={left},
legend style={at={(1.0,1.0)}, anchor=north west, draw=white!60.0!black, row sep=0.027cm},
clip mode=individual,
]

\addplot [dash dot,very thick,dia1, mark options={solid, draw=\tcolorshade, line width=0.5pt}]
table {%
0.0     1.34221314682493
0.20185596580163762     1.33409311861651
0.26601054957955494     1.33163242652029
0.4678399874118748     1.32409984631628
0.5320028392220363     1.32175700638126
0.7338491076444759     1.31442390757258
0.7980690667767155     1.31201472254329
1.0     1.30429587761818       
};
\addlegendentry{{\scriptsize $\vartheta^{\text{(2)}}$}}

\addplot [dash dot,very thick,dia2, mark options={solid, draw=\tcolorshade, line width=0.5pt}]
table {%
0.0     1.278757852652
0.20188173903445528     1.27240918320321
0.26605928202680235     1.27056088688652
0.4679374876636629     1.2650587560059
0.5320879598842401     1.26343974498674
0.7339746338835688     1.25852424617659
0.7980839091891314     1.25695928372052
1.0     1.25232661527326
};
\addlegendentry{{\scriptsize $\vartheta^{\text{(2)}}$}}

\addplot [dash dot,very thick,dia3, mark options={solid, draw=\tcolorshade, line width=0.5pt}]
table {%
0.0     1.26099560756839
0.20190324845516802     1.25166617432102
0.2660875373703188     1.24894666730084
0.46800322529752464     1.24087088868083
0.5321739397598275     1.23844898510093
0.7342675658521483     1.23117422988934
0.7983984910765181     1.22896712440975
1.0     1.22262594270668
};
\addlegendentry{{\scriptsize $\vartheta^{\text{(2)}}$}}

\addplot [dash dot,very thick,dia4, mark options={solid, draw=\tcolorshade, line width=0.5pt}]
table {%
0.0     1.25065895394802
0.20186026294538864     1.23956839199955
0.2660249927613069     1.23628549612733
0.4679104598128649     1.22657788043256
0.5320936936786088     1.22369005490518
0.7339539566238789     1.21510902546084
0.798081695581439     1.212441957211
1.0     1.20448980393222
};
\addlegendentry{{\scriptsize $\vartheta^{\text{(2)}}$}}

\addplot [dash dot,very thick,dia5, mark options={solid, draw=\tcolorshade, line width=0.5pt}]
table {%
0.0     1.24111188550907
0.20189714475417825     1.226656384883
0.2660788310235616     1.22237563349988
0.46797588809359275     1.20967462697717
0.5321551554811583     1.20585785592513
0.7340370820244232     1.19445565338882
0.7982149706466702     1.19103999762429
1.0     1.18090473113939  
};
\addlegendentry{{\scriptsize $\vartheta^{\text{(2)}}$}}

\addplot [dash dot,very thick,dia6, mark options={solid, draw=\tcolorshade, line width=0.5pt}]
table {%
0.0     1.20067461301757
0.2018536709448879     1.18341175285061
0.26600724520280783     1.17825348595205
0.46787512663220865     1.16295979186995
0.5320493156999452     1.15837081640784
0.7338835068707057     1.14470998938532
0.7980523442315836     1.14054428239122
1.0     1.12808404703122
};
\addlegendentry{{\scriptsize $\vartheta^{\text{(2)}}$}}

\addplot [very thick,dia1, mark options={solid, draw=\tcolorshade, line width=0.5pt}]
table {%
0.0     1.29317556140684
0.20185596580163762     1.19008820160038
0.26601054957955494     1.15560633104063
0.4678399874118748     1.04228302466076
0.5320028392220363     1.00499923484429
0.7338491076444759     0.884867733351752
0.7980690667767155     0.845848260575519
1.0     0.723088692685733 
};
\addlegendentry{{\scriptsize $\vartheta^{\text{(3)}}$}}

\addplot [very thick,dia2, mark options={solid, draw=\tcolorshade, line width=0.5pt}]
table {%
0.0     1.25197028263624
0.20188173903445528     1.16987247800855
0.26605928202680235     1.14218039927974
0.4679374876636629     1.04934344348233
0.5320879598842401     1.01845435926102
0.7339746338835688     0.916625415794917
0.7980839091891314     0.882782745267699
1.0     0.774542774770387
};
\addlegendentry{{\scriptsize $\vartheta^{\text{(3)}}$}}

\addplot [very thick,dia3, mark options={solid, draw=\tcolorshade, line width=0.5pt}]
table {%
0.0     1.24351583729471
0.20190324845516802     1.15455372350212
0.2660875373703188     1.12505300687962
0.46800322529752464     1.02753244071154
0.5321739397598275     0.995244008192943
0.7342675658521483     0.889955653164276
0.7983984910765181     0.855613060735612
1.0     0.747758308451389
};
\addlegendentry{{\scriptsize $\vartheta^{\text{(3)}}$}}

\addplot [very thick,dia4, mark options={solid, draw=\tcolorshade, line width=0.5pt}]
table {%
0.0     1.27467795704153
0.20186026294538864     1.18170667033549
0.2660249927613069     1.15084040454281
0.4679104598128649     1.04946689959468
0.5320936936786088     1.01614245055052
0.7339539566238789     0.908434483480362
0.798081695581439     0.872710592994382
1.0     0.7604744496818
};
\addlegendentry{{\scriptsize $\vartheta^{\text{(3)}}$}}

\addplot [very thick,dia5, mark options={solid, draw=\tcolorshade, line width=0.5pt}]
table {%
0.0     1.30650186690935
0.20189714475417825     1.20014205438409
0.2660788310235616     1.16509521627481
0.46797588809359275     1.05209042474821
0.5321551554811583     1.01530153927547
0.7340370820244232     0.898197755970564
0.7982149706466702     0.860734802703801
1.0     0.743769107555335
};
\addlegendentry{{\scriptsize $\vartheta^{\text{(3)}}$}}

\addplot [very thick,dia6, mark options={solid, draw=\tcolorshade, line width=0.5pt}]
table {%
0.0     1.30518913912605
0.2018536709448879     1.18979802727435
0.26600724520280783     1.15237373851029
0.46787512663220865     1.03314851798613
0.5320493156999452     0.99488972136154
0.7338835068707057     0.874709249660542
0.7980523442315836     0.836643638453209
1.0     0.718214551465722
};
\addlegendentry{{\scriptsize $\vartheta^{\text{(3)}}$}}

\end{axis}
\end{tikzpicture}}
\put( 8.9, 0.1){(d)}
}

\put(-0.3,0.0){
\put(1.0,0.0){\includegraphics[width=13.5cm]{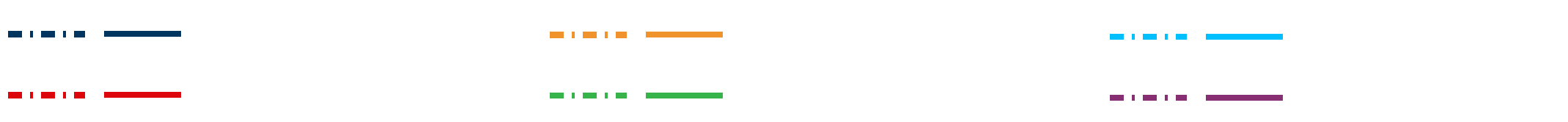}}
\put(2.7,0.68){\small with $\phi^{(2)}_{\text{con}} = 0\,$kPa}
\put(2.7,0.09){\small with $\phi^{(2)}_{\text{con}} = 10\,$kPa}
\put(7.4,0.68){\small with $\phi^{(2)}_{\text{con}} = 20\,$kPa}
\put(7.4,0.09){\small with $\phi^{(2)}_{\text{con}} = 30\,$kPa}
\put(12.2,0.68){\small with $\phi^{(2)}_{\text{con}} = 40\,$kPa}
\put(12.2,0.09){\small with $\phi^{(2)}_{\text{con}} = 50\,$kPa}

\linethickness{0.4mm}
\put(0.9,1.13){\color{dia_grau_hell}\line(1,0){14.32}}
\put(15.2,1.13){\color{dia_grau_hell}\line(0,-1){1.25}}
\put(15.2,-0.1){\color{dia_grau_hell}\line(-1,0){14.32}}
\put(0.9,-0.1){\color{dia_grau_hell}\line(0,1){1.25}}
}

\end{picture}
\caption{Comparison of results from optimizations with different convergence values $\phi^{(2)}_{\text{con}}$ for the driving force  $\phi^{(2)}$. Distribution of
(a)~circumferential Cauchy stress $\sigma^{\text{(cir)}}$,
(b)~axial stress/~driving force $\sigma^{\text{(ax)}}$/~$\phi^{(2)}$,
(c)~radial driving force $\phi^{(3)}$ which directly influences the stresses in circumferential direction, and
(d)~growth factors $\vartheta^{(2)}$ (axial) and $\vartheta^{(3)}$ (radial).
Optimizations show convincing results independent from the convergence values $\phi^{(2)}_{\text{con}}$, however some small gradient is still visible in the distributions of circumferential stresses.
\label{fig:variation_phi}}
\end{figure}
\clearpage

\section{Growth and Active Response with Trace of Cauchy Stress as Driving Force}
\label{sec:act2}
\begin{figure}[!h]
\unitlength1cm
\begin{picture}(15.5,7.0)
\put( -0.1, 0.0){
\put( 0.0, 0.0){
\begin{tikzpicture}
\begin{axis}[
width=5.2cm,
height=7.5cm,
xlabel={{Radial position in $\mu$m}},
xmin=80, xmax=104,
xtick={80, 84, 88, 92, 96, 100, 104},
extra x ticks={0.0},
extra x tick style={%
    grid=major,
},
ylabel={{Stress in kPa}},
ymin=-12, ymax=52,
ytick={-10, 0, 10, 20, 30, 40, 50},
axis background/.style={fill=white!89.80392156862746!black},
axis line style={white!60.0!black},
axis x line*=bottom,
axis y line*=left,
tick align=outside,
tick pos=left,
scaled x ticks=false,
x grid style={white},
xmajorgrids,
x tick label style={
  font=\small,
  /pgf/number format/.cd,
  set decimal separator={.},
    fixed,
    fixed zerofill,
    precision=0,
  /tikz/.cd
},
xlabel style={at={(0.5,-0.1)},font=\small},
ylabel style={at={(-0.18,0.5)},font=\small},
y grid style={white},
ymajorgrids,
y tick label style={
  font=\small,
},
legend cell align={left},
legend style={at={(0.0,0.6)}, anchor=north west, draw=white!60.0!black, row sep=-0.08cm},
clip mode=individual,
]

\addplot [dashed, very thick,rub_green, mark options={solid, draw=\tcolorshade, line width=0.5pt}]
table {%
   80.0147561352955        30.7869728605090
   84.5834291753679        35.8496189799990     
   86.0357786068390        37.3376116316182      
   90.6044516469115        41.6104591381038     
   92.0567865565290        42.8595656899939      
   96.6252075451752        46.3685767216093     
   98.0788896216651        47.4260539383016    
   102.648158111052        50.1576606747691        
};
\addlegendentry{{\scriptsize $\sigma^{\text{(cir)}}$}}

\addplot [dashed, very thick,red, mark options={solid, draw=\tcolorshade, line width=0.5pt}]
table {%
   80.0147561352955       -11.4512206917237 
   84.5834291753679       -8.13746898913026     
   86.0357786068390       -7.04713447183815     
   90.6044516469115       -3.56591055566148     
   92.0567865565290       -2.42660296141613     
   96.6252075451752        1.15556163967543     
   98.0788896216651        2.36155254146225     
   102.648158111052        5.93520774106017                        
};
\addlegendentry{{\scriptsize $\sigma^{\text{(ax)}}$}}

\addplot [very thick,rub_green, mark options={solid, draw=\tcolorshade, line width=0.5pt}]
table {%
   80.0147561352955        37.9584851198603 
   84.5834291753679        35.2991152847807     
   86.0357786068390        34.5062086226423      
   90.6044516469115        32.1198375932654     
   92.0567865565290        31.4110992032437    
   96.6252075451752        29.2941058728913     
   98.0788896216651        28.6754962364277    
   102.648158111052        26.9167791049746                              
};
\addlegendentry{{\scriptsize $\sigma^{\text{(cir)}}_{\text{gr}}$}}

\addplot [very thick,red, mark options={solid, draw=\tcolorshade, line width=0.5pt}]
table {%
   80.0147561352955      -0.120570875993241  
   84.5834291753679      -3.755031858704541E-002
   86.0357786068390      -2.404607740377351E-002   
   90.6044516469115      -4.124626802858161E-002
   92.0567865565290      -5.803032775377131E-002
   96.6252075451752      -0.146375375559695     
   98.0788896216651      -0.170355014969055 
   102.648158111052      -0.148011059196398                       
};
\addlegendentry{{\scriptsize $\sigma^{\text{(ax)}}_{\text{gr}}$}}

\end{axis}
\end{tikzpicture}}

\put( 0.6, 0.0){(a)}
\put( 5.1, 0.0){
\begin{tikzpicture}
\begin{axis}[
width=5.2cm,
height=7.5cm,
xlabel={{Radial position in $\mu$m}},
xmin=80, xmax=104,
xtick={80, 84, 88, 92, 96, 100, 104},
extra x ticks={0.0},
extra x tick style={%
    grid=major,
},
ylabel={{Driving Force in kPa}},
ymin=-12, ymax=60,
ytick={-10, 0, 10, 20, 30, 40, 50, 60},
axis background/.style={fill=white!89.80392156862746!black},
axis line style={white!60.0!black},
axis x line*=bottom,
axis y line*=left,
tick align=outside,
tick pos=left,
scaled x ticks=false,
x grid style={white},
xmajorgrids,
x tick label style={
  font=\small,
  /pgf/number format/.cd,
  set decimal separator={.},
    fixed,
    fixed zerofill,
    precision=0,
  /tikz/.cd
},
xlabel style={at={(0.5,-0.1)},font=\small},
ylabel style={at={(-0.18,0.5)},font=\small},
y grid style={white},
ymajorgrids,
y tick label style={
  font=\small,
},
legend cell align={left},
legend style={at={(0.0,1.0)}, anchor=north west, draw=white!60.0!black, row sep=-0.08cm},
clip mode=individual,
]

\addplot [dashed, very thick,rub_green, mark options={solid, draw=\tcolorshade, line width=0.5pt}]
table {%
   80.0147561352955        6.71435511168554 
   84.5834291753679        17.6477519173427     
   86.0357786068390        21.0435294617977      
   90.6044516469115        31.3173516567155     
   92.0567865565290        34.5072960535701     
   96.6252075451752        44.0178256568734     
   98.0788896216651        47.0858217751474     
   102.648158111052        55.6059949499891                    
};
\addlegendentry{{\scriptsize $\phi^{(3)}$}}

\addplot [dashed, very thick,red, mark options={solid, draw=\tcolorshade, line width=0.5pt}]
table {%
   80.0147561352955       -11.4512206917237 
   84.5834291753679       -8.13746898913026     
   86.0357786068390       -7.04713447183815     
   90.6044516469115       -3.56591055566148     
   92.0567865565290       -2.42660296141613     
   96.6252075451752        1.15556163967543     
   98.0788896216651        2.36155254146225     
   102.648158111052        5.93520774106017                
};
\addlegendentry{{\scriptsize $\phi^{(2)}$}}

\addplot [very thick,rub_green, mark options={solid, draw=\tcolorshade, line width=0.5pt}]
table {%
   80.0147561352955        26.1665826558932
   84.5834291753679        26.2345561040533     
   86.0357786068390        26.2458767864399     
   90.6044516469115        26.2314711092887     
   92.0567865565290        26.2178602013052      
   96.6252075451752        26.1515137744144     
   98.0788896216651        26.1422821738374    
   102.648158111052        26.3380760422433                          
};
\addlegendentry{{\scriptsize $\phi^{(3)}_{\text{gr}}$}}

\addplot [very thick,red, mark options={solid, draw=\tcolorshade, line width=0.5pt}]
table {%
   80.0147561352955      -0.120570875993241  
   84.5834291753679      -3.755031858704541E-002
   86.0357786068390      -2.404607740377351E-002   
   90.6044516469115      -4.124626802858161E-002
   92.0567865565290      -5.803032775377131E-002
   96.6252075451752      -0.146375375559695     
   98.0788896216651      -0.170355014969055 
   102.648158111052      -0.148011059196398        
};
\addlegendentry{{\scriptsize $\phi^{(2)}_{\text{gr}}$}}

\end{axis}
\end{tikzpicture}}

\put( 5.7, 0.0){(b)}
\put( 10.2, 0.0){
\begin{tikzpicture}
\begin{axis}[
width=5.2cm,
height=7.5cm,
xlabel={{Radial position in $\mu$m}},
xmin=80, xmax=104,
xtick={80, 84, 88, 92, 96, 100, 104},
extra x ticks={0.0},
extra x tick style={%
    grid=major,
},
ylabel={{Growth [-]}},
ymin=0.8, ymax=1.4,
ytick={0.8, 0.9, 1.0, 1.1, 1.2, 1.3, 1.4},
axis background/.style={fill=white!89.80392156862746!black},
axis line style={white!60.0!black},
axis x line*=bottom,
axis y line*=left,
tick align=outside,
tick pos=left,
scaled x ticks=false,
x grid style={white},
xmajorgrids,
x tick label style={
  font=\small,
  /pgf/number format/.cd,
  set decimal separator={.},
    fixed,
    fixed zerofill,
    precision=0,
  /tikz/.cd
},
y tick label style={
  font=\small,
  /pgf/number format/.cd,
  set decimal separator={.},
    fixed,
    fixed zerofill,
    precision=1,
  /tikz/.cd
},
xlabel style={at={(0.5,-0.1)},font=\small},
ylabel style={at={(-0.20,0.5)},font=\small},
y grid style={white},
ymajorgrids,
y tick label style={
  font=\small,
},
legend cell align={left},
legend style={at={(0.0,1.0)}, anchor=north west, draw=white!60.0!black, row sep=-0.08cm},
clip mode=individual,
]

\addplot [very thick,rub_green, mark options={solid, draw=\tcolorshade, line width=0.5pt}]
table {%
   80.0147561352955       0.850879522858480 
   84.5834291753679       0.907944050539717     
   86.0357786068390       0.929679233572474    
   90.6044516469115        1.01476955660020     
   92.0567865565290        1.04662453982735      
   96.6252075451752        1.16832444163656     
   98.0788896216651        1.21349395686997      
   102.648158111052        1.38017819905725                              
};
\addlegendentry{{\footnotesize $\vartheta^{\text{(3)}}$}}

\addplot [very thick,red, mark options={solid, draw=\tcolorshade, line width=0.5pt}]
table {%
   80.0147561352955       0.852280601961248   
   84.5834291753679       0.878659648106528     
   86.0357786068390       0.887981738279590   
   90.6044516469115       0.921907166402487     
   92.0567865565290       0.934242390401448      
   96.6252075451752       0.980296345558802     
   98.0788896216651       0.997354606348669    
   102.648158111052        1.06109562162731                      
};
\addlegendentry{{\footnotesize $\vartheta^{\text{(2)}}$}}

\end{axis}
\end{tikzpicture}}

\put( 10.8, 0.0){(c)}
}
\end{picture}
\caption{Distribution of 
(a)~Cauchy stresses $\sigma^{\text{(ax)}}$ and $\sigma^{\text{(cir)}}$, 
(b)~driving forces $\phi^{(2)}$ and $\phi^{(3)}$, and
(c)~growth values $\vartheta^{(2)}$ and $\vartheta^{(3)}$
in circumferential (green) and axial (red) direction over the wall thickness for the fully active material model.
Artery is loaded with an intravascular pressure of 120mmHg. 
Dashed lines show results before growth, solid lines show results after growth. 
The new growth model also generates a homogenized stress distribution for $\sigma^{\text{(ax)}}$ when smooth muscle contraction is activated.
Contrarily, $\sigma^{\text{(cir)}}$ still shows a significant gradient over the wall thickness caused by $\text{tr}(\sigma)$ as driving force $\phi^{(3)}$ which includes the stress in radial direction as contributor to the driving force.
\label{fig:opt_act_app}}
\end{figure}
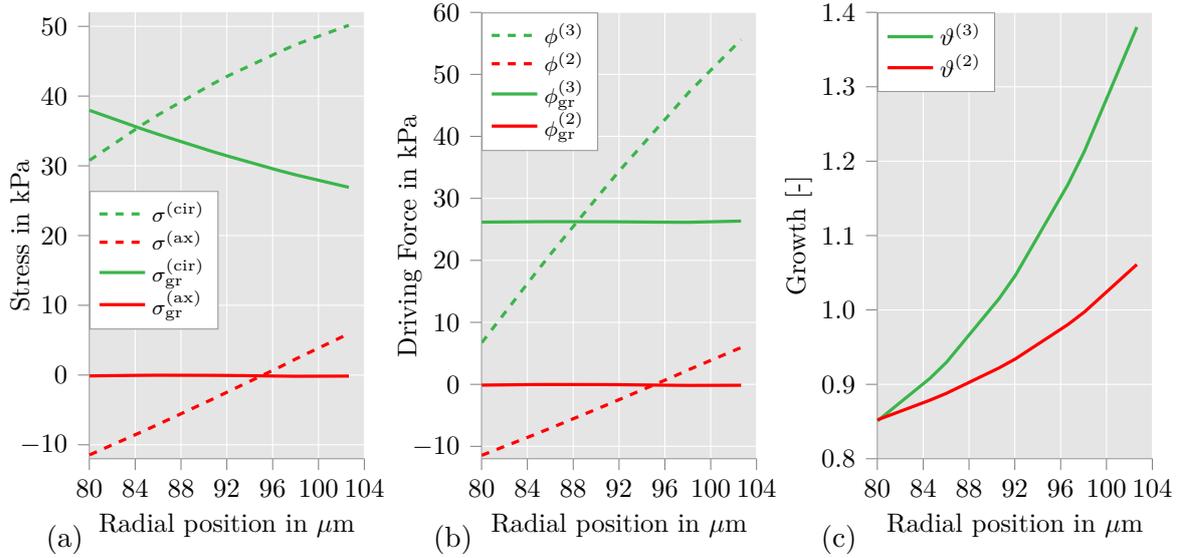

\end{appendix}

\end{document}